\documentclass[screen=true, acmsmall]{acmart}
\usepackage[utf8]{inputenc}
\usepackage[T1]{fontenc}

\newcommand{\packageGraphicx}{\usepackage{graphicx}}
\newcommand{\packageHyperref}{\usepackage{hyperref}}
\newcommand{\renewrmdefault}{\renewcommand{\rmdefault}{ptm}}
\newcommand{\packageRelsize}{\usepackage{relsize}}
\newcommand{\packageAmsmath}{\usepackage{amsmath}}
\newcommand{\packageMathabx}{\usepackage{mathabx}}
\newcommand{\packageWasysym}{
  \let\leftmoon\relax \let\rightmoon\relax \let\fullmoon\relax \let\newmoon\relax \let\diameter\relax
  \usepackage[nointegrals]{wasysym}}
\newcommand{\packageTxfonts}{
  \let\widering\relax
  \let\oldwidebar\widebar
  \let\widebar\relax
  \usepackage{newtxmath}
  \ifx\widebar\relax
    \let\widebar\oldwidebar
  \fi
}
\newcommand{\packageTextcomp}{\usepackage{textcomp}}
\newcommand{\packageFramed}{\usepackage{framed}}
\newcommand{\packageHyphenat}{\usepackage[htt]{hyphenat}}
\newcommand{\packageColor}{\usepackage[usenames,dvipsnames]{color}}
\newcommand{\doHypersetup}{\hypersetup{bookmarks=true,bookmarksopen=true,bookmarksnumbered=true}}
\newcommand{\packageTocstyle}{}
\newcommand{\packageCJK}{\IfFileExists{CJK.sty}{\usepackage{CJK}}{}}
\renewcommand\packageColor\relax
\renewcommand\packageTocstyle\relax
\renewcommand\packageMathabx{\ifx\bigtimes\undefined \usepackage{mathabx} \else \relax \fi}
\renewcommand\packageTxfonts\relax

\renewcommand{\renewrmdefault}{}

\packageGraphicx
\packageHyperref
\renewrmdefault
\packageRelsize
\packageAmsmath
\packageMathabx
\packageWasysym
\packageTxfonts
\packageTextcomp
\packageFramed
\packageHyphenat
\packageColor
\doHypersetup
\packageTocstyle
\packageCJK

\newcommand{\sectionNewpage}{}

\newcommand{\preDoc}{}
\newcommand{\postDoc}{}

\newcommand{\BookRef}[2]{\emph{#2}}
\newcommand{\ChapRef}[2]{\SecRef{#1}{#2}}
\newcommand{\SecRef}[2]{section~#1}
\newcommand{\PartRef}[2]{part~#1}
\newcommand{\BookRefUC}[2]{\BookRef{#1}{#2}}
\newcommand{\ChapRefUC}[2]{\SecRefUC{#1}{#2}}
\newcommand{\SecRefUC}[2]{Section~#1}
\newcommand{\PartRefUC}[2]{Part~#1}

\newcommand{\BookRefLocal}[3]{\hyperref[#1]{\BookRef{#2}{#3}}}
\newcommand{\ChapRefLocal}[3]{\hyperref[#1]{\ChapRef{#2}{#3}}}
\newcommand{\SecRefLocal}[3]{\hyperref[#1]{\SecRef{#2}{#3}}}
\newcommand{\PartRefLocal}[3]{\hyperref[#1]{\PartRef{#2}{#3}}}
\newcommand{\BookRefLocalUC}[3]{\hyperref[#1]{\BookRefUC{#2}{#3}}}
\newcommand{\ChapRefLocalUC}[3]{\hyperref[#1]{\ChapRefUC{#2}{#3}}}
\newcommand{\SecRefLocalUC}[3]{\hyperref[#1]{\SecRefUC{#2}{#3}}}
\newcommand{\PartRefLocalUC}[3]{\hyperref[#1]{\PartRefUC{#2}{#3}}}

\newcommand{\BookRefUN}[1]{\BookRef{}{#1}}

\newcommand{\SecRefUN}[1]{``#1''}

\newcommand{\BookRefLocalUN}[2]{\hyperref[#1]{\BookRefUN{#2}}}

\newcommand{\SecRefLocalUN}[2]{\hyperref[#1]{\SecRefUN{#2}}}

\newcommand{\SectionNumberLink}[2]{\hyperref[#1]{#2}}

\newcommand{\Scribtexttt}[1]{{\texttt{#1}}}

\newcommand{\incolorbox}[2]{{\fboxrule=0pt\fboxsep=0pt\protect\colorbox{#1}{#2}}}

\newcommand{\Smaller}[1]{\textsmaller{#1}}

\newcommand{\planetName}[1]{PLane\hspace{-0.1ex}T}

\newcommand{\Stttextmore}{{\fontencoding{T1}\selectfont>}}
\newcommand{\Stttextless}{{\fontencoding{T1}\selectfont<}}

\makeatletter
\def\empty@finalstrut#1{%
  \unskip\ifhmode\nobreak\fi\vrule\@width\z@\@height\z@\@depth\z@}
\def\no@strut{\global\setbox\@arstrutbox\hbox{%
    \vrule \@height\z@
           \@depth\z@
           \@width\z@}%
    \gdef\@endpbox{\empty@finalstrut\@arstrutbox\par\egroup\hfil}%
}%
\def\yes@strut{\global\setbox\@arstrutbox\hbox{%
    \vrule \@height\arraystretch \ht\strutbox
           \@depth\arraystretch \dp\strutbox
           \@width\z@}%
    \gdef\@endpbox{\@finalstrut\@arstrutbox\par\egroup\hfil}%
}%
\def\@mkpream#1{\@firstamptrue\@lastchclass6
  \let\@preamble\@empty\def\empty@preamble{\add@ins}%
  \let\protect\@unexpandable@protect
  \let\@sharp\relax\let\add@ins\relax
  \let\@startpbox\relax\let\@endpbox\relax
  \@expast{#1}%
  \expandafter\@tfor \expandafter
    \@nextchar \expandafter:\expandafter=\reserved@a\do
       {\@testpach\@nextchar
    \ifcase \@chclass \@classz \or \@classi \or \@classii \or \@classiii
      \or \@classiv \or\@classv \fi\@lastchclass\@chclass}%
  \ifcase \@lastchclass \@acol
      \or \or \@preamerr \@ne\or \@preamerr \tw@\or \or \@acol \fi}
\def\@addamp{%
  \if@firstamp
    \@firstampfalse
    \edef\empty@preamble{\add@ins}%
  \else
    \edef\@preamble{\@preamble &}%
    \edef\empty@preamble{\expandafter\noexpand\empty@preamble &\add@ins}%
  \fi}
\newif\iftw@hlines \tw@hlinesfalse
\def\@xhline{\ifx\reserved@a\hline
               \tw@hlinestrue
             \else\ifx\reserved@a\Hline
               \tw@hlinestrue
             \else
               \tw@hlinesfalse
             \fi\fi
      \iftw@hlines
        \aftergroup\do@after
      \fi
      \ifnum0=`{\fi}%
}
\def\do@after{\emptyrow[\the\doublerulesep]}
\def\emptyrow{\noalign\bgroup\@ifnextchar[\@emptyrow{\@emptyrow[\z@]}}
\def\@emptyrow[#1]{\no@strut\gdef\add@ins{\vrule \@height\z@ \@depth#1 \@width\z@}\egroup%
\empty@preamble\\
\noalign{\yes@strut\gdef\add@ins{\vrule \@height\z@ \@depth\z@ \@width\z@}}%
}
\def\tabrow#1{\noalign\bgroup\@ifnextchar[{\@tabrow{#1}}{\@tabrow{#1}[]}}
\def\@tabrow#1[#2]{\no@strut\egroup#1\ifx.#2.\\\else\\[#2]\fi\noalign{\yes@strut}}
\def\endpltstabular{\crcr\egroup\egroup \egroup}
\expandafter \let \csname endpltstabular*\endcsname = \endpltstabular
\def\pltstabular{\let\@halignto\@empty\@pltstabular}
\@namedef{pltstabular*}#1{\def\@halignto{to#1}\@pltstabular}
\def\@pltstabular{\leavevmode \bgroup \let\@acol\@tabacol
   \let\@classz\@tabclassz
   \let\@classiv\@tabclassiv \let\\\@tabularcr\@stabarray}
\def\@stabarray{\m@th\@ifnextchar[\@sarray{\@sarray[c]}}
\def\@sarray[#1]#2{%
  \bgroup
  \setbox\@arstrutbox\hbox{%
    \vrule \@height\arraystretch\ht\strutbox
           \@depth\arraystretch \dp\strutbox
           \@width\z@}%
  \@mkpream{#2}%
  \edef\@preamble{%
    \ialign \noexpand\@halignto
      \bgroup \@arstrut \@preamble \tabskip\z@skip \cr}%
  \let\@startpbox\@@startpbox \let\@endpbox\@@endpbox
  \let\tabularnewline\\%
    \let\@sharp##%
    \set@typeset@protect
    \lineskip\z@skip\baselineskip\z@skip
    \@preamble}

\makeatother

\newlength{\stabLeft}

\newenvironment{SingleColumn}{\begin{list}{}{\topsep=0pt\partopsep=0pt%
\listparindent=0pt\itemindent=0pt\labelwidth=0pt\leftmargin=0pt\rightmargin=0pt%
\itemsep=0pt\parsep=0pt}\item}{\end{list}}

\newcommand{\Sendabbrev}[1]{#1\@}

\newenvironment{SInsetFlow}{\begin{quote}}{\end{quote}}

\newcommand{\SCodePreSkip}{\vskip\abovedisplayskip}
\newcommand{\SCodePostSkip}{\vskip\belowdisplayskip}
\newenvironment{SCodeFlow}{\SCodePreSkip\begin{list}{}{\topsep=0pt\partopsep=0pt%
\listparindent=0pt\itemindent=0pt\labelwidth=0pt\leftmargin=2ex\rightmargin=2ex%
\itemsep=0pt\parsep=0pt}\item}{\end{list}\SCodePostSkip}

\newcommand{\SVInsetPreSkip}{\vskip\abovedisplayskip}
\newcommand{\SVInsetPostSkip}{\vskip\belowdisplayskip}

\newcommand{\titleAndVersionAndAuthors}[3]{\title{#1\\{\normalsize \SVersionBefore{}#2}}\author{#3}\maketitle}

\newcommand{\titleAndEmptyVersionAndAuthors}[3]{\title{#1}\author{#3}\maketitle}

\newcommand{\SAuthor}[1]{#1}
\newcommand{\SAuthorSep}[1]{\qquad}
\newcommand{\SVersionBefore}[1]{Version }

\newcommand{\SNumberOfAuthors}[1]{}

\let\SOriginalthesubsection\thesubsection
\let\SOriginalthesubsubsection\thesubsubsection

\newcommand{\Ssection}[2]{\section[#1]{#2}\let\thesubsection\SOriginalthesubsection}
\newcommand{\Ssubsection}[2]{\subsection[#1]{#2}\let\thesubsubsection\SOriginalthesubsubsection}

\newcommand{\Ssectionstar}[1]{\section*{#1}\renewcommand*\thesubsection{\arabic{subsection}}\setcounter{subsection}{0}}

\newcommand{\Ssectionstarx}[2]{\Ssectionstar{#2}\phantomsection\addcontentsline{toc}{section}{#1}}

\newcounter{GrouperTemp}

\newcommand{\Snolinkurl}[1]{\nolinkurl{#1}}

\newcommand{\SAuthorinfo}[4]{#1}
\newcommand{\SAuthorPlace}[1]{#1}
\newcommand{\SAuthorEmail}[1]{#1}
\newcommand{\SAuthorOrcid}[1]{#1}

\newcommand{\SConferenceInfo}[2]{}
\newcommand{\SCopyrightYear}[1]{}
\newcommand{\SCopyrightData}[1]{}
\newcommand{\Sdoi}[1]{}

\newcommand{\SCategory}[3]{}
\newcommand{\SCategoryPlus}[4]{}
\newcommand{\STerms}[1]{}
\newcommand{\SKeywords}[1]{}

\newcommand{\AutobibLink}[1]{\color{ACMPurple}{#1}}
\usepackage{savesym}
\savesymbol{iint}
\savesymbol{iiint}
\savesymbol{dddot}
\savesymbol{ddddot}
\savesymbol{overleftrightarrow}
\savesymbol{underrightarrow}
\savesymbol{underleftarrow}
\savesymbol{underleftrightarrow}
\usepackage{amsmath}
\restoresymbol{AMS}{iint}
\restoresymbol{AMS}{iiint}
\restoresymbol{AMS}{dddot}
\restoresymbol{AMS}{ddddot}
\restoresymbol{AMS}{overleftrightarrow}
\restoresymbol{AMS}{underrightarrow}
\restoresymbol{AMS}{underleftarrow}
\restoresymbol{AMS}{underleftrightarrow}
\savesymbol{ulcorner}
\savesymbol{urcorner}
\savesymbol{llcorner}
\savesymbol{lrcorner}
\usepackage{amsfonts}
\restoresymbol{AMS}{ulcorner}
\restoresymbol{AMS}{urcorner}
\restoresymbol{AMS}{llcorner}
\restoresymbol{AMS}{lrcorner}
\newcommand{\Iidentity}[1]{#1}

\newenvironment{AutoBibliography}{\begin{small}}{\end{small}}
\newcommand{\Autobibentry}[1]{\hspace{0.05\linewidth}\parbox[t]{0.95\linewidth}{\parindent=-0.05\linewidth#1\vspace{1.0ex}}}
\newcommand{\Autobibtarget}[1]{\phantomsection#1}

\usepackage{calc}
\newlength{\ABcollength}

\newcommand{\Autobibref}[1]{#1}

\providecommand{\AutobibLink}[1]{#1}

\newcommand{\pseudodoi}[1]{#1}

\newcommand{\NoteBox}[1]{\footnote{#1}}
\newcommand{\NoteContent}[1]{#1}

\newcommand{\FootnoteRef}[1]{}
\newcommand{\FootnoteTarget}[1]{}

\newcommand{\FootnoteBlockContent}[1]{}

\newcommand{\SColorize}[2]{\color{#1}{#2}}

\newcommand{\SHyphen}[1]{#1}

\newcommand{\inColor}[2]{{\SHyphen{\Scribtexttt{\SColorize{#1}{#2}}}}}
\definecolor{PaleBlue}{rgb}{0.90,0.90,1.0}
\definecolor{LightGray}{rgb}{0.90,0.90,0.90}
\definecolor{CommentColor}{rgb}{0.76,0.45,0.12}
\definecolor{ParenColor}{rgb}{0.52,0.24,0.14}
\definecolor{IdentifierColor}{rgb}{0.15,0.15,0.50}
\definecolor{ResultColor}{rgb}{0.0,0.0,0.69}
\definecolor{ValueColor}{rgb}{0.13,0.55,0.13}
\definecolor{OutputColor}{rgb}{0.59,0.00,0.59}

\newcommand{\RktCmt}[1]{\inColor{CommentColor}{#1}}
\newcommand{\RktPn}[1]{\inColor{ParenColor}{#1}}
\newcommand{\RktInBG}[1]{\inColor{ParenColor}{#1}}
\newcommand{\RktSym}[1]{\inColor{IdentifierColor}{#1}}

\newcommand{\RktVal}[1]{\inColor{ValueColor}{#1}}

\newcommand{\RktModLink}[1]{\inColor{blue}{#1}}
\newcommand{\RktRes}[1]{\inColor{ResultColor}{#1}}
\newcommand{\RktOut}[1]{\inColor{OutputColor}{#1}}
\newcommand{\RktMeta}[1]{\inColor{IdentifierColor}{#1}}
\newcommand{\RktMod}[1]{\inColor{black}{#1}}
\newcommand{\RktRdr}[1]{\inColor{black}{#1}}
\newcommand{\RktVarCol}[1]{\inColor{IdentifierColor}{#1}}
\newcommand{\RktVar}[1]{{\RktVarCol{\textsl{#1}}}}
\newcommand{\RktErrCol}[1]{\inColor{red}{#1}}
\newcommand{\RktErr}[1]{{\RktErrCol{\textrm{\textit{#1}}}}}

\newcommand{\RktIn}[1]{\incolorbox{LightGray}{\RktInBG{#1}}}

\newenvironment{RktBlk}{}{}

\newcommand{\RBackgroundLabel}[1]{}

\renewcommand{\titleAndVersionAndAuthors}[3]{\title{#1}#3\maketitle}
\renewcommand{\titleAndEmptyVersionAndAuthors}[3]{\titleAndVersionAndAuthors{#1}{#2}{#3}}

\def\SAuthor#1{\SAutoAuthor#1\SAutoAuthorDone{#1}}
\def\SAutoAuthorDone#1{}
\def\SAutoAuthor{\futurelet\next\SAutoAuthorX}
\def\SAutoAuthorX{\ifx\next\SAuthorinfo \let\Snext\relax \else \let\Snext\SToAuthorDone \fi \Snext}
\def\SToAuthorDone{\futurelet\next\SToAuthorDoneX}
\def\SToAuthorDoneX#1{\ifx\next\SAutoAuthorDone \let\Snext\SAddAuthorInfo \else \let\Snext\SToAuthorDone \fi \Snext}
\newcommand{\SAddAuthorInfo}[1]{\SAuthorinfo{#1}{}{}}

\renewcommand{\SAuthorinfo}[4]{\author{#1}{#2}{#3}{#4}}
\renewcommand{\SAuthorSep}[1]{}
\renewcommand{\SAuthorOrcid}[1]{\orcid{#1}}
\renewcommand{\SAuthorPlace}[1]{\affiliation{#1}}
\renewcommand{\SAuthorEmail}[1]{\email{#1}}

\renewcommand{\SConferenceInfo}[2]{\conferenceinfo{#1}{#2}}
\renewcommand{\SCopyrightYear}[1]{\copyrightyear{#1}}
\renewcommand{\SCopyrightData}[1]{\copyrightdata{#1}}

\renewcommand{\SCategory}[3]{\category{#1}{#2}{#3}}
\renewcommand{\SCategoryPlus}[4]{\category{#1}{#2}{#3}[#4]}
\renewcommand{\STerms}[1]{\terms{#1}}
\renewcommand{\SKeywords}[1]{\keywords{#1}}

\usepackage{simplebnf}
\usepackage{longtable}

\NewCommandCopy{\oldbnfgrammar}{\bnfgrammar}
\NewCommandCopy{\endoldbnfgrammar}{\endbnfgrammar}

\RenewDocumentCommand{\bnfexpr}{m}{\textit{#1}}
\RenewDocumentEnvironment{bnfgrammar}{O{lrclr} O{\|\|} O{\|}}
{\begin{oldbnfgrammar}[#1][#2][#3]}
{\end{oldbnfgrammar}}

\usepackage{semantic}
\newcommand{\lcstyle}[1]{\textnormal{\textbf{#1}}}
\reservestyle{\lc}{\lcstyle}
\lc{module,begin,set!,define,new-frames,assignment,jump,jump-if,compare,+,-,
rsp,rbp,rax,r12,r13,r14,r15,with-label}

\usepackage[breakable]{tcolorbox}

\newcommand{\LangBoxLabel}[1]{
\vspace{-4ex}
\begin{flushright}
  \begin{tcolorbox}[halign=flush right,boxsep=2pt,boxrule=.4pt,left=0pt,right=0pt,top=0pt,bottom=-1pt,middle=1pt,arc=1pt,outer arc=1pt,tcbox width=auto limited,capture=hbox]
  \smaller{#1}
  \end{tcolorbox}
\end{flushright}
}

\newcommand{\LangBox}[2]{
\begin{tcolorbox}[title=#1,breakable,size=minimal,boxsep=0mm,detach title,after
 upper={\vspace{-4ex}\tcbtitle\vspace{-6pt}}]
 #2
\end{tcolorbox}
}

\setcopyright{none}
\acmDOI{}
\acmISBN{}
\acmPrice{}
\acmBooktitle{Workshop on Scheme and Functional Programming (Scheme)}
\begin{document}
\preDoc

\acmConference{Scheme and Functional Programming Workshop (Scheme)}{September 11{--}16, 2022}{Ljubljana, Slovenia}

\acmYear{2022}

\acmMonth{9}

\begin{abstract}We present the design and implementation of a macro{-}embedding of a family of
compiler intermediate languages, from a Scheme{-}like language to x86{-}64, into
Racket.
This embedding is used as part of a testing framework for a compilers course to
derive interpreters for all the intermediate languages.
The embedding implements features including safe, functional abstractions as
well as unsafe assembly features, and the interactions between the two at
various intermediate stages.

This paper aims to demonstrate language{-}oriented techniques and abstractions
for implementing (1) a large family of languages and (2) interoperability
between low{-} and high{-}level languages.
The primary strength of this approach is the high degree of code reuse and
interoperability compared to implementing each interpreter separately.
The design emphasizes modularity and compositionality of an open set of language
features by local macro expansion into a single host language, rather than
implementing a language pre{-}defined by a closed set of features.
This enables reuse from both the host language (Racket) and between intermediate
languages, and enables interoperability between high{-} and low{-}level features,
simplifying development of the intermediate language semantics.
It also facilitates extending or redefining individual language features in
intermediate languages, and exposing multiple interfaces to the embedded
languages.\end{abstract}\titleAndEmptyVersionAndAuthors{Macro{-}embedding Compiler Intermediate Languages in Racket}{}{\SNumberOfAuthors{1}\SAuthor{\SAuthorinfo{William J. Bowman}{\SAuthorOrcid{\Iidentity{0000-0002-6402-4840}}}{\SAuthorPlace{\department{Computer Science}\institution{University of British Columbia}\city{Vancouver}\state{British Columbia}\country{Canada}}}{\SAuthorEmail{wjb@williamjbowman.com}}}}
\label{t:x28part_x22Macrox2dembeddingx5fCompilerx5fIntermediatex5fLanguagesx5finx5fRacketx22x29}

\noindent 

\noindent

\sectionNewpage

\Ssection{Introduction}{Introduction}\label{t:x28part_x22secx3aintrox22x29}

Recently, we faced an engineering challenge: we wanted to design and implement
interpreters for a family of 126 languages.
These languages share considerable structure and features sets, and many are
nearly idententical; changes to one language are not independent of any other.
Some are high level, functional, safe languages, while some are low level,
unsafe and equivalent to a subset of x86{-}64 assembly (x64).
Most are somewhere in between{---}intermediate languages in which safe, high{-}level
features interoperable with unsafe, low{-}level features.
The language definitions may change as the specification evolves, so code
maintence is a primary concern.

Our users could use these interpreters to test an optimizing nanopass
compiler\Autobibref{~(\hyperref[t:x28autobib_x22Abdulaziz_GhuloumAn_Incremental_Approach_to_Compiler_ConstructionIn_Procx2e_Scheme_Workshop2006httpx3ax2fx2fscheme2006x2ecsx2euchicagox2eedux2f11x2dghuloumx2epdfx22x29]{\AutobibLink{Ghuloum}} \hyperref[t:x28autobib_x22Abdulaziz_GhuloumAn_Incremental_Approach_to_Compiler_ConstructionIn_Procx2e_Scheme_Workshop2006httpx3ax2fx2fscheme2006x2ecsx2euchicagox2eedux2f11x2dghuloumx2epdfx22x29]{\AutobibLink{2006}}; \hyperref[t:x28autobib_x22Andrew_Wx2e_Keep_and_Rx2e_Kent_DybvigA_nanopass_framework_for_commercial_compiler_developmentIn_Procx2e_International_Conference_on_Functional_Programming_x28ICFPx292013doix3a10x2e1145x2f2500365x2e2500618x22x29]{\AutobibLink{Keep and Dybvig}} \hyperref[t:x28autobib_x22Andrew_Wx2e_Keep_and_Rx2e_Kent_DybvigA_nanopass_framework_for_commercial_compiler_developmentIn_Procx2e_International_Conference_on_Functional_Programming_x28ICFPx292013doix3a10x2e1145x2f2500365x2e2500618x22x29]{\AutobibLink{2013}})} from a small Scheme{-}like language to
x64.
At the time, the users relied on a combination of fragile syntactic tests and
end{-}to{-}end tests of the entire compiler.
But syntactic tests would often yield false test failure when the compiler
produced an equivalent but synctically different program, and end{-}to{-}end tests
failed to support debugging since a test failure could not be attributed to a
particular compiler pass.
Interpreters that provide the ground truth value of a program for each
compiler pass would enable resiliant tests and better debugging.
The users should not need to know anything about the interpreters other than
their interface: they take a (possibly valid) program, and produce its value or
an error.
They would also need debugging output and metrics such as reasonable error
messages and the ability to distinguish whether the provided program was invalid
or the test failed because the value produced was unexpected.

Our context is the University of British Columbia{'}s undergraduate compilers
course (CPSC 411).\NoteBox{\NoteContent{For additional context, the course material is publically available online at
\href{https://github.com/cpsc411/}{\Snolinkurl{https://github.com/cpsc411/}}, with the most recent spring 2022
iteration deployed at \href{https://www.students.cs.ubc.ca/~cs-411/2021w2/}{\Snolinkurl{https://www.students.cs.ubc.ca/~cs-411/2021w2/}}.}}
Our users are the students and the course staff, who use the interpreters to
test the compilers and provide feedback.
The nanopass structure of the compiler means the compiler includes many passes,
up to about 30 by the end of the course, and each pass has separate intermediate
languages specifying the source and target of the pass.
Students develop the compiler incremenetally in 10 milestones, with new
languages and changes to existing languages each milestone.

Our first question is an engineering question: how should we implement these
interpreters?
The constraints described above prevent us from doing the obvious thing: writing
each interpreter as a function that folds over the syntax of the language and
computes the value of the input program.
Language{-}oriented programming\Autobibref{~(\hyperref[t:x28autobib_x22Matthias_Felleisenx2c_Robert_Bruce_Findlerx2c_Matthew_Flattx2c_Shriram_Krishnamurthix2c_Eli_Barzilayx2c_Jay_Mccarthyx2c_and_Sam_Tobinx2dHochstadtA_Programmable_Programing_LanguageCommunications_of_the_ACM_61x283x29x2c_ppx2e_62x2dx2d712018doix3a10x2e1145x2f3127323x22x29]{\AutobibLink{Felleisen et al\Sendabbrev{.}}} \hyperref[t:x28autobib_x22Matthias_Felleisenx2c_Robert_Bruce_Findlerx2c_Matthew_Flattx2c_Shriram_Krishnamurthix2c_Eli_Barzilayx2c_Jay_Mccarthyx2c_and_Sam_Tobinx2dHochstadtA_Programmable_Programing_LanguageCommunications_of_the_ACM_61x283x29x2c_ppx2e_62x2dx2d712018doix3a10x2e1145x2f3127323x22x29]{\AutobibLink{2018}}; \hyperref[t:x28autobib_x22Matthias_Felleisenx2c_Robert_Bruce_Findlerx2c_Matthew_Flattx2c_Shriram_Krishnamurthix2c_Eli_Barzilayx2c_Jay_Ax2e_McCarthyx2c_and_Sam_Tobinx2dHochstadtThe_Racket_ManifestoIn_Procx2e_SNAPL2015doix3a10x2e4230x2fLIPIcsx2eSNAPLx2e2015x2e113x22x29]{\AutobibLink{Felleisen et al\Sendabbrev{.}}} \hyperref[t:x28autobib_x22Matthias_Felleisenx2c_Robert_Bruce_Findlerx2c_Matthew_Flattx2c_Shriram_Krishnamurthix2c_Eli_Barzilayx2c_Jay_Ax2e_McCarthyx2c_and_Sam_Tobinx2dHochstadtThe_Racket_ManifestoIn_Procx2e_SNAPL2015doix3a10x2e4230x2fLIPIcsx2eSNAPLx2e2015x2e113x22x29]{\AutobibLink{2015}}; \hyperref[t:x28autobib_x22Sam_Tobinx2dHochstadtx2c_Vincent_Stx2dAmourx2c_Ryan_Culpepperx2c_Matthew_Flattx2c_and_Matthias_FelleisenLanguages_As_LibrariesIn_Procx2e_International_Conference_on_Programming_Language_Design_and_Implementation_x28PLDIx292011doix3a10x2e1145x2f1993498x2e1993514x22x29]{\AutobibLink{Tobin{-}Hochstadt et al\Sendabbrev{.}}} \hyperref[t:x28autobib_x22Sam_Tobinx2dHochstadtx2c_Vincent_Stx2dAmourx2c_Ryan_Culpepperx2c_Matthew_Flattx2c_and_Matthias_FelleisenLanguages_As_LibrariesIn_Procx2e_International_Conference_on_Programming_Language_Design_and_Implementation_x28PLDIx292011doix3a10x2e1145x2f1993498x2e1993514x22x29]{\AutobibLink{2011}})} could provide a suitable design for answering our first
question, \emph{if} the existing abstractions apply to low{-}level languages as
well as high{-}level languages.
But this yields a second question: (how) can we implement a family of compiler
intermediate languages, from low{-} to high{-}level languages, as a macro{-}embedding
using language{-}oriented programming?
This paper answers this latter question, demonstrating how to embed such
low{-}level abstractions using language{-}oriented programming, particularly in
Racket, so that we can easily develop and maintain our family of interpreters.

Before considering our case study, let{'}s consider the small example language
below.

\Iidentity{\begin{bnfgrammar}
p ::= (\<begin> effect effect ...)
;;

effect ::=
(\<set!> reg  word64)
| (\<set!> reg$_1$ (binop reg$_1$ word32))
;;

reg ::=
\<rax>
| ...
| \<r12>
| \<r13>
| \<r14>
| \<r15>
;;

word32 ::= int32 | reg
;;

word64 ::= word32 | int64
;;

binop ::= + | - | *
\end{bnfgrammar}}

The language is a simple subset of x64 with an s{-}expression syntax, described by
the grammar above.
A program begins with \Iidentity{\<begin>} and is followed by a non{-}empty list of
effects.
Each effect can modify a register, either setting it to a value or to the result
of a binary operation.
We assume every program ends with \Iidentity{\<rax>} set to the result.

We could implement a whole{-}program environment{-}passing interpreter as below.

\LangBox{\LangBoxLabel{\RktModLink{\RktMod{\#lang}}\RktMeta{}\mbox{\hphantom{\Scribtexttt{x}}}\RktMeta{}\RktModLink{\RktSym{racket}}\RktMeta{}}}%
{\begin{SCodeFlow}\begin{RktBlk}\begin{SingleColumn}\begin{RktBlk}\begin{SingleColumn}\Scribtexttt{{\Stttextmore} }\RktPn{(}\RktSym{define}\mbox{\hphantom{\Scribtexttt{x}}}\RktPn{(}\RktSym{interp{-}x64{-}v1}\mbox{\hphantom{\Scribtexttt{x}}}\RktSym{p}\RktPn{)}

\mbox{\hphantom{\Scribtexttt{xx}}}\mbox{\hphantom{\Scribtexttt{xx}}}\RktPn{(}\RktSym{define}\mbox{\hphantom{\Scribtexttt{x}}}\RktPn{(}\RktSym{interp{-}binop}\mbox{\hphantom{\Scribtexttt{x}}}\RktSym{op}\RktPn{)}

\mbox{\hphantom{\Scribtexttt{xx}}}\mbox{\hphantom{\Scribtexttt{xxxx}}}\RktPn{(}\RktSym{match}\mbox{\hphantom{\Scribtexttt{x}}}\RktSym{op}

\mbox{\hphantom{\Scribtexttt{xx}}}\mbox{\hphantom{\Scribtexttt{xxxxxx}}}\RktPn{[}\RktVal{{\textquotesingle}}\RktVal{+}\mbox{\hphantom{\Scribtexttt{x}}}\RktSym{+}\RktPn{]}

\mbox{\hphantom{\Scribtexttt{xx}}}\mbox{\hphantom{\Scribtexttt{xxxxxx}}}\RktPn{[}\RktVal{{\textquotesingle}}\RktVal{\mbox{{-}}}\mbox{\hphantom{\Scribtexttt{x}}}\RktSym{\mbox{{-}}}\RktPn{]}

\mbox{\hphantom{\Scribtexttt{xx}}}\mbox{\hphantom{\Scribtexttt{xxxxxx}}}\RktPn{[}\RktVal{{\textquotesingle}}\RktVal{*}\mbox{\hphantom{\Scribtexttt{x}}}\RktSym{*}\RktPn{]}\RktPn{)}\RktPn{)}

\mbox{\hphantom{\Scribtexttt{xx}}}

\mbox{\hphantom{\Scribtexttt{xx}}}\mbox{\hphantom{\Scribtexttt{xx}}}\RktPn{(}\RktSym{define}\mbox{\hphantom{\Scribtexttt{x}}}\RktPn{(}\RktSym{interp{-}opand}\mbox{\hphantom{\Scribtexttt{x}}}\RktSym{opand}\mbox{\hphantom{\Scribtexttt{x}}}\RktSym{env}\RktPn{)}

\mbox{\hphantom{\Scribtexttt{xx}}}\mbox{\hphantom{\Scribtexttt{xxxx}}}\RktPn{(}\RktSym{match}\mbox{\hphantom{\Scribtexttt{x}}}\RktSym{opand}

\mbox{\hphantom{\Scribtexttt{xx}}}\mbox{\hphantom{\Scribtexttt{xxxxxx}}}\RktPn{[}\RktPn{(}\RktSym{{\hbox{\texttt{?}}}}\mbox{\hphantom{\Scribtexttt{x}}}\RktSym{integer{\hbox{\texttt{?}}}}\RktPn{)}\mbox{\hphantom{\Scribtexttt{x}}}\RktSym{opand}\RktPn{]}

\mbox{\hphantom{\Scribtexttt{xx}}}\mbox{\hphantom{\Scribtexttt{xxxxxx}}}\RktPn{[}\RktPn{(}\RktSym{{\hbox{\texttt{?}}}}\mbox{\hphantom{\Scribtexttt{x}}}\RktSym{symbol{\hbox{\texttt{?}}}}\RktPn{)}\mbox{\hphantom{\Scribtexttt{x}}}\RktPn{(}\RktSym{dict{-}ref}\mbox{\hphantom{\Scribtexttt{x}}}\RktSym{env}\mbox{\hphantom{\Scribtexttt{x}}}\RktSym{opand}\RktPn{)}\RktPn{]}\RktPn{)}\RktPn{)}

\mbox{\hphantom{\Scribtexttt{xx}}}

\mbox{\hphantom{\Scribtexttt{xx}}}\mbox{\hphantom{\Scribtexttt{xx}}}\RktPn{(}\RktSym{define}\mbox{\hphantom{\Scribtexttt{x}}}\RktPn{(}\RktSym{interp{-}effect}\mbox{\hphantom{\Scribtexttt{x}}}\RktSym{effect}\mbox{\hphantom{\Scribtexttt{x}}}\RktSym{env}\RktPn{)}

\mbox{\hphantom{\Scribtexttt{xx}}}\mbox{\hphantom{\Scribtexttt{xxxx}}}\RktPn{(}\RktSym{match}\mbox{\hphantom{\Scribtexttt{x}}}\RktSym{effect}

\mbox{\hphantom{\Scribtexttt{xx}}}\mbox{\hphantom{\Scribtexttt{xxxxxx}}}\RktPn{[}\RktVal{{\textasciigrave}}\RktVal{(}\RktVal{set{\hbox{\texttt{!}}}}\mbox{\hphantom{\Scribtexttt{x}}}\RktRdr{,}\RktSym{reg}\mbox{\hphantom{\Scribtexttt{x}}}\RktVal{(}\RktRdr{,}\RktSym{binop}\mbox{\hphantom{\Scribtexttt{x}}}\RktRdr{,}\RktSym{reg}\mbox{\hphantom{\Scribtexttt{x}}}\RktRdr{,}\RktSym{opand}\RktVal{)}\RktVal{)}

\mbox{\hphantom{\Scribtexttt{xx}}}\mbox{\hphantom{\Scribtexttt{xxxxxxx}}}\RktPn{(}\RktSym{dict{-}set}\mbox{\hphantom{\Scribtexttt{x}}}\RktSym{env}\mbox{\hphantom{\Scribtexttt{x}}}\RktSym{reg}\mbox{\hphantom{\Scribtexttt{x}}}\RktPn{(}\RktPn{(}\RktSym{interp{-}binop}\mbox{\hphantom{\Scribtexttt{x}}}\RktSym{binop}\RktPn{)}

\mbox{\hphantom{\Scribtexttt{xx}}}\mbox{\hphantom{\Scribtexttt{xxxxxxxxxxxxxxxxxxxxxxxxxx}}}\RktPn{(}\RktSym{interp{-}opand}\mbox{\hphantom{\Scribtexttt{x}}}\RktSym{reg}\mbox{\hphantom{\Scribtexttt{x}}}\RktSym{env}\RktPn{)}

\mbox{\hphantom{\Scribtexttt{xx}}}\mbox{\hphantom{\Scribtexttt{xxxxxxxxxxxxxxxxxxxxxxxxxx}}}\RktPn{(}\RktSym{interp{-}opand}\mbox{\hphantom{\Scribtexttt{x}}}\RktSym{opand}\mbox{\hphantom{\Scribtexttt{x}}}\RktSym{env}\RktPn{)}\RktPn{)}\RktPn{)}\RktPn{]}

\mbox{\hphantom{\Scribtexttt{xx}}}\mbox{\hphantom{\Scribtexttt{xxxxxx}}}\RktPn{[}\RktVal{{\textasciigrave}}\RktVal{(}\RktVal{set{\hbox{\texttt{!}}}}\mbox{\hphantom{\Scribtexttt{x}}}\RktRdr{,}\RktSym{reg}\mbox{\hphantom{\Scribtexttt{x}}}\RktRdr{,}\RktSym{opand}\RktVal{)}

\mbox{\hphantom{\Scribtexttt{xx}}}\mbox{\hphantom{\Scribtexttt{xxxxxxx}}}\RktPn{(}\RktSym{dict{-}set}\mbox{\hphantom{\Scribtexttt{x}}}\RktSym{env}\mbox{\hphantom{\Scribtexttt{x}}}\RktSym{reg}\mbox{\hphantom{\Scribtexttt{x}}}\RktPn{(}\RktSym{interp{-}opand}\mbox{\hphantom{\Scribtexttt{x}}}\RktSym{opand}\mbox{\hphantom{\Scribtexttt{x}}}\RktSym{env}\RktPn{)}\RktPn{)}\RktPn{]}\RktPn{)}\RktPn{)}

\mbox{\hphantom{\Scribtexttt{xx}}}

\mbox{\hphantom{\Scribtexttt{xx}}}\mbox{\hphantom{\Scribtexttt{xx}}}\RktPn{(}\RktSym{match}\mbox{\hphantom{\Scribtexttt{x}}}\RktSym{p}

\mbox{\hphantom{\Scribtexttt{xx}}}\mbox{\hphantom{\Scribtexttt{xxxx}}}\RktPn{[}\RktVal{{\textasciigrave}}\RktVal{(}\RktVal{begin}\mbox{\hphantom{\Scribtexttt{x}}}\RktRdr{,}\RktSym{list{-}of{-}effect}\mbox{\hphantom{\Scribtexttt{x}}}\RktVal{{\hbox{\texttt{.}}}{\hbox{\texttt{.}}}{\hbox{\texttt{.}}}}\RktVal{)}

\mbox{\hphantom{\Scribtexttt{xx}}}\mbox{\hphantom{\Scribtexttt{xxxxx}}}\RktPn{(}\RktSym{interp{-}opand}\mbox{\hphantom{\Scribtexttt{x}}}\RktVal{{\textquotesingle}}\RktVal{rax}\mbox{\hphantom{\Scribtexttt{x}}}\RktPn{(}\RktSym{for/fold}\mbox{\hphantom{\Scribtexttt{x}}}\RktPn{(}\RktPn{[}\RktSym{env}\mbox{\hphantom{\Scribtexttt{x}}}\RktSym{empty}\RktPn{]}\RktPn{)}

\mbox{\hphantom{\Scribtexttt{xx}}}\mbox{\hphantom{\Scribtexttt{xxxxxxxxxxxxxxxxxxxxxxxxxxxxxxxxxx}}}\RktPn{(}\RktPn{[}\RktSym{effect}\mbox{\hphantom{\Scribtexttt{x}}}\RktSym{list{-}of{-}effect}\RktPn{]}\RktPn{)}

\mbox{\hphantom{\Scribtexttt{xx}}}\mbox{\hphantom{\Scribtexttt{xxxxxxxxxxxxxxxxxxxxxxxxxx}}}\RktPn{(}\RktSym{interp{-}effect}\mbox{\hphantom{\Scribtexttt{x}}}\RktSym{effect}\mbox{\hphantom{\Scribtexttt{x}}}\RktSym{env}\RktPn{)}\RktPn{)}\RktPn{)}\RktPn{]}\RktPn{)}\RktPn{)}\end{SingleColumn}\end{RktBlk}\end{SingleColumn}\end{RktBlk}\end{SCodeFlow}}

This interpreter is a fold over the list of effects, which ends by dereferencing
the value of the register \RktVal{{\textquotesingle}}\RktVal{rax} from the final environment.
Initially, the environment is empty.
Each effect is interpretered by updating the environment, which maps registers
to values.

\LangBox{\LangBoxLabel{\RktModLink{\RktMod{\#lang}}\RktMeta{}\mbox{\hphantom{\Scribtexttt{x}}}\RktMeta{}\RktModLink{\RktSym{racket}}\RktMeta{}}}%
{\begin{SCodeFlow}\begin{RktBlk}\begin{SingleColumn}\Scribtexttt{{\Stttextmore} }\RktPn{(}\RktSym{interp{-}x64{-}v1}\mbox{\hphantom{\Scribtexttt{x}}}\RktVal{{\textquotesingle}}\RktVal{(}\RktVal{begin}\mbox{\hphantom{\Scribtexttt{x}}}\RktVal{(}\RktVal{set{\hbox{\texttt{!}}}}\mbox{\hphantom{\Scribtexttt{x}}}\RktVal{rax}\mbox{\hphantom{\Scribtexttt{x}}}\RktVal{15}\RktVal{)}\RktVal{)}\RktPn{)}

\RktRes{15}\end{SingleColumn}\end{RktBlk}\end{SCodeFlow}}

The interpreter is straightforward to implement, once, for a single small
language.
But we need a different approach to scale to developing and maintaining
hundreds of interpreters.
Many of the intermediate languages include registers, assignment statements to
registers, integer operands, etc{---}we do not want to copy/paste, but must design
for sharing the implementation of features between interpreters.

The key change we make to our design is to macro{-}embed each \emph{language
feature} into a common host language, rather than implementing each
\emph{language} separately.
We then derive an interpreter for any given language using the host language{'}s
\RktSym{eval}, with the set of languages features imported.

For the above example language, we could start by implementing the \RktSym{set{\hbox{\texttt{!}}}}
feature as follows.

\LangBox{\LangBoxLabel{\RktModLink{\RktMod{\#lang}}\RktMeta{}\mbox{\hphantom{\Scribtexttt{x}}}\RktMeta{}\RktModLink{\RktSym{racket}}\RktMeta{}}}%
{\begin{SCodeFlow}\begin{RktBlk}\begin{SingleColumn}\Scribtexttt{{\Stttextmore} }\RktPn{(}\RktSym{define}\mbox{\hphantom{\Scribtexttt{x}}}\RktSym{env}\mbox{\hphantom{\Scribtexttt{x}}}\RktPn{(}\RktSym{make{-}hash}\RktPn{)}\RktPn{)}

\begin{RktBlk}\begin{SingleColumn}\Scribtexttt{{\Stttextmore} }\RktPn{(}\RktSym{define{-}syntax}\mbox{\hphantom{\Scribtexttt{x}}}\RktPn{(}\RktSym{set{\hbox{\texttt{!}}}}\mbox{\hphantom{\Scribtexttt{x}}}\RktSym{stx}\RktPn{)}

\mbox{\hphantom{\Scribtexttt{xx}}}\mbox{\hphantom{\Scribtexttt{xx}}}\RktPn{(}\RktSym{syntax{-}parse}\mbox{\hphantom{\Scribtexttt{x}}}\RktSym{stx}

\mbox{\hphantom{\Scribtexttt{xx}}}\mbox{\hphantom{\Scribtexttt{xxxx}}}\RktPn{[}\RktPn{(}\RktSym{set{\hbox{\texttt{!}}}}\mbox{\hphantom{\Scribtexttt{x}}}\RktSym{reg}\mbox{\hphantom{\Scribtexttt{x}}}\RktPn{(}\RktSym{binop}\mbox{\hphantom{\Scribtexttt{x}}}\RktSym{{\char`\_}}\mbox{\hphantom{\Scribtexttt{x}}}\RktSym{opand}\RktPn{)}\RktPn{)}

\mbox{\hphantom{\Scribtexttt{xx}}}\mbox{\hphantom{\Scribtexttt{xxxxx}}}\RktRdr{\#{\textasciigrave}}\RktPn{(}\RktSym{dict{-}set{\hbox{\texttt{!}}}}\mbox{\hphantom{\Scribtexttt{x}}}\RktSym{env}\mbox{\hphantom{\Scribtexttt{x}}}\RktVal{{\textquotesingle}}\RktVal{reg}\mbox{\hphantom{\Scribtexttt{x}}}\RktPn{(}\RktSym{binop}\mbox{\hphantom{\Scribtexttt{x}}}\RktPn{(}\RktSym{interp{-}opand}\mbox{\hphantom{\Scribtexttt{x}}}\RktVal{{\textquotesingle}}\RktVal{reg}\RktPn{)}

\mbox{\hphantom{\Scribtexttt{xx}}}\mbox{\hphantom{\Scribtexttt{xxxxxxxxxxxxxxxxxxxxxxxxxxxxxxxxxx}}}\RktPn{(}\RktSym{interp{-}opand}\mbox{\hphantom{\Scribtexttt{x}}}\RktSym{opand}\RktPn{)}\RktPn{)}\RktPn{)}\RktPn{]}

\mbox{\hphantom{\Scribtexttt{xx}}}\mbox{\hphantom{\Scribtexttt{xxxx}}}\RktPn{[}\RktPn{(}\RktSym{set{\hbox{\texttt{!}}}}\mbox{\hphantom{\Scribtexttt{x}}}\RktSym{reg}\mbox{\hphantom{\Scribtexttt{x}}}\RktSym{opand}\RktPn{)}

\mbox{\hphantom{\Scribtexttt{xx}}}\mbox{\hphantom{\Scribtexttt{xxxxx}}}\RktRdr{\#{\textasciigrave}}\RktPn{(}\RktSym{dict{-}set{\hbox{\texttt{!}}}}\mbox{\hphantom{\Scribtexttt{x}}}\RktSym{env}\mbox{\hphantom{\Scribtexttt{x}}}\RktVal{{\textquotesingle}}\RktVal{reg}\mbox{\hphantom{\Scribtexttt{x}}}\RktPn{(}\RktSym{interp{-}opand}\mbox{\hphantom{\Scribtexttt{x}}}\RktSym{opand}\RktPn{)}\RktPn{)}\RktPn{]}\RktPn{)}\RktPn{)}\end{SingleColumn}\end{RktBlk}\end{SingleColumn}\end{RktBlk}\end{SCodeFlow}}

In this example, we define as a macro the \RktSym{set{\hbox{\texttt{!}}}} language feature,
without the rest of the language.
We first define the environment as a mutable hash table at run time in the host
language.
Since the syntax of the program is traversed implicitly in the macro expander,
we cannot pass additional arguments during the syntax traversal, so we use a
mutable environment.
The macro \RktSym{set{\hbox{\texttt{!}}}}, defined using \RktSym{define{-}syntax}, is implicitly
called during compile time by the macro expander whenever it encounters the
defined identifier \RktSym{set{\hbox{\texttt{!}}}}.
Each macro transforms syntax into new syntax at compile time, so we require
slightly different abstractions to work over syntax objects rather than over
quoted lists.
The macro{'}s definition receives as input a single syntax object, representing
the entire syntax of the call to the macro, including its own name.
We use \RktSym{syntax{-}parse}\Autobibref{~(\hyperref[t:x28autobib_x22Ryan_CulpepperFortifying_macrosJournal_of_Functional_Programming_x28JFPx29_22x284x2d5x29x2c_ppx2e_439x2dx2d4762012doix3a10x2e1017x2fS0956796812000275x22x29]{\AutobibLink{Culpepper}} \hyperref[t:x28autobib_x22Ryan_CulpepperFortifying_macrosJournal_of_Functional_Programming_x28JFPx29_22x284x2d5x29x2c_ppx2e_439x2dx2d4762012doix3a10x2e1017x2fS0956796812000275x22x29]{\AutobibLink{2012}}; \hyperref[t:x28autobib_x22Ryan_Culpepper_and_Matthias_FelleisenFortifying_macrosIn_Procx2e_International_Conference_on_Functional_Programming_x28ICFPx292010doix3a10x2e1145x2f1932681x2e1863577x22x29]{\AutobibLink{Culpepper and Felleisen}} \hyperref[t:x28autobib_x22Ryan_Culpepper_and_Matthias_FelleisenFortifying_macrosIn_Procx2e_International_Conference_on_Functional_Programming_x28ICFPx292010doix3a10x2e1145x2f1932681x2e1863577x22x29]{\AutobibLink{2010}})} for
pattern matching over syntax objects, and use \RktSym{quasisyntax} (written
\RktInBG{\RktIn{\#{\textasciigrave}}}) to construct syntax objects.
We destructure the input syntax expecting \RktSym{set{\hbox{\texttt{!}}}} in operator position
(which is unused in the definition), a register on the left{-}hand side, and
either a binary operation or an operand on the right{-}hand side.
The syntax template constructed with \RktInBG{\RktIn{\#{\textasciigrave}}} can refer to bound syntax
pattern variables, whose bound names are automatically replaced by their value
in the template without the need to unquote.

The definition of \RktSym{set{\hbox{\texttt{!}}}} is essentially similar to the earlier
interpreter, implementing assignment as an update to the environment.
The key difference is that we generate a run{-}time expression that updates the
environment, rather than immediately executing the update to the environment.
We must explicitly transform the register operand from a raw piece of syntax
into a symbol by quoting it in the generated syntax.
But we do not need to explicitly transform binary operations such as \RktSym{+},
since they are already defined and have sufficiently similar semantics in the
host language.

The implementation relies on \RktSym{interp{-}opand}, defined below, to expand
operands.

\noindent \LangBox{\LangBoxLabel{\RktModLink{\RktMod{\#lang}}\RktMeta{}\mbox{\hphantom{\Scribtexttt{x}}}\RktMeta{}\RktModLink{\RktSym{racket}}\RktMeta{}}}%
{\begin{SCodeFlow}\begin{RktBlk}\begin{SingleColumn}\begin{RktBlk}\begin{SingleColumn}\Scribtexttt{{\Stttextmore} }\RktPn{(}\RktSym{define{-}syntax}\mbox{\hphantom{\Scribtexttt{x}}}\RktPn{(}\RktSym{interp{-}opand}\mbox{\hphantom{\Scribtexttt{x}}}\RktSym{stx}\RktPn{)}

\mbox{\hphantom{\Scribtexttt{xx}}}\mbox{\hphantom{\Scribtexttt{xx}}}\RktPn{(}\RktSym{syntax{-}parse}\mbox{\hphantom{\Scribtexttt{x}}}\RktSym{stx}

\mbox{\hphantom{\Scribtexttt{xx}}}\mbox{\hphantom{\Scribtexttt{xxxx}}}\RktPn{[}\RktPn{(}\RktSym{{\char`\_}}\mbox{\hphantom{\Scribtexttt{x}}}\RktSym{reg{\hbox{\texttt{:}}}id}\RktPn{)}

\mbox{\hphantom{\Scribtexttt{xx}}}\mbox{\hphantom{\Scribtexttt{xxxxx}}}\RktRdr{\#{\textasciigrave}}\RktPn{(}\RktSym{dict{-}ref}\mbox{\hphantom{\Scribtexttt{x}}}\RktSym{env}\mbox{\hphantom{\Scribtexttt{x}}}\RktVal{{\textquotesingle}}\RktVal{reg}\RktPn{)}\RktPn{]}

\mbox{\hphantom{\Scribtexttt{xx}}}\mbox{\hphantom{\Scribtexttt{xxxx}}}\RktPn{[}\RktPn{(}\RktSym{{\char`\_}}\mbox{\hphantom{\Scribtexttt{x}}}\RktSym{int{\hbox{\texttt{:}}}integer}\RktPn{)}

\mbox{\hphantom{\Scribtexttt{xx}}}\mbox{\hphantom{\Scribtexttt{xxxxx}}}\RktRdr{\#{\textasciigrave}}\RktSym{int}\RktPn{]}\RktPn{)}\RktPn{)}\end{SingleColumn}\end{RktBlk}\end{SingleColumn}\end{RktBlk}\end{SCodeFlow}}

\RktSym{interp{-}opand} has two cases: the pattern \RktSym{reg{\hbox{\texttt{:}}}id} matches a raw
identifier and binds the syntax pattern variable \RktSym{reg}, while the pattern
\RktSym{int{\hbox{\texttt{:}}}integer} matches an integer literal and binds the syntax pattern
variable \RktSym{int}.
Identifiers are assumed to be registers and expand to dereference from the
environment.
The register is quoted, transforming the raw identifier into a symbol.
Integers expand to themselves since the object language and the host language
share integer literal syntax.
Following a common convention, we use \RktSym{{\char`\_}} to bind the irrelevant name of
the operator (\RktSym{interp{-}opand}) in the patterns.

To get a full interpreter from this feature, we need to combine it with the
other features (\RktSym{begin}, \RktSym{+}, etc), and provide an entry point
that provides the value interpretation of programs by dereferencing
\RktVal{{\textquotesingle}}\RktVal{rax} from the environment as the final operation.
We can reuse the host language \RktSym{begin} and binary operations, as their
semantics are close enough, and add the following entry point.

\LangBox{\LangBoxLabel{\RktModLink{\RktMod{\#lang}}\RktMeta{}\mbox{\hphantom{\Scribtexttt{x}}}\RktMeta{}\RktModLink{\RktSym{racket}}\RktMeta{}}}%
{\begin{SCodeFlow}\begin{RktBlk}\begin{SingleColumn}\begin{RktBlk}\begin{SingleColumn}\Scribtexttt{{\Stttextmore} }\RktPn{(}\RktSym{define{-}syntax}\mbox{\hphantom{\Scribtexttt{x}}}\RktPn{(}\RktSym{interp{-}x64{-}v2}\mbox{\hphantom{\Scribtexttt{x}}}\RktSym{stx}\RktPn{)}

\mbox{\hphantom{\Scribtexttt{xx}}}\mbox{\hphantom{\Scribtexttt{xx}}}\RktPn{(}\RktSym{syntax{-}parse}\mbox{\hphantom{\Scribtexttt{x}}}\RktSym{stx}

\mbox{\hphantom{\Scribtexttt{xx}}}\mbox{\hphantom{\Scribtexttt{xxxx}}}\RktPn{[}\RktPn{(}\RktSym{{\char`\_}}\mbox{\hphantom{\Scribtexttt{x}}}\RktSym{stx}\RktPn{)}

\mbox{\hphantom{\Scribtexttt{xx}}}\mbox{\hphantom{\Scribtexttt{xxxxx}}}\RktRdr{\#{\textasciigrave}}\RktPn{(}\RktSym{begin}\mbox{\hphantom{\Scribtexttt{x}}}\RktSym{stx}\mbox{\hphantom{\Scribtexttt{x}}}\RktPn{(}\RktSym{dict{-}ref}\mbox{\hphantom{\Scribtexttt{x}}}\RktSym{env}\mbox{\hphantom{\Scribtexttt{x}}}\RktVal{{\textquotesingle}}\RktVal{rax}\RktPn{)}\RktPn{)}\RktPn{]}\RktPn{)}\RktPn{)}\end{SingleColumn}\end{RktBlk}

\begin{RktBlk}\begin{SingleColumn}\Scribtexttt{{\Stttextmore} }\RktPn{(}\RktSym{interp{-}x64{-}v2}

\mbox{\hphantom{\Scribtexttt{xx}}}\mbox{\hphantom{\Scribtexttt{x}}}\RktPn{(}\RktSym{begin}

\mbox{\hphantom{\Scribtexttt{xx}}}\mbox{\hphantom{\Scribtexttt{xxx}}}\RktPn{(}\RktSym{set{\hbox{\texttt{!}}}}\mbox{\hphantom{\Scribtexttt{x}}}\RktSym{rax}\mbox{\hphantom{\Scribtexttt{x}}}\RktVal{15}\RktPn{)}\RktPn{)}\RktPn{)}\end{SingleColumn}\end{RktBlk}

\RktRes{15}\end{SingleColumn}\end{RktBlk}\end{SCodeFlow}}

Our entry point, \RktSym{interp{-}x64{-}v2}, is a macro that uses the host language
\RktSym{begin} to sequentially execute the original syntax and then dereference
the symbol \RktVal{{\textquotesingle}}\RktVal{rax} from the environment.
The macro expander proceeds outside{-}in, so it will expand all the macros in
\RktSym{stx} after expanding a call to \RktSym{interp{-}x64{-}v2}.
Unlike when we used the procedure \RktSym{interp{-}x64{-}v1}, we do not need to
explicitly quote the input program \RktPn{(}\RktSym{begin}\Scribtexttt{ }\RktPn{(}\RktSym{set{\hbox{\texttt{!}}}}\Scribtexttt{ }\RktSym{rax}\Scribtexttt{ }\RktVal{15}\RktPn{)}\RktPn{)} since the
expander syntax{-}quotes the syntax of the call automatically before passing that
syntax object to the macro.

This is the basic idea behind language{-}oriented design\Autobibref{~(\hyperref[t:x28autobib_x22Matthias_Felleisenx2c_Robert_Bruce_Findlerx2c_Matthew_Flattx2c_Shriram_Krishnamurthix2c_Eli_Barzilayx2c_Jay_Mccarthyx2c_and_Sam_Tobinx2dHochstadtA_Programmable_Programing_LanguageCommunications_of_the_ACM_61x283x29x2c_ppx2e_62x2dx2d712018doix3a10x2e1145x2f3127323x22x29]{\AutobibLink{Felleisen et al\Sendabbrev{.}}} \hyperref[t:x28autobib_x22Matthias_Felleisenx2c_Robert_Bruce_Findlerx2c_Matthew_Flattx2c_Shriram_Krishnamurthix2c_Eli_Barzilayx2c_Jay_Mccarthyx2c_and_Sam_Tobinx2dHochstadtA_Programmable_Programing_LanguageCommunications_of_the_ACM_61x283x29x2c_ppx2e_62x2dx2d712018doix3a10x2e1145x2f3127323x22x29]{\AutobibLink{2018}}; \hyperref[t:x28autobib_x22Matthias_Felleisenx2c_Robert_Bruce_Findlerx2c_Matthew_Flattx2c_Shriram_Krishnamurthix2c_Eli_Barzilayx2c_Jay_Ax2e_McCarthyx2c_and_Sam_Tobinx2dHochstadtThe_Racket_ManifestoIn_Procx2e_SNAPL2015doix3a10x2e4230x2fLIPIcsx2eSNAPLx2e2015x2e113x22x29]{\AutobibLink{Felleisen et al\Sendabbrev{.}}} \hyperref[t:x28autobib_x22Matthias_Felleisenx2c_Robert_Bruce_Findlerx2c_Matthew_Flattx2c_Shriram_Krishnamurthix2c_Eli_Barzilayx2c_Jay_Ax2e_McCarthyx2c_and_Sam_Tobinx2dHochstadtThe_Racket_ManifestoIn_Procx2e_SNAPL2015doix3a10x2e4230x2fLIPIcsx2eSNAPLx2e2015x2e113x22x29]{\AutobibLink{2015}}; \hyperref[t:x28autobib_x22Sam_Tobinx2dHochstadtx2c_Vincent_Stx2dAmourx2c_Ryan_Culpepperx2c_Matthew_Flattx2c_and_Matthias_FelleisenLanguages_As_LibrariesIn_Procx2e_International_Conference_on_Programming_Language_Design_and_Implementation_x28PLDIx292011doix3a10x2e1145x2f1993498x2e1993514x22x29]{\AutobibLink{Tobin{-}Hochstadt et al\Sendabbrev{.}}} \hyperref[t:x28autobib_x22Sam_Tobinx2dHochstadtx2c_Vincent_Stx2dAmourx2c_Ryan_Culpepperx2c_Matthew_Flattx2c_and_Matthias_FelleisenLanguages_As_LibrariesIn_Procx2e_International_Conference_on_Programming_Language_Design_and_Implementation_x28PLDIx292011doix3a10x2e1145x2f1993498x2e1993514x22x29]{\AutobibLink{2011}})}.
Rather than writing the language as a closed set of features, a language is a
collection of individual features open to extension.
We implement the features by embedding into a common host language, then we can
use the module system to export, import, extend, redefine and mix the features
into a desired language.
If we{'}re careful about how we design the embedding, we can easily support
interoperability and extension.
The above implementation is not particularly careful.
We completely redefine \RktSym{set{\hbox{\texttt{!}}}}, so we{'}re unable to interoperate with
Racket{'}s \RktSym{set{\hbox{\texttt{!}}}}, and registers are only embedded as part of
\RktSym{set{\hbox{\texttt{!}}}}, not as their own feature, so do not work correctly in other
contexts.
We explore a better implementation in the rest of the paper.

Language{-}oriented programming in Racket has been used to implement
general{-}purpose languages such as Typed Racket\Autobibref{~(\hyperref[t:x28autobib_x22Sam_Tobinx2dHochstadt_and_Matthias_FelleisenThe_Design_and_Implementation_of_Typed_Schemex3a_From_Scripts_to_ProgramsCoRR_absx2f1106x2e25752011doix3a10x2e48550x2farXivx2e1106x2e2575x22x29]{\AutobibLink{Tobin{-}Hochstadt and Felleisen}} \hyperref[t:x28autobib_x22Sam_Tobinx2dHochstadt_and_Matthias_FelleisenThe_design_and_implementation_of_typed_schemeIn_Procx2e_Symposium_on_Principles_of_Programming_Languages_x28POPLx292008doix3a10x2e1145x2f1328438x2e1328486x22x29]{\AutobibLink{2008}}, \hyperref[t:x28autobib_x22Sam_Tobinx2dHochstadt_and_Matthias_FelleisenThe_Design_and_Implementation_of_Typed_Schemex3a_From_Scripts_to_ProgramsCoRR_absx2f1106x2e25752011doix3a10x2e48550x2farXivx2e1106x2e2575x22x29]{\AutobibLink{2011}})} and domain{-}specific languages such as
Scribble\Autobibref{~(\hyperref[t:x28autobib_x22Matthew_Flattx2c_Eli_Barzilayx2c_and_Robert_Bruce_FindlerScribblex3a_Closing_the_Book_on_Ad_Hoc_Documentation_ToolsIn_Procx2e_International_Conference_on_Functional_Programming_x28ICFPx292009doix3a10x2e1145x2f1596550x2e1596569x22x29]{\AutobibLink{Flatt et al\Sendabbrev{.}}} \hyperref[t:x28autobib_x22Matthew_Flattx2c_Eli_Barzilayx2c_and_Robert_Bruce_FindlerScribblex3a_Closing_the_Book_on_Ad_Hoc_Documentation_ToolsIn_Procx2e_International_Conference_on_Functional_Programming_x28ICFPx292009doix3a10x2e1145x2f1596550x2e1596569x22x29]{\AutobibLink{2009}})}.
However, these canonical examples are high{-}level languages.

This approach is not obviously suited to macro{-}embedding assembly language, with
its global state, labeled code, and unstructured control flow.
However, Racket{'}s language{-}extension facilities{---}namely the module system, the
static interposition points, run{-}time interposition features, macros designed
for extension, and \RktSym{eval} parameterized by a namespace{---}support this use
case and our engineering constraints well.
All interpreters share a single {\textasciitilde}600 LOC file that implements the embedding.
The embedding is largely modular, so individual language \emph{features} can be
changed once, locally, for all languages, with some exceptions.
Embedding most high{-}level features is straightforward by borrowing from Racket.
Embedding into a shared host language simplifies implementing intermediate
languages, where low{-} and high{-}level features blend together.

This language{-}oriented approach has added benefits beyond achieving our
engineering goals.
It simplifies using the intermediate languages in Racket{'}s toolchain (such as
the REPL and IDE) and enables interoperating with Racket (useful for in{-}class
demonstrations and typesetting notes).
Trying to define macros locally and compositionally{---}instead of, \emph{e.g.},
analyzing, transforming, or executing a whole program{---}exposed some bad design
decisions in our intermediate languages, enabling us to improve the languages.

In this article, we walk through the design and implementation of embedding key
interesting language features, and of the interfaces to our interpreters.
We describe which parts of the Racket language{-}extension API are particularly
useful, and discuss limitations in our current design, implementation, and
Racket abstractions.

The implementation of our embedding can be installed locally using the following
command.\NoteBox{\NoteContent{The software is archived at \href{https://doi.org/10.5281/zenodo.7051910}{\Snolinkurl{https://doi.org/10.5281/zenodo.7051910}}.}}

\begin{SInsetFlow}\Smaller{\Scribtexttt{raco pkg install https{\hbox{\texttt{:}}}//github{\hbox{\texttt{.}}}com/cpsc411/cpsc411{-}pub{\hbox{\texttt{.}}}git{\hbox{\texttt{?}}}path=cpsc411{-}lib\#hashlang{-}x64}}\end{SInsetFlow}

We typeset all interactive examples run in a REPL in a grey box, with a language
name indicating in which language the examples must be run.
Typeset code that does not appear in a box is an excerpt from the implementation
of the languages, is not expected to run on its own, and may be simplified to
elide irrelevant details.
Most of these excerpts are from a single file, found online at
\href{https://github.com/cpsc411/cpsc411-pub/blob/hashlang-x64/cpsc411-lib/cpsc411/langs/base.rkt}{\Snolinkurl{https://github.com/cpsc411/cpsc411-pub/blob/hashlang-x64/cpsc411-lib/cpsc411/langs/base.rkt}}.

\sectionNewpage

\Ssection{x64}{x64}\label{t:x28part_x22secx3ax64x22x29}

We start with the embedding of the lowest{-}level features from x64, in
particular, global state (registers and memory), pointer arithmetic, and labels
and jumps.
These features are present not only in the target language, but in many of the
intermediate languages.

The syntax of our target language, a parenthesized subset of x64, is given
below.
\Iidentity{\begin{bnfgrammar}
p ::= (\<begin> effect ...)
;;

effect ::=
 (\<set!> addr word32)
 | (\<set!> reg  word64)
 | (\<set!> reg$_1$ (binop reg$_1$ word32))
 || (\<set!> reg$_1$ (binop reg$_1$ addr))
 | (\<with-label> label effect)
 | (\<jump> trg)
 || (\<compare> reg opand)
 | (\<jump-if> relop label)
;;

reg ::= \<rsp>
| \<rbp>
| \<rax>
| ...
| \<r12>
| \<r13>
| \<r14>
| \<r15>
;;

addr ::= (reg \<+> int32) | (reg \<+> reg) | (rbp \<-> int32)
;;

word32 ::= int32 | label | reg
;;

word64 ::= word32 | int64 | addr
;;

opand ::= int64 | reg
;;

trg ::= reg | label
\end{bnfgrammar}}

Every program is a sequence of statements, which have a one{-}to{-}one
correspondence with some x64 instruction.
Labels must be symbols whose string representation is accepted by the assembler.
The displacement{-} and index{-}mode offsets in addresses must be divisible by 8, as
we assume all addresses are word aligned.
We use the small{-}code model, so labels are 32{-}bit words.
The register operand of a binary operation must be the same on the right{-}hand
side and the left{-}hand side, to suit x64, but we duplicate the operand in our
syntax to simplify the interpretation of \RktSym{set{\hbox{\texttt{!}}}} expressions as a Racket
\RktSym{set{\hbox{\texttt{!}}}}.

Our run{-}time system follows various conventions from the System V
ABI\NoteBox{\NoteContent{\href{https://github.com/hjl-tools/x86-psABI/wiki/x86-64-psABI-1.0.pdf}{\Snolinkurl{https://github.com/hjl-tools/x86-psABI/wiki/x86-64-psABI-1.0.pdf}}}},
which we introduce as they become relevant.
The program must also end by jumping to the special \Scribtexttt{done} label with the
result in the register \Scribtexttt{rax}.

We can write everyone{'}s favourite function, factorial, in \RktModLink{\RktMod{\#lang}}\RktMeta{}\mbox{\hphantom{\Scribtexttt{x}}}\RktMeta{}\RktModLink{\RktSym{cpsc411/hashlangs/base}}\RktMeta{} (the language that exposes the majority of the macro
embedding on which all the interpreters are based) as follows.

\noindent \LangBox{\LangBoxLabel{\RktModLink{\RktMod{\#lang}}\RktMeta{}\mbox{\hphantom{\Scribtexttt{x}}}\RktMeta{}\RktModLink{\RktSym{cpsc411/hashlangs/base}}\RktMeta{}}}%
{\begin{SCodeFlow}\begin{RktBlk}\begin{SingleColumn}\begin{RktBlk}\begin{SingleColumn}\Scribtexttt{{\Stttextmore} }\RktPn{(}\RktSym{begin}

\mbox{\hphantom{\Scribtexttt{xx}}}\mbox{\hphantom{\Scribtexttt{xx}}}\RktPn{(}\RktSym{set{\hbox{\texttt{!}}}}\mbox{\hphantom{\Scribtexttt{x}}}\RktSym{r15}\mbox{\hphantom{\Scribtexttt{x}}}\RktVal{5}\RktPn{)}

\mbox{\hphantom{\Scribtexttt{xx}}}\mbox{\hphantom{\Scribtexttt{xx}}}\RktPn{(}\RktSym{set{\hbox{\texttt{!}}}}\mbox{\hphantom{\Scribtexttt{x}}}\RktSym{r14}\mbox{\hphantom{\Scribtexttt{x}}}\RktVal{1}\RktPn{)}

\mbox{\hphantom{\Scribtexttt{xx}}}\mbox{\hphantom{\Scribtexttt{xx}}}\RktPn{(}\RktSym{with{-}label}\mbox{\hphantom{\Scribtexttt{x}}}\RktSym{fact}

\mbox{\hphantom{\Scribtexttt{xx}}}\mbox{\hphantom{\Scribtexttt{xxxx}}}\RktPn{(}\RktSym{compare}\mbox{\hphantom{\Scribtexttt{x}}}\RktSym{r15}\mbox{\hphantom{\Scribtexttt{x}}}\RktVal{0}\RktPn{)}\RktPn{)}

\mbox{\hphantom{\Scribtexttt{xx}}}\mbox{\hphantom{\Scribtexttt{xx}}}\RktPn{(}\RktSym{jump{-}if}\mbox{\hphantom{\Scribtexttt{x}}}\RktSym{=}\mbox{\hphantom{\Scribtexttt{x}}}\RktSym{end}\RktPn{)}

\mbox{\hphantom{\Scribtexttt{xx}}}\mbox{\hphantom{\Scribtexttt{xx}}}\RktPn{(}\RktSym{set{\hbox{\texttt{!}}}}\mbox{\hphantom{\Scribtexttt{x}}}\RktSym{r14}\mbox{\hphantom{\Scribtexttt{x}}}\RktPn{(}\RktSym{*}\mbox{\hphantom{\Scribtexttt{x}}}\RktSym{r14}\mbox{\hphantom{\Scribtexttt{x}}}\RktSym{r15}\RktPn{)}\RktPn{)}

\mbox{\hphantom{\Scribtexttt{xx}}}\mbox{\hphantom{\Scribtexttt{xx}}}\RktPn{(}\RktSym{set{\hbox{\texttt{!}}}}\mbox{\hphantom{\Scribtexttt{x}}}\RktSym{r15}\mbox{\hphantom{\Scribtexttt{x}}}\RktPn{(}\RktSym{+}\mbox{\hphantom{\Scribtexttt{x}}}\RktSym{r15}\mbox{\hphantom{\Scribtexttt{x}}}\RktVal{\mbox{{-}1}}\RktPn{)}\RktPn{)}

\mbox{\hphantom{\Scribtexttt{xx}}}\mbox{\hphantom{\Scribtexttt{xx}}}\RktPn{(}\RktSym{jump}\mbox{\hphantom{\Scribtexttt{x}}}\RktSym{fact}\RktPn{)}

\mbox{\hphantom{\Scribtexttt{xx}}}\mbox{\hphantom{\Scribtexttt{xx}}}\RktPn{(}\RktSym{with{-}label}\mbox{\hphantom{\Scribtexttt{x}}}\RktSym{end}

\mbox{\hphantom{\Scribtexttt{xx}}}\mbox{\hphantom{\Scribtexttt{xxxx}}}\RktPn{(}\RktSym{set{\hbox{\texttt{!}}}}\mbox{\hphantom{\Scribtexttt{x}}}\RktSym{rax}\mbox{\hphantom{\Scribtexttt{x}}}\RktSym{r14}\RktPn{)}\RktPn{)}

\mbox{\hphantom{\Scribtexttt{xx}}}\mbox{\hphantom{\Scribtexttt{xx}}}\RktPn{(}\RktSym{jump}\mbox{\hphantom{\Scribtexttt{x}}}\RktSym{done}\RktPn{)}\RktPn{)}\end{SingleColumn}\end{RktBlk}

\RktRes{120}\end{SingleColumn}\end{RktBlk}\end{SCodeFlow}}

Programs in this \RktModLink{\RktMod{\#lang}} are embedded into Racket, so we can interoperate at
run time, although these interoperability semantics are not well documented nor
the primary goal.

\noindent \LangBox{\LangBoxLabel{\RktModLink{\RktMod{\#lang}}\RktMeta{}\mbox{\hphantom{\Scribtexttt{x}}}\RktMeta{}\RktModLink{\RktSym{cpsc411/hashlangs/base}}\RktMeta{}}}%
{\begin{SCodeFlow}\begin{RktBlk}\begin{SingleColumn}\Scribtexttt{{\Stttextmore} }\RktPn{(}\RktSym{require}\mbox{\hphantom{\Scribtexttt{x}}}\RktPn{(}\RktSym{only{-}in}\mbox{\hphantom{\Scribtexttt{x}}}\RktSym{racket/base}\mbox{\hphantom{\Scribtexttt{x}}}\RktSym{displayln}\mbox{\hphantom{\Scribtexttt{x}}}\RktSym{define}\RktPn{)}\RktPn{)}

\Scribtexttt{{\Stttextmore} }\RktPn{(}\RktSym{displayln}\mbox{\hphantom{\Scribtexttt{x}}}\RktSym{rax}\RktPn{)}

\RktOut{120}

\begin{RktBlk}\begin{SingleColumn}\Scribtexttt{{\Stttextmore} }\RktPn{(}\RktSym{define}\mbox{\hphantom{\Scribtexttt{x}}}\RktPn{(}\RktSym{fact}\mbox{\hphantom{\Scribtexttt{x}}}\RktSym{x}\RktPn{)}

\mbox{\hphantom{\Scribtexttt{xx}}}\mbox{\hphantom{\Scribtexttt{xx}}}\RktPn{(}\RktSym{begin}

\mbox{\hphantom{\Scribtexttt{xx}}}\mbox{\hphantom{\Scribtexttt{xxxx}}}\RktPn{(}\RktSym{set{\hbox{\texttt{!}}}}\mbox{\hphantom{\Scribtexttt{x}}}\RktSym{r15}\mbox{\hphantom{\Scribtexttt{x}}}\RktSym{x}\RktPn{)}

\mbox{\hphantom{\Scribtexttt{xx}}}\mbox{\hphantom{\Scribtexttt{xxxx}}}\RktPn{(}\RktSym{set{\hbox{\texttt{!}}}}\mbox{\hphantom{\Scribtexttt{x}}}\RktSym{r14}\mbox{\hphantom{\Scribtexttt{x}}}\RktVal{1}\RktPn{)}

\mbox{\hphantom{\Scribtexttt{xx}}}\mbox{\hphantom{\Scribtexttt{xxxx}}}\RktPn{(}\RktSym{with{-}label}\mbox{\hphantom{\Scribtexttt{x}}}\RktSym{fact}

\mbox{\hphantom{\Scribtexttt{xx}}}\mbox{\hphantom{\Scribtexttt{xxxxxx}}}\RktPn{(}\RktSym{compare}\mbox{\hphantom{\Scribtexttt{x}}}\RktSym{r15}\mbox{\hphantom{\Scribtexttt{x}}}\RktVal{0}\RktPn{)}\RktPn{)}

\mbox{\hphantom{\Scribtexttt{xx}}}\mbox{\hphantom{\Scribtexttt{xxxx}}}\RktPn{(}\RktSym{jump{-}if}\mbox{\hphantom{\Scribtexttt{x}}}\RktSym{=}\mbox{\hphantom{\Scribtexttt{x}}}\RktSym{end}\RktPn{)}

\mbox{\hphantom{\Scribtexttt{xx}}}\mbox{\hphantom{\Scribtexttt{xxxx}}}\RktPn{(}\RktSym{set{\hbox{\texttt{!}}}}\mbox{\hphantom{\Scribtexttt{x}}}\RktSym{r14}\mbox{\hphantom{\Scribtexttt{x}}}\RktPn{(}\RktSym{*}\mbox{\hphantom{\Scribtexttt{x}}}\RktSym{r14}\mbox{\hphantom{\Scribtexttt{x}}}\RktSym{r15}\RktPn{)}\RktPn{)}

\mbox{\hphantom{\Scribtexttt{xx}}}\mbox{\hphantom{\Scribtexttt{xxxx}}}\RktPn{(}\RktSym{set{\hbox{\texttt{!}}}}\mbox{\hphantom{\Scribtexttt{x}}}\RktSym{r15}\mbox{\hphantom{\Scribtexttt{x}}}\RktPn{(}\RktSym{+}\mbox{\hphantom{\Scribtexttt{x}}}\RktSym{r15}\mbox{\hphantom{\Scribtexttt{x}}}\RktVal{\mbox{{-}1}}\RktPn{)}\RktPn{)}

\mbox{\hphantom{\Scribtexttt{xx}}}\mbox{\hphantom{\Scribtexttt{xxxx}}}\RktPn{(}\RktSym{jump}\mbox{\hphantom{\Scribtexttt{x}}}\RktSym{fact}\RktPn{)}

\mbox{\hphantom{\Scribtexttt{xx}}}\mbox{\hphantom{\Scribtexttt{xxxx}}}\RktPn{(}\RktSym{with{-}label}\mbox{\hphantom{\Scribtexttt{x}}}\RktSym{end}

\mbox{\hphantom{\Scribtexttt{xx}}}\mbox{\hphantom{\Scribtexttt{xxxxxx}}}\RktPn{(}\RktSym{set{\hbox{\texttt{!}}}}\mbox{\hphantom{\Scribtexttt{x}}}\RktSym{rax}\mbox{\hphantom{\Scribtexttt{x}}}\RktSym{r14}\RktPn{)}\RktPn{)}

\mbox{\hphantom{\Scribtexttt{xx}}}\mbox{\hphantom{\Scribtexttt{xxxx}}}\RktPn{(}\RktSym{jump}\mbox{\hphantom{\Scribtexttt{x}}}\RktSym{done}\RktPn{)}\RktPn{)}\RktPn{)}\end{SingleColumn}\end{RktBlk}

\Scribtexttt{{\Stttextmore} }\RktPn{(}\RktSym{fact}\mbox{\hphantom{\Scribtexttt{x}}}\RktVal{6}\RktPn{)}

\RktRes{720}\end{SingleColumn}\end{RktBlk}\end{SCodeFlow}}

\Ssubsection{Registers}{Registers}\label{t:x28part_x22Registersx22x29}

Registers should be modelled as global variables, provided by the language.
However, Racket does not allow a variable provided by a module to be modified by
another module using \RktSym{set{\hbox{\texttt{!}}}}, so we cannot define registers as in
\RktPn{(}\RktSym{define}\Scribtexttt{ }\RktSym{rax}\Scribtexttt{ }\RktPn{(}\RktSym{void}\RktPn{)}\RktPn{)}.
We could define the registers in the user{'}s module by non{-}hygienically
introducing them in \RktSym{\#\%module{-}begin} (the module{-}level interposition
point), but breaking hygiene is complicated and fragile so should be avoided,
and introducing module{-}level state that is supposed to be global doesn{'}t allow
interoperation with Racket.

Instead, we define each register as a \RktSym{box}.
The \RktSym{box} is exported, and the contents of the \RktSym{box} is modified
instead of the variable itself.
Normally, a user creates a box as in \RktPn{(}\RktSym{define}\Scribtexttt{ }\RktSym{x}\Scribtexttt{ }\RktPn{(}\RktSym{box}\Scribtexttt{ }\RktVal{0}\RktPn{)}\RktPn{)}, assigns to a
box using \RktPn{(}\RktSym{set{-}box{\hbox{\texttt{!}}}}\Scribtexttt{ }\RktSym{x}\Scribtexttt{ }\RktVal{1}\RktPn{)}, and dereferences a box explicitly using
\RktPn{(}\RktSym{unbox}\Scribtexttt{ }\RktSym{x}\RktPn{)}.
To embed registers as boxes, we want to transparently transform any use of
\RktSym{set{\hbox{\texttt{!}}}} on a register, and any use of a register, into one of these
operations.

Thankfully, Racket{'}s \RktSym{set{\hbox{\texttt{!}}}} was designed with extension in mind.
It cooperates with any identifier that is defined as a variable{-}like
transformer using \RktSym{make{-}variable{-}like{-}transformer}.
Such macros have two definitions attached: one that is used when they appear on
their own (like a reference to a variable), and one that is used by
\RktSym{set{\hbox{\texttt{!}}}} during macro expansion when the variable{-}like macro appears on the
left{-}hand side of a \RktSym{set{\hbox{\texttt{!}}}}.
For example, the definition of the register \RktSym{rax} is given below.

\begin{SCodeFlow}\begin{RktBlk}\begin{SingleColumn}\RktPn{(}\RktSym{define}\mbox{\hphantom{\Scribtexttt{x}}}\RktVar{{\char`\_}rax}\mbox{\hphantom{\Scribtexttt{x}}}\RktPn{(}\RktSym{box}\mbox{\hphantom{\Scribtexttt{x}}}\RktPn{(}\RktSym{void}\RktPn{)}\RktPn{)}\RktPn{)}

\mbox{\hphantom{\Scribtexttt{x}}}

\RktPn{(}\RktSym{define{-}syntax}\mbox{\hphantom{\Scribtexttt{x}}}\RktSym{rax}

\mbox{\hphantom{\Scribtexttt{xx}}}\RktPn{(}\RktSym{make{-}variable{-}like{-}transformer}

\mbox{\hphantom{\Scribtexttt{xxx}}}\RktRdr{\#{\textasciigrave}}\RktPn{(}\RktSym{unbox}\mbox{\hphantom{\Scribtexttt{x}}}\RktVar{{\char`\_}rax}\RktPn{)}

\mbox{\hphantom{\Scribtexttt{xxx}}}\RktPn{(}\RktSym{lambda}\mbox{\hphantom{\Scribtexttt{x}}}\RktPn{(}\RktSym{stx}\RktPn{)}

\mbox{\hphantom{\Scribtexttt{xxxxx}}}\RktPn{(}\RktSym{syntax{-}parse}\mbox{\hphantom{\Scribtexttt{x}}}\RktSym{stx}

\mbox{\hphantom{\Scribtexttt{xxxxxxx}}}\RktPn{[}\RktPn{(}\RktSym{set{\hbox{\texttt{!}}}}\mbox{\hphantom{\Scribtexttt{x}}}\RktSym{{\char`\_}}\mbox{\hphantom{\Scribtexttt{x}}}\RktSym{v}\RktPn{)}

\mbox{\hphantom{\Scribtexttt{xxxxxxxx}}}\RktRdr{\#{\textasciigrave}}\RktPn{(}\RktSym{set{-}box{\hbox{\texttt{!}}}}\mbox{\hphantom{\Scribtexttt{x}}}\RktVar{{\char`\_}rax}\mbox{\hphantom{\Scribtexttt{x}}}\RktSym{v}\RktPn{)}\RktPn{]}\RktPn{)}\RktPn{)}\RktPn{)}\RktPn{)}\end{SingleColumn}\end{RktBlk}\end{SCodeFlow}

For each register, we create two definitions: a run{-}time representation, and a
compile{-}time transformer for references of and assignments to the surface syntax
name for the register.
The run{-}time representation is prefixed by an underscore, and is defined as a
\RktSym{box}, initialized either to \RktPn{(}\RktSym{void}\RktPn{)} or a value define by the
run{-}time system.
The compile{-}time transformer uses \RktSym{make{-}variable{-}like{-}transformer} to
transform references, using the first argument, and assignments, using the
second argument.

When the defined syntax, \RktSym{rax} in this case, appears as a reference on
its own and not in operator position, the transformer unconditionally expands to
the first argument, in this case generating the syntax object representing the
run{-}time expression \RktPn{(}\RktSym{unbox}\Scribtexttt{ }\RktVar{{\char`\_}rax}\RktPn{)}.

When the syntax appears as the left{-}hand side of a \RktSym{set{\hbox{\texttt{!}}}}, the expander
delegates to the second argument of \RktSym{make{-}variable{-}like{-}transformer} to
expand the entire \RktSym{set{\hbox{\texttt{!}}}} expression.
We know that the syntax of the expression must be \RktPn{(}\RktSym{set{\hbox{\texttt{!}}}}\Scribtexttt{ }\RktSym{rax}\Scribtexttt{ }\RktSym{v}\RktPn{)} for this
transformer to be called, and we only need the syntax of \RktSym{v}.
We rewrite the syntax \RktPn{(}\RktSym{set{\hbox{\texttt{!}}}}\Scribtexttt{ }\RktSym{rax}\Scribtexttt{ }\RktSym{v}\RktPn{)} to the syntax object representing
the run{-}time expression \RktPn{(}\RktSym{set{-}box{\hbox{\texttt{!}}}}\Scribtexttt{ }\RktVar{{\char`\_}rax}\Scribtexttt{ }\RktSym{v}\RktPn{)}.

Now, all references to \RktSym{rax} as a variable implicitly expand to
\RktPn{(}\RktSym{unbox}\Scribtexttt{ }\RktVar{{\char`\_}rax}\RktPn{)}, and all instances of \RktPn{(}\RktSym{set{\hbox{\texttt{!}}}}\Scribtexttt{ }\RktSym{rax}\Scribtexttt{ }\RktSym{v}\RktPn{)} expand to
\RktPn{(}\RktSym{set{-}box{\hbox{\texttt{!}}}}\Scribtexttt{ }\RktVar{{\char`\_}rax}\Scribtexttt{ }\RktSym{v}\RktPn{)}.
Further, the embedding of this register is local{---}we only define the language
feature \RktSym{rax}, and did not need to modify any existing definitions.

\RktSym{make{-}variable{-}like{-}transformer} is not necessary to implement registers,
but is a useful abstraction of the pattern, and is already designed to cooperate
with \RktSym{set{\hbox{\texttt{!}}}}.
We could alternatively redefine \RktSym{set{\hbox{\texttt{!}}}}, as we did in \ChapRefLocalUC{t:x28part_x22secx3aintrox22x29}{1}{Introduction},
but using variable{-}like transformers instead means we can implement the register
as its own feature, separate from \RktSym{set{\hbox{\texttt{!}}}}, isolating the embedding of the
register rather than spreading it across two definitions (the register as a
reference, and a separate extension of \RktSym{set{\hbox{\texttt{!}}}}).

\Ssubsection{Memory and Pointer Operations}{Memory and Pointer Operations}\label{t:x28part_x22Memoryx5fandx5fPointerx5fOperationsx22x29}

To model memory, we create two vectors: \RktSym{stack} and \RktSym{memory}.
Similar to registers, we must extend the behaviour of \RktSym{set{\hbox{\texttt{!}}}} to support
setting memory addresses, and define the interpretation of dereferencing
addresses.
Unfortunately, since addresses of the form \RktPn{(}\RktSym{rbp}\Scribtexttt{ }\RktSym{\mbox{{-}}}\Scribtexttt{ }\RktSym{offset}\RktPn{)} are compound
rather than identifiers, we cannot use \RktSym{make{-}variable{-}like{-}transformer}.
This suggests the need for a new abstraction, which we return to in
\ChapRefLocalUC{t:x28part_x22secx3adiscussionx22x29}{6}{Discussion}.

We start with the stack, since it has more structure.
The stack pointer is required to live in \RktSym{rbp}\NoteBox{\NoteContent{Actually, it holds
the frame base pointer, but we haven{'}t introduced the frame yet, so they{'}re the
same for now.}}, and this register is only allowed to be used in a particular,
stack{-}like manner, specified by the compiler.
\RktSym{rbp} can be decremented by a word{-}aligned integer{-}literal offset.
Its value can be read directly, or a value can be read from it at a word{-}aligned
offset.
The stack pointer starts at the very end of the stack, and stack allocation
should grow backwards towards the front.

Like with other registers, we create a box \RktVar{{\char`\_}rbp} to hold the current
stack pointer.
However, unlike the other registers, we define \RktSym{rbp} as a custom macro,
rather than as a variable{-}like transformer.

\begin{SCodeFlow}\begin{RktBlk}\begin{SingleColumn}\RktPn{(}\RktSym{define{-}syntax}\mbox{\hphantom{\Scribtexttt{x}}}\RktPn{(}\RktSym{rbp}\mbox{\hphantom{\Scribtexttt{x}}}\RktSym{stx}\RktPn{)}

\mbox{\hphantom{\Scribtexttt{xx}}}\RktPn{(}\RktSym{syntax{-}parse}\mbox{\hphantom{\Scribtexttt{x}}}\RktSym{stx}

\mbox{\hphantom{\Scribtexttt{xxxx}}}\RktPn{[}\RktSym{{\hbox{\texttt{:}}}id}

\mbox{\hphantom{\Scribtexttt{xxxxx}}}\RktRdr{\#{\textquotesingle}}\RktPn{(}\RktSym{unbox}\mbox{\hphantom{\Scribtexttt{x}}}\RktVar{{\char`\_}rbp}\RktPn{)}\RktPn{]}

\mbox{\hphantom{\Scribtexttt{xxxx}}}\RktPn{[}\RktPn{(}\RktSym{rbp}\mbox{\hphantom{\Scribtexttt{x}}}\RktPn{(}\RktSym{{\textasciitilde}datum}\mbox{\hphantom{\Scribtexttt{x}}}\RktSym{\mbox{{-}}}\RktPn{)}\mbox{\hphantom{\Scribtexttt{x}}}\RktSym{offset{\hbox{\texttt{:}}}integer}\RktPn{)}

\mbox{\hphantom{\Scribtexttt{xxxxx}}}\RktRdr{\#{\textasciigrave}}\RktPn{(}\RktSym{vector{-}ref}\mbox{\hphantom{\Scribtexttt{x}}}\RktSym{stack}\mbox{\hphantom{\Scribtexttt{x}}}\RktPn{(}\RktSym{\mbox{{-}}}\mbox{\hphantom{\Scribtexttt{x}}}\RktPn{(}\RktSym{unbox}\mbox{\hphantom{\Scribtexttt{x}}}\RktVar{{\char`\_}rbp}\RktPn{)}\mbox{\hphantom{\Scribtexttt{x}}}\RktSym{offset}\RktPn{)}\RktPn{)}\RktPn{]}\RktPn{)}\RktPn{)}\end{SingleColumn}\end{RktBlk}\end{SCodeFlow}

The first clause only matches when \RktSym{rbp} appears by itself as an
identifier, not in operator position, and binds no pattern variable.
The second clause matches \RktSym{rbp} in operator position, followed by a
displacement{-}mode offset, indicated by the literal symbol \RktSym{\mbox{{-}}} and an
integer literal.
Referencing \RktSym{rbp} in non{-}operator position produces the current stack
pointer, but the syntax \RktPn{(}\RktSym{rbp}\Scribtexttt{ }\RktSym{\mbox{{-}}}\Scribtexttt{ }\RktSym{offset}\RktPn{)} (when not used as the left{-}hand
side of a \RktSym{set{\hbox{\texttt{!}}}}) dereferences the address at the current stack pointer
minus the offset.

\LangBox{\LangBoxLabel{\RktModLink{\RktMod{\#lang}}\RktMeta{}\mbox{\hphantom{\Scribtexttt{x}}}\RktMeta{}\RktModLink{\RktSym{cpsc411/hashlangs/base}}\RktMeta{}}}%
{\begin{SCodeFlow}\begin{RktBlk}\begin{SingleColumn}\Scribtexttt{{\Stttextmore} }\RktPn{(}\RktSym{require}\mbox{\hphantom{\Scribtexttt{x}}}\RktPn{(}\RktSym{only{-}in}\mbox{\hphantom{\Scribtexttt{x}}}\RktSym{racket/base}\mbox{\hphantom{\Scribtexttt{x}}}\RktSym{vector{-}length}\RktPn{)}\RktPn{)}

\Scribtexttt{{\Stttextmore} }\RktPn{(}\RktSym{vector{-}length}\mbox{\hphantom{\Scribtexttt{x}}}\RktSym{stack}\RktPn{)}

\RktRes{12960}

\Scribtexttt{{\Stttextmore} }\RktPn{(}\RktSym{rbp}\mbox{\hphantom{\Scribtexttt{x}}}\RktSym{\mbox{{-}}}\mbox{\hphantom{\Scribtexttt{x}}}\RktVal{0}\RktPn{)}

\RktRes{{\textquotesingle}uninit}

\Scribtexttt{{\Stttextmore} }\RktSym{rbp}

\RktRes{12959}

\Scribtexttt{{\Stttextmore} }\RktPn{(}\RktSym{rbp}\mbox{\hphantom{\Scribtexttt{x}}}\RktSym{+}\mbox{\hphantom{\Scribtexttt{x}}}\RktVal{0}\RktPn{)}

\RktErr{eval:10:0: rbp: expected the literal symbol {\textasciigrave}{-}{\textquotesingle}}

\RktErr{}\mbox{\hphantom{\Scribtexttt{xx}}}\RktErr{at: +}

\RktErr{}\mbox{\hphantom{\Scribtexttt{xx}}}\RktErr{in: (rbp + 0)}

\Scribtexttt{{\Stttextmore} }\RktPn{(}\RktSym{rbp}\mbox{\hphantom{\Scribtexttt{x}}}\RktSym{\mbox{{-}}}\mbox{\hphantom{\Scribtexttt{x}}}\RktSym{rax}\RktPn{)}

\RktErr{eval:11:0: rbp: expected integer}

\RktErr{}\mbox{\hphantom{\Scribtexttt{xx}}}\RktErr{at: rax}

\RktErr{}\mbox{\hphantom{\Scribtexttt{xx}}}\RktErr{in: (rbp {-} rax)}\end{SingleColumn}\end{RktBlk}\end{SCodeFlow}}

When dereferencing the address at \RktPn{(}\RktSym{rbp}\Scribtexttt{ }\RktSym{\mbox{{-}}}\Scribtexttt{ }\RktVal{0}\RktPn{)}, we get an uninitialized
value (since we never set that address).
But when referencing \RktSym{rbp}, we get the address of the end of the stack.
Other syntax for \RktSym{rbp} is invalid.
We have not spent much effort on error reporting yet, but \RktSym{syntax{-}parse}
occassionally makes good error reporting easy.

This macro is only used when the address appears as a value; we implement
assigning to the stack separately.
The implementation for assignment to the stack is entirely in \RktSym{set{\hbox{\texttt{!}}}},
which must detect what kind of left{-}hand side is being used, and expand
depending on the kind of address.

There are three cases in \RktSym{set{\hbox{\texttt{!}}}} for stack addresses: the stack pointer is
incremented, the stack pointer is updated, or some stack address is assigned.

\begin{SCodeFlow}\begin{RktBlk}\begin{SingleColumn}\RktPn{(}\RktSym{define{-}syntax}\mbox{\hphantom{\Scribtexttt{x}}}\RktPn{(}\RktSym{set{\hbox{\texttt{!}}}}\mbox{\hphantom{\Scribtexttt{x}}}\RktSym{stx}\RktPn{)}

\mbox{\hphantom{\Scribtexttt{xx}}}\RktPn{(}\RktSym{syntax{-}parse}\mbox{\hphantom{\Scribtexttt{x}}}\RktSym{stx}

\mbox{\hphantom{\Scribtexttt{xxxx}}}\RktCmt{;}\RktCmt{~}\RktCmt{Stack pointer increment}

\mbox{\hphantom{\Scribtexttt{xxxx}}}\RktPn{[}\RktPn{(}\RktSym{set{\hbox{\texttt{!}}}}\mbox{\hphantom{\Scribtexttt{x}}}\RktPn{(}\RktSym{{\textasciitilde}literal}\mbox{\hphantom{\Scribtexttt{x}}}\RktSym{rbp}\RktPn{)}\mbox{\hphantom{\Scribtexttt{x}}}\RktPn{(}\RktSym{binop}\mbox{\hphantom{\Scribtexttt{x}}}\RktPn{(}\RktSym{{\textasciitilde}literal}\mbox{\hphantom{\Scribtexttt{x}}}\RktSym{rbp}\RktPn{)}\mbox{\hphantom{\Scribtexttt{x}}}\RktSym{v}\RktPn{)}\RktPn{)}

\mbox{\hphantom{\Scribtexttt{xxxxx}}}\RktRdr{\#{\textasciigrave}}\RktPn{(}\RktSym{set{-}box{\hbox{\texttt{!}}}}\mbox{\hphantom{\Scribtexttt{x}}}\RktVar{{\char`\_}rbp}\mbox{\hphantom{\Scribtexttt{x}}}\RktPn{(}\RktSym{binop}\mbox{\hphantom{\Scribtexttt{x}}}\RktSym{rbp}\mbox{\hphantom{\Scribtexttt{x}}}\RktSym{v}\RktPn{)}\RktPn{)}\RktPn{]}

\mbox{\hphantom{\Scribtexttt{xxxx}}}\RktCmt{;}\RktCmt{~}\RktCmt{Stack pointer update}

\mbox{\hphantom{\Scribtexttt{xxxx}}}\RktPn{[}\RktPn{(}\RktSym{set{\hbox{\texttt{!}}}}\mbox{\hphantom{\Scribtexttt{x}}}\RktPn{(}\RktSym{{\textasciitilde}literal}\mbox{\hphantom{\Scribtexttt{x}}}\RktSym{rbp}\RktPn{)}\mbox{\hphantom{\Scribtexttt{x}}}\RktSym{v}\RktPn{)}

\mbox{\hphantom{\Scribtexttt{xxxxx}}}\RktRdr{\#{\textasciigrave}}\RktPn{(}\RktSym{set{-}box{\hbox{\texttt{!}}}}\mbox{\hphantom{\Scribtexttt{x}}}\RktVar{{\char`\_}rbp}\mbox{\hphantom{\Scribtexttt{x}}}\RktSym{v}\RktPn{)}\RktPn{]}

\mbox{\hphantom{\Scribtexttt{xxxx}}}\RktCmt{;}\RktCmt{~}\RktCmt{Stack slot update}

\mbox{\hphantom{\Scribtexttt{xxxx}}}\RktPn{[}\RktPn{(}\RktSym{set{\hbox{\texttt{!}}}}\mbox{\hphantom{\Scribtexttt{x}}}\RktPn{(}\RktPn{(}\RktSym{{\textasciitilde}literal}\mbox{\hphantom{\Scribtexttt{x}}}\RktSym{rbp}\RktPn{)}\mbox{\hphantom{\Scribtexttt{x}}}\RktSym{\mbox{{-}}}\mbox{\hphantom{\Scribtexttt{x}}}\RktSym{offset}\RktPn{)}\mbox{\hphantom{\Scribtexttt{x}}}\RktSym{v2}\RktPn{)}

\mbox{\hphantom{\Scribtexttt{xxxxx}}}\RktRdr{\#{\textasciigrave}}\RktPn{(}\RktSym{vector{-}set{\hbox{\texttt{!}}}}\mbox{\hphantom{\Scribtexttt{x}}}\RktSym{stack}\mbox{\hphantom{\Scribtexttt{x}}}\RktPn{(}\RktSym{\mbox{{-}}}\mbox{\hphantom{\Scribtexttt{x}}}\RktPn{(}\RktSym{unbox}\mbox{\hphantom{\Scribtexttt{x}}}\RktVar{{\char`\_}rbp}\RktPn{)}\mbox{\hphantom{\Scribtexttt{x}}}\RktSym{offset}\RktPn{)}\mbox{\hphantom{\Scribtexttt{x}}}\RktSym{v2}\RktPn{)}\RktPn{]}

\mbox{\hphantom{\Scribtexttt{xxxx}}}\RktSym{{\hbox{\texttt{.}}}{\hbox{\texttt{.}}}{\hbox{\texttt{.}}}{\hbox{\texttt{.}}}}\RktPn{)}\RktPn{)}\end{SingleColumn}\end{RktBlk}\end{SCodeFlow}

This definition uses the syntax class \RktSym{{\textasciitilde}literal} to match only when the
left{-}hand side is or includes the bound identifier \RktSym{rbp}; these clauses
are only triggered when \RktSym{set{\hbox{\texttt{!}}}} is used with the stack pointer, not an
arbitrary other register.
The first two cases only use \RktSym{rbp} as the left{-}hand side, so could be
supported by \RktSym{make{-}variable{-}like{-}transformer}.
These clauses expand to update \RktSym{rbp}, modifying the current stack
pointer.
The third clause uses a compound left{-}hand side.
Since we need this compound left{-}hand side (as well as the earlier compound
dereferences), we cannot use \RktSym{make{-}variable{-}like{-}transformer}, and must
spread our implementation of \RktSym{rbp} between \RktSym{set{\hbox{\texttt{!}}}} and a custom
tranformer for \RktSym{rbp}.
The final clause expands to modify a location on the stack, computed as the
current stack pointer minus the offset.

\LangBox{\LangBoxLabel{\RktModLink{\RktMod{\#lang}}\RktMeta{}\mbox{\hphantom{\Scribtexttt{x}}}\RktMeta{}\RktModLink{\RktSym{cpsc411/hashlangs/base}}\RktMeta{}}}%
{\begin{SCodeFlow}\begin{RktBlk}\begin{SingleColumn}\Scribtexttt{{\Stttextmore} }\RktPn{(}\RktSym{set{\hbox{\texttt{!}}}}\mbox{\hphantom{\Scribtexttt{x}}}\RktPn{(}\RktSym{rbp}\mbox{\hphantom{\Scribtexttt{x}}}\RktSym{\mbox{{-}}}\mbox{\hphantom{\Scribtexttt{x}}}\RktVal{0}\RktPn{)}\mbox{\hphantom{\Scribtexttt{x}}}\RktVal{5}\RktPn{)}

\Scribtexttt{{\Stttextmore} }\RktPn{(}\RktSym{rbp}\mbox{\hphantom{\Scribtexttt{x}}}\RktSym{\mbox{{-}}}\mbox{\hphantom{\Scribtexttt{x}}}\RktVal{0}\RktPn{)}

\RktRes{5}\end{SingleColumn}\end{RktBlk}\end{SCodeFlow}}

To implement the heap, we use another vector, called \RktSym{memory}.
Unlike the stack, memory addresses can end up in any register, so we cannot
implement the register as its own transformer or special case any particular
register like we did with \RktSym{rbp}.
Instead, we must detect the general form of the address operands in the
implementation of \RktSym{set{\hbox{\texttt{!}}}}.
There are two cases to consider: assigning to memory (the left{-}hand side
has the form \RktPn{(}\RktSym{reg}\Scribtexttt{ }\RktSym{binop}\Scribtexttt{ }\RktSym{offset}\RktPn{)}), or reading from memory (the
right{-}hand side has the form \RktPn{(}\RktSym{reg}\Scribtexttt{ }\RktSym{binop}\Scribtexttt{ }\RktSym{offset}\RktPn{)}).
In each case, we generate an instruction using either \RktSym{vector{-}set{\hbox{\texttt{!}}}} or
\RktSym{vector{-}ref} accessing \RktSym{memory} at the location base plus or minus
the offset.

\begin{SCodeFlow}\begin{RktBlk}\begin{SingleColumn}\RktPn{(}\RktSym{begin{-}for{-}syntax}

\mbox{\hphantom{\Scribtexttt{xx}}}\RktPn{(}\RktSym{define{-}syntax{-}class}\mbox{\hphantom{\Scribtexttt{x}}}\RktSym{addr{-}op}

\mbox{\hphantom{\Scribtexttt{xxxx}}}\RktPn{(}\RktSym{pattern}\mbox{\hphantom{\Scribtexttt{x}}}\RktPn{(}\RktSym{{\textasciitilde}or}\mbox{\hphantom{\Scribtexttt{x}}}\RktPn{(}\RktSym{{\textasciitilde}datum}\mbox{\hphantom{\Scribtexttt{x}}}\RktSym{+}\RktPn{)}\mbox{\hphantom{\Scribtexttt{x}}}\RktPn{(}\RktSym{{\textasciitilde}datum}\mbox{\hphantom{\Scribtexttt{x}}}\RktSym{\mbox{{-}}}\RktPn{)}\RktPn{)}\RktPn{)}\RktPn{)}

\mbox{\hphantom{\Scribtexttt{x}}}

\mbox{\hphantom{\Scribtexttt{xx}}}\RktPn{(}\RktSym{define{-}syntax{-}class}\mbox{\hphantom{\Scribtexttt{x}}}\RktSym{not{-}rbp}

\mbox{\hphantom{\Scribtexttt{xxxx}}}\RktPn{(}\RktSym{pattern}\mbox{\hphantom{\Scribtexttt{x}}}\RktPn{(}\RktSym{{\textasciitilde}not}\mbox{\hphantom{\Scribtexttt{x}}}\RktPn{(}\RktSym{{\textasciitilde}literal}\mbox{\hphantom{\Scribtexttt{x}}}\RktSym{rbp}\RktPn{)}\RktPn{)}\RktPn{)}\RktPn{)}\RktPn{)}

\mbox{\hphantom{\Scribtexttt{x}}}

\RktPn{(}\RktSym{define{-}syntax}\mbox{\hphantom{\Scribtexttt{x}}}\RktPn{(}\RktSym{set{\hbox{\texttt{!}}}}\mbox{\hphantom{\Scribtexttt{x}}}\RktSym{stx}\RktPn{)}

\mbox{\hphantom{\Scribtexttt{xx}}}\RktPn{(}\RktSym{syntax{-}parse}\mbox{\hphantom{\Scribtexttt{x}}}\RktSym{stx}

\mbox{\hphantom{\Scribtexttt{xxxx}}}\RktSym{{\hbox{\texttt{.}}}{\hbox{\texttt{.}}}{\hbox{\texttt{.}}}{\hbox{\texttt{.}}}}

\mbox{\hphantom{\Scribtexttt{xxxx}}}\RktCmt{;}\RktCmt{~}\RktCmt{Assign to memory}

\mbox{\hphantom{\Scribtexttt{xxxx}}}\RktPn{[}\RktPn{(}\RktSym{set{\hbox{\texttt{!}}}}\mbox{\hphantom{\Scribtexttt{x}}}\RktPn{(}\RktSym{base{\hbox{\texttt{:}}}not{-}rbp}\mbox{\hphantom{\Scribtexttt{x}}}\RktSym{op{\hbox{\texttt{:}}}addr{-}op}\mbox{\hphantom{\Scribtexttt{x}}}\RktSym{offset}\RktPn{)}\mbox{\hphantom{\Scribtexttt{x}}}\RktSym{value}\RktPn{)}

\mbox{\hphantom{\Scribtexttt{xxxxx}}}\RktRdr{\#{\textasciigrave}}\RktPn{(}\RktSym{vector{-}set{\hbox{\texttt{!}}}}\mbox{\hphantom{\Scribtexttt{x}}}\RktSym{memory}\mbox{\hphantom{\Scribtexttt{x}}}\RktPn{(}\RktSym{op}\mbox{\hphantom{\Scribtexttt{x}}}\RktSym{base}\mbox{\hphantom{\Scribtexttt{x}}}\RktSym{offset}\RktPn{)}\mbox{\hphantom{\Scribtexttt{x}}}\RktSym{value}\RktPn{)}\RktPn{]}

\mbox{\hphantom{\Scribtexttt{xxxx}}}\RktCmt{;}\RktCmt{~}\RktCmt{Read from memory}

\mbox{\hphantom{\Scribtexttt{xxxx}}}\RktPn{[}\RktPn{(}\RktSym{set{\hbox{\texttt{!}}}}\mbox{\hphantom{\Scribtexttt{x}}}\RktSym{v1{\hbox{\texttt{:}}}id}\mbox{\hphantom{\Scribtexttt{x}}}\RktPn{(}\RktSym{base{\hbox{\texttt{:}}}not{-}rbp}\mbox{\hphantom{\Scribtexttt{x}}}\RktSym{op{\hbox{\texttt{:}}}addr{-}op}\mbox{\hphantom{\Scribtexttt{x}}}\RktSym{offset}\RktPn{)}\RktPn{)}

\mbox{\hphantom{\Scribtexttt{xxxxx}}}\RktRdr{\#{\textasciigrave}}\RktPn{(}\RktSym{r{\hbox{\texttt{:}}}set{\hbox{\texttt{!}}}}\mbox{\hphantom{\Scribtexttt{x}}}\RktSym{v1}\mbox{\hphantom{\Scribtexttt{x}}}\RktPn{(}\RktSym{vector{-}ref}\mbox{\hphantom{\Scribtexttt{x}}}\RktSym{memory}\mbox{\hphantom{\Scribtexttt{x}}}\RktPn{(}\RktSym{op}\mbox{\hphantom{\Scribtexttt{x}}}\RktSym{base}\mbox{\hphantom{\Scribtexttt{x}}}\RktSym{offset}\RktPn{)}\RktPn{)}\RktPn{)}\RktPn{]}

\mbox{\hphantom{\Scribtexttt{xxxx}}}\RktCmt{;}\RktCmt{~}\RktCmt{Else use Racket{\textquotesingle}s set{\hbox{\texttt{!}}}}

\mbox{\hphantom{\Scribtexttt{xxxx}}}\RktPn{[}\RktPn{(}\RktSym{set{\hbox{\texttt{!}}}}\mbox{\hphantom{\Scribtexttt{x}}}\RktSym{v1}\mbox{\hphantom{\Scribtexttt{x}}}\RktSym{v2}\RktPn{)}

\mbox{\hphantom{\Scribtexttt{xxxxx}}}\RktRdr{\#{\textasciigrave}}\RktPn{(}\RktSym{r{\hbox{\texttt{:}}}set{\hbox{\texttt{!}}}}\mbox{\hphantom{\Scribtexttt{x}}}\RktSym{v1}\mbox{\hphantom{\Scribtexttt{x}}}\RktSym{v2}\RktPn{)}\RktPn{]}\RktPn{)}\RktPn{)}\end{SingleColumn}\end{RktBlk}\end{SCodeFlow}

We use two syntax classes, which extend the pattern language of
\RktSym{syntax{-}parse}, to detect the index{-}mode operand.
The base must not be the identifier bound to the stack pointer register, and
there should be an address operator (either the literal syntax \RktSym{+} or
\RktSym{\mbox{{-}}}) between the base and the offset.
We reimport Racket primitives that get redefined by our embedding with the
prefix \RktSym{r{\hbox{\texttt{:}}}}, so \RktSym{r{\hbox{\texttt{:}}}set{\hbox{\texttt{!}}}} refers to Racket{'}s implementation of
\RktSym{set{\hbox{\texttt{!}}}}.

\LangBox{\LangBoxLabel{\RktModLink{\RktMod{\#lang}}\RktMeta{}\mbox{\hphantom{\Scribtexttt{x}}}\RktMeta{}\RktModLink{\RktSym{cpsc411/hashlangs/base}}\RktMeta{}}}%
{\begin{SCodeFlow}\begin{RktBlk}\begin{SingleColumn}\Scribtexttt{{\Stttextmore} }\RktPn{(}\RktSym{vector{-}length}\mbox{\hphantom{\Scribtexttt{x}}}\RktSym{memory}\RktPn{)}

\RktRes{10000}

\Scribtexttt{{\Stttextmore} }\RktPn{(}\RktSym{vector{-}ref}\mbox{\hphantom{\Scribtexttt{x}}}\RktSym{memory}\mbox{\hphantom{\Scribtexttt{x}}}\RktVal{0}\RktPn{)}

\RktRes{{\textquotesingle}unalloced}

\Scribtexttt{{\Stttextmore} }\RktPn{(}\RktSym{set{\hbox{\texttt{!}}}}\mbox{\hphantom{\Scribtexttt{x}}}\RktPn{(}\RktSym{r12}\mbox{\hphantom{\Scribtexttt{x}}}\RktSym{+}\mbox{\hphantom{\Scribtexttt{x}}}\RktVal{0}\RktPn{)}\mbox{\hphantom{\Scribtexttt{x}}}\RktVal{5}\RktPn{)}

\Scribtexttt{{\Stttextmore} }\RktPn{(}\RktSym{set{\hbox{\texttt{!}}}}\mbox{\hphantom{\Scribtexttt{x}}}\RktSym{rax}\mbox{\hphantom{\Scribtexttt{x}}}\RktPn{(}\RktSym{r12}\mbox{\hphantom{\Scribtexttt{x}}}\RktSym{+}\mbox{\hphantom{\Scribtexttt{x}}}\RktVal{0}\RktPn{)}\RktPn{)}

\Scribtexttt{{\Stttextmore} }\RktSym{rax}

\RktRes{5}

\Scribtexttt{{\Stttextmore} }\RktPn{(}\RktSym{vector{-}ref}\mbox{\hphantom{\Scribtexttt{x}}}\RktSym{memory}\mbox{\hphantom{\Scribtexttt{x}}}\RktVal{0}\RktPn{)}

\RktRes{5}\end{SingleColumn}\end{RktBlk}\end{SCodeFlow}}

By default, our run{-}time system puts a pointer to unallocated heap memory in
\RktSym{r12}, which the compiler uses to implement allocation.

\Ssubsection{Labels and Jumps}{Labels and Jumps}\label{t:x28part_x22Labelsx5fandx5fJumpsx22x29}

We model labelled instructions as procedures with no arguments that are
immediately called, and jumps as procedure calls that must not return.

All statements appear under a \RktSym{begin}.
To implement labels, we redefine \RktSym{begin} to transform the sequence of
instructions into a \RktSym{letrec}, binding each label to a procedure.

This definition is the first we{'}ve seen that uses syntax quasiquotation with
unquoting.
The syntax unquote \RktInBG{\RktIn{\#,}} splices into a syntax template a value computed
by running a compile{-}time expression.
This works analogously to unquote \RktInBG{\RktIn{,}} over quasiquoted lists \RktInBG{\RktIn{{\textasciigrave}}}.
Similarly, syntax splicing unquote \RktInBG{\RktIn{\#,@}} splices a list of values into
the template.

\begin{SCodeFlow}\begin{RktBlk}\begin{SingleColumn}\RktPn{(}\RktSym{begin{-}for{-}syntax}

\mbox{\hphantom{\Scribtexttt{xx}}}\RktPn{(}\RktSym{define}\mbox{\hphantom{\Scribtexttt{x}}}\RktPn{(}\RktSym{labelify}\mbox{\hphantom{\Scribtexttt{x}}}\RktSym{defs}\mbox{\hphantom{\Scribtexttt{x}}}\RktSym{effects}\RktPn{)}

\mbox{\hphantom{\Scribtexttt{xxxx}}}\RktPn{(}\RktSym{match}\mbox{\hphantom{\Scribtexttt{x}}}\RktSym{effects}

\mbox{\hphantom{\Scribtexttt{xxxxxx}}}\RktPn{[}\RktVal{{\textquotesingle}}\RktVal{(}\RktVal{)}\mbox{\hphantom{\Scribtexttt{x}}}\RktPn{(}\RktSym{values}\mbox{\hphantom{\Scribtexttt{x}}}\RktSym{defs}\mbox{\hphantom{\Scribtexttt{x}}}\RktVal{{\textquotesingle}}\RktVal{(}\RktVal{)}\RktPn{)}\RktPn{]}

\mbox{\hphantom{\Scribtexttt{xxxxxx}}}\RktPn{[}\RktPn{(}\RktSym{cons}\mbox{\hphantom{\Scribtexttt{x}}}\RktSym{effect}\mbox{\hphantom{\Scribtexttt{x}}}\RktSym{effects}\RktPn{)}

\mbox{\hphantom{\Scribtexttt{xxxxxxx}}}\RktPn{(}\RktSym{syntax{-}parse}\mbox{\hphantom{\Scribtexttt{x}}}\RktSym{effect}

\mbox{\hphantom{\Scribtexttt{xxxxxxxxx}}}\RktPn{\#{\hbox{\texttt{:}}}literals}\mbox{\hphantom{\Scribtexttt{x}}}\RktPn{(}\RktSym{with{-}label}\mbox{\hphantom{\Scribtexttt{x}}}\RktSym{begin}\RktPn{)}

\mbox{\hphantom{\Scribtexttt{xxxxxxxxx}}}\RktPn{[}\RktPn{(}\RktSym{with{-}label}\mbox{\hphantom{\Scribtexttt{x}}}\RktSym{label}\mbox{\hphantom{\Scribtexttt{x}}}\RktSym{effect}\RktPn{)}

\mbox{\hphantom{\Scribtexttt{xxxxxxxxxx}}}\RktPn{(}\RktSym{let{-}values}\mbox{\hphantom{\Scribtexttt{x}}}\RktPn{(}\RktPn{[}\RktPn{(}\RktSym{defs}\mbox{\hphantom{\Scribtexttt{x}}}\RktSym{effects{\char'136}}\RktPn{)}

\mbox{\hphantom{\Scribtexttt{xxxxxxxxxxxxxxxxxxxxxxxx}}}\RktPn{(}\RktSym{labelify}\mbox{\hphantom{\Scribtexttt{x}}}\RktSym{defs}\mbox{\hphantom{\Scribtexttt{x}}}\RktPn{(}\RktSym{cons}\mbox{\hphantom{\Scribtexttt{x}}}\RktRdr{\#{\textquotesingle}}\RktSym{effect}\mbox{\hphantom{\Scribtexttt{x}}}\RktSym{effects}\RktPn{)}\RktPn{)}\RktPn{]}\RktPn{)}

\mbox{\hphantom{\Scribtexttt{xxxxxxxxxxxx}}}\RktPn{(}\RktSym{values}

\mbox{\hphantom{\Scribtexttt{xxxxxxxxxxxxx}}}\RktPn{(}\RktSym{cons}\mbox{\hphantom{\Scribtexttt{x}}}\RktRdr{\#{\textasciigrave}}\RktPn{[}\RktSym{label}\mbox{\hphantom{\Scribtexttt{x}}}\RktPn{(}\RktSym{lambda}\mbox{\hphantom{\Scribtexttt{x}}}\RktPn{(}\RktPn{)}\mbox{\hphantom{\Scribtexttt{x}}}\RktPn{(}\RktSym{r{\hbox{\texttt{:}}}begin}\mbox{\hphantom{\Scribtexttt{x}}}\RktRdr{\#,@}\RktSym{effects{\char'136}}\RktPn{)}\RktPn{)}\RktPn{]}\mbox{\hphantom{\Scribtexttt{x}}}\RktSym{defs}\RktPn{)}

\mbox{\hphantom{\Scribtexttt{xxxxxxxxxxxxx}}}\RktPn{(}\RktSym{list}\mbox{\hphantom{\Scribtexttt{x}}}\RktRdr{\#{\textasciigrave}}\RktPn{(}\RktSym{jump}\mbox{\hphantom{\Scribtexttt{x}}}\RktSym{label}\RktPn{)}\RktPn{)}\RktPn{)}\RktPn{)}\RktPn{]}

\mbox{\hphantom{\Scribtexttt{xxxxxxxxx}}}\RktPn{[}\RktPn{(}\RktSym{begin}\mbox{\hphantom{\Scribtexttt{x}}}\RktSym{effects1}\mbox{\hphantom{\Scribtexttt{x}}}\RktSym{{\hbox{\texttt{.}}}{\hbox{\texttt{.}}}{\hbox{\texttt{.}}}}\RktPn{)}

\mbox{\hphantom{\Scribtexttt{xxxxxxxxxx}}}\RktPn{(}\RktSym{labelify}\mbox{\hphantom{\Scribtexttt{x}}}\RktSym{defs}\mbox{\hphantom{\Scribtexttt{x}}}\RktPn{(}\RktSym{append}\mbox{\hphantom{\Scribtexttt{x}}}\RktPn{(}\RktSym{attribute}\mbox{\hphantom{\Scribtexttt{x}}}\RktSym{effects1}\RktPn{)}\mbox{\hphantom{\Scribtexttt{x}}}\RktSym{effects}\RktPn{)}\RktPn{)}\RktPn{]}

\mbox{\hphantom{\Scribtexttt{xxxxxxxxx}}}\RktPn{[}\RktSym{effect}

\mbox{\hphantom{\Scribtexttt{xxxxxxxxxx}}}\RktPn{(}\RktSym{let{-}values}\mbox{\hphantom{\Scribtexttt{x}}}\RktPn{(}\RktPn{[}\RktPn{(}\RktSym{defs}\mbox{\hphantom{\Scribtexttt{x}}}\RktSym{effects{\char'136}}\RktPn{)}\mbox{\hphantom{\Scribtexttt{x}}}\RktPn{(}\RktSym{labelify}\mbox{\hphantom{\Scribtexttt{x}}}\RktSym{defs}\mbox{\hphantom{\Scribtexttt{x}}}\RktSym{effects}\RktPn{)}\RktPn{]}\RktPn{)}

\mbox{\hphantom{\Scribtexttt{xxxxxxxxxxxx}}}\RktPn{(}\RktSym{values}\mbox{\hphantom{\Scribtexttt{x}}}\RktSym{defs}\mbox{\hphantom{\Scribtexttt{x}}}\RktRdr{\#{\textasciigrave}}\RktPn{(}\RktSym{effect}\mbox{\hphantom{\Scribtexttt{x}}}\RktRdr{\#,@}\RktSym{effects{\char'136}}\RktPn{)}\RktPn{)}\RktPn{)}\RktPn{]}\RktPn{)}\RktPn{]}\RktPn{)}\RktPn{)}\RktPn{)}

\mbox{\hphantom{\Scribtexttt{x}}}

\RktPn{(}\RktSym{define{-}syntax}\mbox{\hphantom{\Scribtexttt{x}}}\RktPn{(}\RktSym{begin}\mbox{\hphantom{\Scribtexttt{x}}}\RktSym{stx}\RktPn{)}

\mbox{\hphantom{\Scribtexttt{xx}}}\RktPn{(}\RktSym{let{-}values}\mbox{\hphantom{\Scribtexttt{x}}}\RktPn{(}\RktPn{[}\RktPn{(}\RktSym{defs}\mbox{\hphantom{\Scribtexttt{x}}}\RktSym{effects}\RktPn{)}\mbox{\hphantom{\Scribtexttt{x}}}\RktPn{(}\RktSym{labelify}\mbox{\hphantom{\Scribtexttt{x}}}\RktVal{{\textquotesingle}}\RktVal{(}\RktVal{)}\mbox{\hphantom{\Scribtexttt{x}}}\RktPn{(}\RktSym{cdr}\mbox{\hphantom{\Scribtexttt{x}}}\RktPn{(}\RktSym{syntax{-}{\Stttextmore}list}\mbox{\hphantom{\Scribtexttt{x}}}\RktSym{stx}\RktPn{)}\RktPn{)}\RktPn{)}\RktPn{]}\RktPn{)}

\mbox{\hphantom{\Scribtexttt{xxxx}}}\RktRdr{\#{\textasciigrave}}\RktPn{(}\RktSym{letrec}\mbox{\hphantom{\Scribtexttt{x}}}\RktRdr{\#,}\RktSym{defs}\mbox{\hphantom{\Scribtexttt{x}}}\RktPn{(}\RktSym{r{\hbox{\texttt{:}}}begin}\mbox{\hphantom{\Scribtexttt{x}}}\RktRdr{\#,@}\RktSym{effects}\RktPn{)}\RktPn{)}\RktPn{)}\RktPn{)}\end{SingleColumn}\end{RktBlk}\end{SCodeFlow}

The macro \RktSym{begin} does not immediately pattern match on its input, but
deligates to a compile{-}time procedure \RktSym{labelify}.
This procedure takes the list of statement as syntax objects (created by
converting the syntax object to a list of syntax objects using
\RktSym{syntax{-}{\Stttextmore}list}), without the leading \RktSym{begin} operator (dropped by
calling \RktSym{cdr}).
The procedure traverses the list of instructions \RktPn{(}\RktSym{cons}\Scribtexttt{ }\RktSym{effect}\Scribtexttt{ }\RktSym{effects}\RktPn{)},
collecting a list of definitions to bind with \RktSym{letrec} and a list of
instructions to execute in the body of the \RktSym{letrec}.
When \RktSym{effect} is a labeled instruction \RktPn{(}\RktSym{with{-}label}\Scribtexttt{ }\RktSym{label}\Scribtexttt{ }\RktSym{effect{\char'136}}\RktPn{)}, we create a binding for a procedure definition \RktRdr{\#{\textasciigrave}}\RktPn{[}\RktSym{label}\Scribtexttt{ }\RktPn{(}\RktSym{lambda}\Scribtexttt{ }\RktPn{(}\RktPn{)}\Scribtexttt{ }\RktPn{(}\RktSym{begin}\Scribtexttt{ }\RktSym{effects{\char'136}}\Scribtexttt{ }\RktSym{{\hbox{\texttt{.}}}{\hbox{\texttt{.}}}{\hbox{\texttt{.}}}}\RktPn{)}\RktPn{)}\RktPn{]}, where \RktSym{effects{\char'136}} are the instructions
that remain after recursively labeling the rest of the list of effect, starting
with the labelled one, \RktPn{(}\RktSym{cons}\Scribtexttt{ }\RktRdr{\#{\textquotesingle}}\RktSym{effect{\char'136}}\mbox{\hphantom{\Scribtexttt{xx}}}\RktSym{effects}\RktPn{)}.
We prepend the new definitions to those returned by the recursive call, and
replace the instruction sequence with a jump to the new definition \RktPn{(}\RktSym{list}\Scribtexttt{ }\RktRdr{\#{\textasciigrave}}\RktPn{(}\RktSym{jump}\Scribtexttt{ }\RktSym{label}\RktPn{)}\RktPn{)}.
When we encounter a nested \RktSym{begin}, we prepend all the instructions in
the body to the current \RktSym{effects} and recur.
All other instructions are left in place.
At the end of the loop, we bind all the labels in the scope of the list of
instructions, expanding to (roughly) \RktRdr{\#{\textasciigrave}}\RktPn{(}\RktSym{letrec}\Scribtexttt{ }\RktPn{(}\RktSym{defs}\Scribtexttt{ }\RktSym{{\hbox{\texttt{.}}}{\hbox{\texttt{.}}}{\hbox{\texttt{.}}}}\RktPn{)}\Scribtexttt{ }\RktPn{(}\RktSym{r{\hbox{\texttt{:}}}begin}\Scribtexttt{ }\RktSym{effects{\char'136}}\Scribtexttt{ }\RktSym{{\hbox{\texttt{.}}}{\hbox{\texttt{.}}}{\hbox{\texttt{.}}}}\RktPn{)}\RktPn{)}.

To embed conditional jumps, we create a mutable \RktSym{eq{\hbox{\texttt{?}}}}{-}comparable
hash table mapping each comparison procedure to a flag.
The \RktSym{compare} instruction is modelled as a procedure, which updates each
flag using each comparison procedure, and \RktSym{jump{-}if} jumps when the
comparison procedure{'}s flag is set.

\begin{SCodeFlow}\begin{RktBlk}\begin{SingleColumn}\RktPn{(}\RktSym{define}\mbox{\hphantom{\Scribtexttt{x}}}\RktSym{flags}

\mbox{\hphantom{\Scribtexttt{xx}}}\RktPn{(}\RktSym{make{-}hasheq}

\mbox{\hphantom{\Scribtexttt{xxx}}}\RktVal{{\textasciigrave}}\RktVal{(}\RktVal{(}\RktRdr{,}\RktSym{{\hbox{\texttt{!}}}=}\mbox{\hphantom{\Scribtexttt{x}}}\RktVal{{\hbox{\texttt{.}}} }\RktVal{\#f}\RktVal{)}\mbox{\hphantom{\Scribtexttt{x}}}\RktVal{(}\RktRdr{,}\RktSym{=}\mbox{\hphantom{\Scribtexttt{x}}}\RktVal{{\hbox{\texttt{.}}} }\RktVal{\#f}\RktVal{)}\mbox{\hphantom{\Scribtexttt{x}}}\RktVal{(}\RktRdr{,}\RktSym{{\Stttextless}}\mbox{\hphantom{\Scribtexttt{x}}}\RktVal{{\hbox{\texttt{.}}} }\RktVal{\#f}\RktVal{)}\mbox{\hphantom{\Scribtexttt{x}}}\RktVal{(}\RktRdr{,}\RktSym{{\Stttextless}=}\mbox{\hphantom{\Scribtexttt{x}}}\RktVal{{\hbox{\texttt{.}}} }\RktVal{\#f}\RktVal{)}\mbox{\hphantom{\Scribtexttt{x}}}\RktVal{(}\RktRdr{,}\RktSym{{\Stttextmore}}\mbox{\hphantom{\Scribtexttt{x}}}\RktVal{{\hbox{\texttt{.}}} }\RktVal{\#f}\RktVal{)}\mbox{\hphantom{\Scribtexttt{x}}}\RktVal{(}\RktRdr{,}\RktSym{{\Stttextmore}=}\mbox{\hphantom{\Scribtexttt{x}}}\RktVal{{\hbox{\texttt{.}}} }\RktVal{\#f}\RktVal{)}\RktVal{)}\RktPn{)}\RktPn{)}

\mbox{\hphantom{\Scribtexttt{x}}}

\RktPn{(}\RktSym{define}\mbox{\hphantom{\Scribtexttt{x}}}\RktPn{(}\RktSym{compare}\mbox{\hphantom{\Scribtexttt{x}}}\RktSym{v1}\mbox{\hphantom{\Scribtexttt{x}}}\RktSym{v2}\RktPn{)}

\mbox{\hphantom{\Scribtexttt{xx}}}\RktPn{(}\RktSym{for{-}each}\mbox{\hphantom{\Scribtexttt{x}}}\RktPn{(}\RktSym{lambda}\mbox{\hphantom{\Scribtexttt{x}}}\RktPn{(}\RktSym{cmp}\RktPn{)}\mbox{\hphantom{\Scribtexttt{x}}}\RktPn{(}\RktSym{hash{-}set{\hbox{\texttt{!}}}}\mbox{\hphantom{\Scribtexttt{x}}}\RktSym{flags}\mbox{\hphantom{\Scribtexttt{x}}}\RktSym{cmp}\mbox{\hphantom{\Scribtexttt{x}}}\RktPn{(}\RktSym{cmp}\mbox{\hphantom{\Scribtexttt{x}}}\RktSym{v1}\mbox{\hphantom{\Scribtexttt{x}}}\RktSym{v2}\RktPn{)}\RktPn{)}\RktPn{)}

\mbox{\hphantom{\Scribtexttt{xxxxxxxxxxxx}}}\RktPn{(}\RktSym{list}\mbox{\hphantom{\Scribtexttt{x}}}\RktSym{{\hbox{\texttt{!}}}=}\mbox{\hphantom{\Scribtexttt{x}}}\RktSym{=}\mbox{\hphantom{\Scribtexttt{x}}}\RktSym{{\Stttextless}}\mbox{\hphantom{\Scribtexttt{x}}}\RktSym{{\Stttextless}=}\mbox{\hphantom{\Scribtexttt{x}}}\RktSym{{\Stttextmore}}\mbox{\hphantom{\Scribtexttt{x}}}\RktSym{{\Stttextmore}=}\RktPn{)}\RktPn{)}\RktPn{)}

\mbox{\hphantom{\Scribtexttt{x}}}

\RktPn{(}\RktSym{define}\mbox{\hphantom{\Scribtexttt{x}}}\RktPn{(}\RktSym{jump{-}if}\mbox{\hphantom{\Scribtexttt{x}}}\RktSym{cmp{-}proc}\mbox{\hphantom{\Scribtexttt{x}}}\RktSym{d}\RktPn{)}\mbox{\hphantom{\Scribtexttt{x}}}\RktPn{(}\RktSym{when}\mbox{\hphantom{\Scribtexttt{x}}}\RktPn{(}\RktSym{hash{-}ref}\mbox{\hphantom{\Scribtexttt{x}}}\RktSym{flags}\mbox{\hphantom{\Scribtexttt{x}}}\RktSym{cmp{-}proc}\RktPn{)}\mbox{\hphantom{\Scribtexttt{x}}}\RktPn{(}\RktSym{d}\RktPn{)}\RktPn{)}\RktPn{)}\end{SingleColumn}\end{RktBlk}\end{SCodeFlow}

This implementation uses each procedure pointer as a key in the table.
This can be fragile, as the pointer for a procedure can be unpredictable, but
this local use should not be a problem.
We could avoid the fragility with extra indirection, using symbols as keys in
the flag table and mapping those symbols to the correct procedure, or use
\RktSym{eval}.
However, in either case, \RktSym{jump{-}if} would need to be a macro that quotes
ones of its arguments.
The other solutions also introduce more code and indirection to reason about.

The implementation of jumps assumes that jumps never return.
To ensure they never return, we rely on the program always explicitly exiting by
jumping to \RktSym{done}, as required by the run{-}time system.
To implement \RktSym{done}, we store an escape continuation at the start of the
module, and \RktSym{done} invokes that continuation, ending the whole
computation with the value of \RktSym{rax}.
The return{-}address register is initialized to \RktSym{done}, so intermediate
languages that implement procedures that follow the calling convention also exit
properly.

\begin{SCodeFlow}\begin{RktBlk}\begin{SingleColumn}\RktPn{(}\RktSym{define}\mbox{\hphantom{\Scribtexttt{x}}}\RktSym{exit{-}cont}\mbox{\hphantom{\Scribtexttt{x}}}\RktPn{(}\RktSym{box}\mbox{\hphantom{\Scribtexttt{x}}}\RktPn{(}\RktSym{lambda}\mbox{\hphantom{\Scribtexttt{x}}}\RktSym{{\char`\_}}\mbox{\hphantom{\Scribtexttt{x}}}\RktPn{(}\RktSym{r{\hbox{\texttt{:}}}error}\mbox{\hphantom{\Scribtexttt{x}}}\RktVal{"halt not initialized"}\RktPn{)}\RktPn{)}\RktPn{)}\RktPn{)}

\mbox{\hphantom{\Scribtexttt{x}}}

\RktPn{(}\RktSym{define}\mbox{\hphantom{\Scribtexttt{x}}}\RktPn{(}\RktSym{done}\RktPn{)}\mbox{\hphantom{\Scribtexttt{x}}}\RktPn{(}\RktPn{(}\RktSym{unbox}\mbox{\hphantom{\Scribtexttt{x}}}\RktSym{exit{-}cont}\RktPn{)}\mbox{\hphantom{\Scribtexttt{x}}}\RktPn{(}\RktSym{unbox}\mbox{\hphantom{\Scribtexttt{x}}}\RktVar{{\char`\_}rax}\RktPn{)}\RktPn{)}\RktPn{)}

\mbox{\hphantom{\Scribtexttt{x}}}

\RktPn{(}\RktSym{define{-}syntax}\mbox{\hphantom{\Scribtexttt{x}}}\RktPn{(}\RktSym{boundary}\mbox{\hphantom{\Scribtexttt{x}}}\RktSym{stx}\RktPn{)}

\mbox{\hphantom{\Scribtexttt{xx}}}\RktPn{(}\RktSym{syntax{-}parse}\mbox{\hphantom{\Scribtexttt{x}}}\RktSym{stx}

\mbox{\hphantom{\Scribtexttt{xxxx}}}\RktPn{[}\RktPn{(}\RktSym{{\char`\_}}\mbox{\hphantom{\Scribtexttt{x}}}\RktSym{term}\RktPn{)}

\mbox{\hphantom{\Scribtexttt{xxxxx}}}\RktRdr{\#{\textasciigrave}}\RktPn{(}\RktSym{let/ec}\mbox{\hphantom{\Scribtexttt{x}}}\RktSym{k}

\mbox{\hphantom{\Scribtexttt{xxxxxxxxx}}}\RktPn{(}\RktSym{set{-}box{\hbox{\texttt{!}}}}\mbox{\hphantom{\Scribtexttt{x}}}\RktSym{exit{-}cont}\mbox{\hphantom{\Scribtexttt{x}}}\RktSym{k}\RktPn{)}

\mbox{\hphantom{\Scribtexttt{xxxxxxxxx}}}\RktSym{term}\RktPn{)}\RktPn{]}\RktPn{)}\RktPn{)}\end{SingleColumn}\end{RktBlk}\end{SCodeFlow}

The \RktSym{boundary} form must be installed around any term that uses
labels and jumps.
The \RktSym{boundary} is implicitly installed around terms at the module level.
We must instead use \RktSym{let/cc} around interactive terms via
\RktSym{\#\%top{-}interaction} (the interposition point that lets our language be
used from Racket{'}s REPL) since the continuation lives past the local scope
during REPL interactions.
The \RktSym{boundary} form is also exposed to enable Racket interoperability.
Without an explicit boundary, terms may not return to Racket properly.
The \RktSym{boundary} is similar to the interoperability semantics by
\Autobibref{\hyperref[t:x28autobib_x22Daniel_Pattersonx2c_Jamie_Percontix2c_Christos_Dimoulasx2c_and_Amal_AhmedFunTALx3a_Reasonably_Mixing_a_Functional_Language_with_AssemblyIn_Procx2e_International_Conference_on_Programming_Language_Design_and_Implementation_x28PLDIx292017doix3a10x2e1145x2f3062341x2e3062347x22x29]{\AutobibLink{Patterson et al\Sendabbrev{.}}}~(\hyperref[t:x28autobib_x22Daniel_Pattersonx2c_Jamie_Percontix2c_Christos_Dimoulasx2c_and_Amal_AhmedFunTALx3a_Reasonably_Mixing_a_Functional_Language_with_AssemblyIn_Procx2e_International_Conference_on_Programming_Language_Design_and_Implementation_x28PLDIx292017doix3a10x2e1145x2f3062341x2e3062347x22x29]{\AutobibLink{2017}})} enabling inline typed assembly in a typed
functional language.
Their semantics enable the value of the return register to flow to the function
language when the \Scribtexttt{halt} instruction is delimited by an explicit boundary
term.

\LangBox{\LangBoxLabel{\RktModLink{\RktMod{\#lang}}\RktMeta{}\mbox{\hphantom{\Scribtexttt{x}}}\RktMeta{}\RktModLink{\RktSym{cpsc411/hashlangs/base}}\RktMeta{}}}%
{\begin{SCodeFlow}\begin{RktBlk}\begin{SingleColumn}\begin{RktBlk}\begin{SingleColumn}\Scribtexttt{{\Stttextmore} }\RktPn{(}\RktSym{define}\mbox{\hphantom{\Scribtexttt{x}}}\RktPn{(}\RktSym{f1}\mbox{\hphantom{\Scribtexttt{x}}}\RktSym{x}\RktPn{)}

\mbox{\hphantom{\Scribtexttt{xx}}}\mbox{\hphantom{\Scribtexttt{xx}}}\RktPn{(}\RktSym{begin}\mbox{\hphantom{\Scribtexttt{x}}}\RktPn{(}\RktSym{set{\hbox{\texttt{!}}}}\mbox{\hphantom{\Scribtexttt{x}}}\RktSym{rax}\mbox{\hphantom{\Scribtexttt{x}}}\RktSym{x}\RktPn{)}\mbox{\hphantom{\Scribtexttt{x}}}\RktPn{(}\RktSym{jump}\mbox{\hphantom{\Scribtexttt{x}}}\RktSym{done}\RktPn{)}\RktPn{)}

\mbox{\hphantom{\Scribtexttt{xx}}}\mbox{\hphantom{\Scribtexttt{xx}}}\RktPn{(}\RktSym{displayln}\mbox{\hphantom{\Scribtexttt{x}}}\RktVal{"After boundary"}\RktPn{)}

\mbox{\hphantom{\Scribtexttt{xx}}}\mbox{\hphantom{\Scribtexttt{xx}}}\RktSym{rax}\RktPn{)}\end{SingleColumn}\end{RktBlk}

\Scribtexttt{{\Stttextmore} }\RktPn{(}\RktSym{f1}\mbox{\hphantom{\Scribtexttt{x}}}\RktVal{5}\RktPn{)}

\RktRes{5}

\begin{RktBlk}\begin{SingleColumn}\Scribtexttt{{\Stttextmore} }\RktPn{(}\RktSym{define}\mbox{\hphantom{\Scribtexttt{x}}}\RktPn{(}\RktSym{f2}\mbox{\hphantom{\Scribtexttt{x}}}\RktSym{x}\RktPn{)}

\mbox{\hphantom{\Scribtexttt{xx}}}\mbox{\hphantom{\Scribtexttt{xx}}}\RktPn{(}\RktSym{boundary}

\mbox{\hphantom{\Scribtexttt{xx}}}\mbox{\hphantom{\Scribtexttt{xxxx}}}\RktPn{(}\RktSym{begin}\mbox{\hphantom{\Scribtexttt{x}}}\RktPn{(}\RktSym{set{\hbox{\texttt{!}}}}\mbox{\hphantom{\Scribtexttt{x}}}\RktSym{rax}\mbox{\hphantom{\Scribtexttt{x}}}\RktSym{x}\RktPn{)}\mbox{\hphantom{\Scribtexttt{x}}}\RktPn{(}\RktSym{jump}\mbox{\hphantom{\Scribtexttt{x}}}\RktSym{done}\RktPn{)}\RktPn{)}\RktPn{)}

\mbox{\hphantom{\Scribtexttt{xx}}}\mbox{\hphantom{\Scribtexttt{xx}}}\RktPn{(}\RktSym{displayln}\mbox{\hphantom{\Scribtexttt{x}}}\RktVal{"After boundary"}\RktPn{)}

\mbox{\hphantom{\Scribtexttt{xx}}}\mbox{\hphantom{\Scribtexttt{xx}}}\RktSym{rax}\RktPn{)}\end{SingleColumn}\end{RktBlk}

\Scribtexttt{{\Stttextmore} }\RktPn{(}\RktSym{f2}\mbox{\hphantom{\Scribtexttt{x}}}\RktVal{5}\RktPn{)}

\RktOut{After boundary}

\RktRes{5}\end{SingleColumn}\end{RktBlk}\end{SCodeFlow}}

The boundary term delimits the halt continuation used for the run{-}time system.
A better implementation might use delimited countinuations explicitly.

\sectionNewpage

\Ssection{Intermediate language abstractions and metadata}{Intermediate language abstractions and metadata}\label{t:x28part_x22secx3ailx22x29}

While interpreting the low{-}level abstractions is one major challenge, another
major challenge occurs at the various intermediate language points where
source{-}like abstractions must interoperate with target{-}like abstractions.
Two examples that make interesting use of Racket{'}s macro system are abstract
locations and frame variables.
We provide a grammar for one of the intermediate languages that includes these
features.

\Iidentity{\begin{bnfgrammar}
p ::= (\<module> info (\<define> label info tail) ... tail)
;;

info ::= ((\<assignment> ((aloc rloc) ...)))
;;

frame ::= (aloc ...)
;;

tail ::=
(\<jump> trg loc ...)
| (\<begin> effect ... tail)
;;

effect ::=
(\<set!> loc triv)
| (\<set!> loc$_1$ (binop loc$_1$ opand))
| (\<begin> effect ... effect)
;;

opand ::=
int64
| loc
;;

triv ::=
opand
| label
;;

loc ::=
rloc
| aloc
;;

rloc ::=
reg
| fvar
;;

trg ::=
label
| loc
\end{bnfgrammar}}

In this intermediate language, the \textbf{module} and procedure definitions are
annotated with metadata \emph{info} describing where abstract locations have
been assigned so far.
The \textbf{assignment} is a map from abstract locations to physical locations,
which is updated in stages during register allocation as variables are assigned
either to register or frame locations, depending on various factors.

Prior to register allocation, each language uses abstract locations \emph{aloc},
which are variables uniquely (within the scope of a procedure) identified by
their name, but whose physical location is handled by the compiler.
Abstract locations look like an arbitrary variable name followed by a "." and a
numeric suffix, such as \RktSym{x{\hbox{\texttt{.}}}1} in \RktPn{(}\RktSym{set{\hbox{\texttt{!}}}}\Scribtexttt{ }\RktSym{x{\hbox{\texttt{.}}}1}\Scribtexttt{ }\RktVal{5}\RktPn{)}.
In the intermediate languages \emph{during} (as it takes several nanopasses to
complete) and just \emph{after} register allocation, some or all abstract
locations should be aliases for physical locations.
However, prior to register allocation and when not yet assigned, abstract
locations must act like undeclared local variables.
Being able to run programs in these intermediate states between register
allocation nanopasses enables tests to fail early.

In many intermediate languages, we abstract away from the stack \emph{per se},
and expose frame variables.
Frame variables allow most of the compiler to avoid some tedium when dealing
with stack locations.
A frame variable is an identifier named "fv" with a numeric suffix, such as the
\RktSym{fv1} in \RktPn{(}\RktSym{set{\hbox{\texttt{!}}}}\Scribtexttt{ }\RktSym{fv1}\Scribtexttt{ }\RktVal{5}\RktPn{)}.
Frame variables are allocated in the enclosing procedure{'}s stack frame in the
word{-}sized slot indicated by the suffix.
For example, \RktSym{fv1} is roughly equivalent to \RktPn{(}\RktSym{rbp}\Scribtexttt{ }\RktSym{\mbox{{-}}}\Scribtexttt{ }\RktVal{8}\RktPn{)}.
However, the exact location on the stack depends on the details of how
procedures and frames are implemented; \RktSym{rbp} normally points to the
base of the enclosing procedure{'}s stack frame, but changes during a non{-}tail
call to point to the callee{'}s frame.
The intermediate languages must map frame variables to stack locations, so that
explicit stack manipulation implicitly manipulates frame variables and procedure
calling conventions are interpreted correctly.

For abstract locations, we can use \RktSym{make{-}variable{-}like{-}transformer} if
the abstract location has been assigned a physical location.
However, the macro must be locally bound, not globally bound, so we use
\RktSym{let{-}syntax} instead of \RktSym{define{-}syntax}.
If an abstract location has not been assigned a physical location, we must
also bind it in a local scope.
We use a local analysis implemented with \RktSym{local{-}expand} (which enables a
macro to locally change expansion order) to detect abstract locations that
appear in \RktSym{set{\hbox{\texttt{!}}}} and bind them as local variables during expansion.\NoteBox{\NoteContent{This analysis would be unnecessary if the intermediate language declared local
variables, suggesting a better language design.}}
We can implement frame variables using \RktSym{make{-}variable{-}like{-}transformer},
although we must break hygiene to get them globally bound.
However, we must also keep track of some uses of the frame base pointer
\RktSym{rbp} to provide a consistent interpretation of frame variables, so this
implementation must interoperate at run time with assignments to \RktSym{rbp}.

\Ssubsection{Abstract Locations}{Abstract Locations}\label{t:x28part_x22Abstractx5fLocationsx22x29}

By default, Racket does not let us assign to an undefined variable, so
\RktPn{(}\RktSym{set{\hbox{\texttt{!}}}}\Scribtexttt{ }\RktSym{x{\hbox{\texttt{.}}}1}\Scribtexttt{ }\RktVal{5}\RktPn{)} is invalid if \RktSym{x{\hbox{\texttt{.}}}1} is not in scope.

\LangBox{\LangBoxLabel{\RktModLink{\RktMod{\#lang}}\RktMeta{}\mbox{\hphantom{\Scribtexttt{x}}}\RktMeta{}\RktModLink{\RktSym{racket}}\RktMeta{}}}%
{\begin{SCodeFlow}\begin{RktBlk}\begin{SingleColumn}\Scribtexttt{{\Stttextmore} }\RktPn{(}\RktSym{set{\hbox{\texttt{!}}}}\mbox{\hphantom{\Scribtexttt{x}}}\RktSym{x{\hbox{\texttt{.}}}1}\mbox{\hphantom{\Scribtexttt{x}}}\RktVal{5}\RktPn{)}

\RktErr{set!: assignment disallowed;}

\RktErr{}\mbox{\hphantom{\Scribtexttt{x}}}\RktErr{cannot set variable before its definition}

\RktErr{}\mbox{\hphantom{\Scribtexttt{xx}}}\RktErr{variable: x.1}

\RktErr{}\mbox{\hphantom{\Scribtexttt{xx}}}\RktErr{in module: top{-}level}

\Scribtexttt{{\Stttextmore} }\RktPn{(}\RktSym{define}\mbox{\hphantom{\Scribtexttt{x}}}\RktSym{x{\hbox{\texttt{.}}}1}\mbox{\hphantom{\Scribtexttt{x}}}\RktPn{(}\RktSym{void}\RktPn{)}\RktPn{)}

\Scribtexttt{{\Stttextmore} }\RktPn{(}\RktSym{set{\hbox{\texttt{!}}}}\mbox{\hphantom{\Scribtexttt{x}}}\RktSym{x{\hbox{\texttt{.}}}1}\mbox{\hphantom{\Scribtexttt{x}}}\RktVal{5}\RktPn{)}\end{SingleColumn}\end{RktBlk}\end{SCodeFlow}}

We could tell Racket to do it anyway, but the scope of the assigned variable may
be unexpected.

\LangBox{\LangBoxLabel{\RktModLink{\RktMod{\#lang}}\RktMeta{}\mbox{\hphantom{\Scribtexttt{x}}}\RktMeta{}\RktModLink{\RktSym{racket}}\RktMeta{}}}%
{\begin{SCodeFlow}\begin{RktBlk}\begin{SingleColumn}\Scribtexttt{{\Stttextmore} }\RktPn{(}\RktSym{compile{-}allow{-}set{\hbox{\texttt{!}}}{-}undefined}\mbox{\hphantom{\Scribtexttt{x}}}\RktVal{\#t}\RktPn{)}

\Scribtexttt{{\Stttextmore} }\RktSym{y{\hbox{\texttt{.}}}2}

\RktErr{y.2: undefined;}

\RktErr{}\mbox{\hphantom{\Scribtexttt{x}}}\RktErr{cannot reference an identifier before its definition}

\RktErr{}\mbox{\hphantom{\Scribtexttt{xx}}}\RktErr{in module: top{-}level}

\Scribtexttt{{\Stttextmore} }\RktPn{(}\RktSym{set{\hbox{\texttt{!}}}}\mbox{\hphantom{\Scribtexttt{x}}}\RktSym{y{\hbox{\texttt{.}}}2}\mbox{\hphantom{\Scribtexttt{x}}}\RktVal{5}\RktPn{)}

\Scribtexttt{{\Stttextmore} }\RktSym{y{\hbox{\texttt{.}}}2}

\RktRes{5}

\Scribtexttt{{\Stttextmore} }\RktPn{(}\RktSym{let}\mbox{\hphantom{\Scribtexttt{x}}}\RktPn{(}\RktPn{)}\mbox{\hphantom{\Scribtexttt{x}}}\RktPn{(}\RktSym{set{\hbox{\texttt{!}}}}\mbox{\hphantom{\Scribtexttt{x}}}\RktSym{z{\hbox{\texttt{.}}}3}\mbox{\hphantom{\Scribtexttt{x}}}\RktVal{5}\RktPn{)}\RktPn{)}

\Scribtexttt{{\Stttextmore} }\RktSym{z{\hbox{\texttt{.}}}3}

\RktRes{5}\end{SingleColumn}\end{RktBlk}\end{SCodeFlow}}

Abstract locations can be freely generated by the compiler and have no defined
or sequential pattern, so we cannot create them ahead of time like we do
registers and frame variables.
Abstract locations are not global variables; they are unique to a procedure{'}s
scope.
Two procedures can refer to \RktSym{x{\hbox{\texttt{.}}}1}, and those are different variables.
As we see in the grammar, programs do not declare what abstract locations are in
scope.
So allowing \RktSym{set{\hbox{\texttt{!}}}} to undefined variables would not implementat abstract
locations correctly.

We must first bind abstract locations.
We want to avoid writing an explicit traversal over the entire program.
This would require either fixing the syntax, which we don{'}t want to do since we
want to reuse this solution across many languages, or writing an abstract tree
traversal, which is silly since we already have a macro expander.
We implement an analysis by locally changing expansion order to fully
expand an expression that references abstract locations, and instrumenting
\RktSym{set{\hbox{\texttt{!}}}} to collect the set of abstract locations that appear on the
left{-}hand side during expansion.
This is sufficient since programs must be well defined{---}an abstract location
must be assigned before it is referenced.

The analysis form \RktSym{do{-}bind{-}locals} is defined below, accompanied by some
compile{-}time definitions for collecting a scope{-}aware set of identifiers.

\begin{SCodeFlow}\begin{RktBlk}\begin{SingleColumn}\RktPn{(}\RktSym{begin{-}for{-}syntax}

\mbox{\hphantom{\Scribtexttt{xx}}}\RktPn{(}\RktSym{require}\mbox{\hphantom{\Scribtexttt{x}}}\RktSym{syntax/id{-}set}\mbox{\hphantom{\Scribtexttt{x}}}\RktSym{racket/set}\RktPn{)}

\mbox{\hphantom{\Scribtexttt{xx}}}\RktPn{(}\RktSym{define}\mbox{\hphantom{\Scribtexttt{x}}}\RktSym{gathered{-}locals}\mbox{\hphantom{\Scribtexttt{x}}}\RktPn{(}\RktSym{mutable{-}free{-}id{-}set}\RktPn{)}\RktPn{)}

\mbox{\hphantom{\Scribtexttt{x}}}

\mbox{\hphantom{\Scribtexttt{xx}}}\RktPn{(}\RktSym{define}\mbox{\hphantom{\Scribtexttt{x}}}\RktPn{(}\RktSym{reset{-}locals{\hbox{\texttt{!}}}}\RktPn{)}

\mbox{\hphantom{\Scribtexttt{xxxx}}}\RktPn{(}\RktSym{set{-}clear{\hbox{\texttt{!}}}}\mbox{\hphantom{\Scribtexttt{x}}}\RktSym{gathered{-}locals}\RktPn{)}\RktPn{)}

\mbox{\hphantom{\Scribtexttt{x}}}

\mbox{\hphantom{\Scribtexttt{xx}}}\RktPn{(}\RktSym{define}\mbox{\hphantom{\Scribtexttt{x}}}\RktPn{(}\RktSym{collect{-}local{\hbox{\texttt{!}}}}\mbox{\hphantom{\Scribtexttt{x}}}\RktSym{id}\RktPn{)}

\mbox{\hphantom{\Scribtexttt{xxxx}}}\RktPn{(}\RktSym{set{-}add{\hbox{\texttt{!}}}}\mbox{\hphantom{\Scribtexttt{x}}}\RktSym{gathered{-}locals}\mbox{\hphantom{\Scribtexttt{x}}}\RktSym{id}\RktPn{)}\RktPn{)}

\mbox{\hphantom{\Scribtexttt{x}}}

\mbox{\hphantom{\Scribtexttt{xx}}}\RktPn{(}\RktSym{define}\mbox{\hphantom{\Scribtexttt{x}}}\RktPn{(}\RktSym{get{-}locals}\RktPn{)}

\mbox{\hphantom{\Scribtexttt{xxxx}}}\RktPn{(}\RktSym{set{-}{\Stttextmore}list}\mbox{\hphantom{\Scribtexttt{x}}}\RktSym{gathered{-}locals}\RktPn{)}\RktPn{)}\RktPn{)}

\mbox{\hphantom{\Scribtexttt{x}}}

\RktPn{(}\RktSym{define{-}syntax}\mbox{\hphantom{\Scribtexttt{x}}}\RktPn{(}\RktSym{do{-}bind{-}locals}\mbox{\hphantom{\Scribtexttt{x}}}\RktSym{stx}\RktPn{)}

\mbox{\hphantom{\Scribtexttt{xx}}}\RktPn{(}\RktSym{syntax{-}parse}\mbox{\hphantom{\Scribtexttt{x}}}\RktSym{stx}

\mbox{\hphantom{\Scribtexttt{xxxx}}}\RktPn{[}\RktPn{(}\RktSym{{\char`\_}}\mbox{\hphantom{\Scribtexttt{x}}}\RktSym{body}\mbox{\hphantom{\Scribtexttt{x}}}\RktSym{except}\mbox{\hphantom{\Scribtexttt{x}}}\RktSym{{\hbox{\texttt{.}}}{\hbox{\texttt{.}}}{\hbox{\texttt{.}}}}\RktPn{)}

\mbox{\hphantom{\Scribtexttt{xxxxx}}}\RktPn{(}\RktSym{reset{-}locals{\hbox{\texttt{!}}}}\RktPn{)}

\mbox{\hphantom{\Scribtexttt{xxxxx}}}\RktPn{(}\RktSym{define}\mbox{\hphantom{\Scribtexttt{x}}}\RktSym{b}\mbox{\hphantom{\Scribtexttt{x}}}\RktPn{(}\RktSym{local{-}expand}\mbox{\hphantom{\Scribtexttt{x}}}\RktRdr{\#{\textquotesingle}}\RktSym{body}\mbox{\hphantom{\Scribtexttt{x}}}\RktVal{{\textquotesingle}}\RktVal{expression}\mbox{\hphantom{\Scribtexttt{x}}}\RktVal{{\textquotesingle}}\RktVal{(}\RktVal{)}\RktPn{)}\RktPn{)}

\mbox{\hphantom{\Scribtexttt{xxxxx}}}\RktRdr{\#{\textasciigrave}}\RktPn{(}\RktSym{let}\mbox{\hphantom{\Scribtexttt{x}}}\RktPn{(}\RktRdr{\#,@}\RktPn{(}\RktSym{for/list}\mbox{\hphantom{\Scribtexttt{x}}}\RktPn{(}\RktPn{[}\RktSym{l}\mbox{\hphantom{\Scribtexttt{x}}}\RktPn{(}\RktSym{get{-}locals}\RktPn{)}\RktPn{]}

\mbox{\hphantom{\Scribtexttt{xxxxxxxxxxxxxxxxxxxxxxxxxxx}}}\RktPn{\#{\hbox{\texttt{:}}}unless}\mbox{\hphantom{\Scribtexttt{x}}}\RktPn{(}\RktSym{set{-}member{\hbox{\texttt{?}}}}\mbox{\hphantom{\Scribtexttt{x}}}\RktPn{(}\RktSym{immutable{-}free{-}id{-}set}

\mbox{\hphantom{\Scribtexttt{xxxxxxxxxxxxxxxxxxxxxxxxxxxxxxxxxxxxxxxxxxxxxxxxxx}}}\RktPn{(}\RktSym{attribute}\mbox{\hphantom{\Scribtexttt{x}}}\RktSym{except}\RktPn{)}\RktPn{)}

\mbox{\hphantom{\Scribtexttt{xxxxxxxxxxxxxxxxxxxxxxxxxxxxxxxxxxxxxxxxxxxxxxxxx}}}\RktSym{l}\RktPn{)}\RktPn{)}

\mbox{\hphantom{\Scribtexttt{xxxxxxxxxxxxxxxxxx}}}\RktRdr{\#{\textasciigrave}}\RktPn{[}\RktRdr{\#,}\RktPn{(}\RktSym{syntax{-}local{-}introduce}\mbox{\hphantom{\Scribtexttt{x}}}\RktPn{(}\RktSym{format{-}id}\mbox{\hphantom{\Scribtexttt{x}}}\RktVal{\#f}\mbox{\hphantom{\Scribtexttt{x}}}\RktVal{"{\textasciitilde}a"}\mbox{\hphantom{\Scribtexttt{x}}}\RktSym{l}\RktPn{)}\RktPn{)}\mbox{\hphantom{\Scribtexttt{x}}}\RktPn{(}\RktSym{void}\RktPn{)}\RktPn{]}\RktPn{)}\RktPn{)}

\mbox{\hphantom{\Scribtexttt{xxxxxxxxx}}}\RktRdr{\#,}\RktSym{b}\RktPn{)}\RktPn{]}\RktPn{)}\RktPn{)}\end{SingleColumn}\end{RktBlk}\end{SCodeFlow}

We wrap the body of any scope{-}introducing form, particularly \textbf{define} and
\textbf{module}, with \RktSym{do{-}bind{-}locals}.
The possibly empty sequence \RktRdr{\#{\textquotesingle}}\RktPn{(}\RktSym{except}\Scribtexttt{ }\RktSym{{\hbox{\texttt{.}}}{\hbox{\texttt{.}}}{\hbox{\texttt{.}}}}\RktPn{)} declares a set of variables
that should not be bound, such as abstract locations that have been assigned
physical locations.
We first fully expand the \RktSym{body} using \RktSym{local{-}expand}.
During this local expansion, all assigned abstract locations are collected into
the set \RktSym{gathered{-}locals}.
Each is bound within the scope of the fully expanded body \RktSym{b}, using
\RktSym{format{-}id} with \RktVal{\#f} to clear the original scope from the
identifier, and using \RktSym{syntax{-}local{-}introduce} to bind it in the current
scope.

To collect the abstract locations, we make a modification to \RktSym{set{\hbox{\texttt{!}}}}.

\begin{SCodeFlow}\begin{RktBlk}\begin{SingleColumn}\RktPn{(}\RktSym{define{-}syntax}\mbox{\hphantom{\Scribtexttt{x}}}\RktPn{(}\RktSym{set{\hbox{\texttt{!}}}}\mbox{\hphantom{\Scribtexttt{x}}}\RktSym{stx}\RktPn{)}

\mbox{\hphantom{\Scribtexttt{xx}}}\RktPn{(}\RktSym{syntax{-}parse}\mbox{\hphantom{\Scribtexttt{x}}}\RktSym{stx}

\mbox{\hphantom{\Scribtexttt{xxxx}}}\RktCmt{;}\RktCmt{~}\RktCmt{Read from memory}

\mbox{\hphantom{\Scribtexttt{xxxx}}}\RktPn{[}\RktPn{(}\RktSym{set{\hbox{\texttt{!}}}}\mbox{\hphantom{\Scribtexttt{x}}}\RktSym{v1{\hbox{\texttt{:}}}id}\mbox{\hphantom{\Scribtexttt{x}}}\RktPn{(}\RktSym{base{\hbox{\texttt{:}}}not{-}rbp}\mbox{\hphantom{\Scribtexttt{x}}}\RktSym{op{\hbox{\texttt{:}}}addr{-}op}\mbox{\hphantom{\Scribtexttt{x}}}\RktSym{offset}\RktPn{)}\RktPn{)}

\mbox{\hphantom{\Scribtexttt{xxxx}}}\RktCmt{;}\RktCmt{~}\RktCmt{Collect used abstract location}

\mbox{\hphantom{\Scribtexttt{xxxxx}}}\RktPn{(}\RktSym{when}\mbox{\hphantom{\Scribtexttt{x}}}\RktPn{(}\RktSym{aloc{\hbox{\texttt{?}}}}\mbox{\hphantom{\Scribtexttt{x}}}\RktPn{(}\RktSym{syntax{-}{\Stttextmore}datum}\mbox{\hphantom{\Scribtexttt{x}}}\RktRdr{\#{\textquotesingle}}\RktSym{v1}\RktPn{)}\RktPn{)}

\mbox{\hphantom{\Scribtexttt{xxxxxxx}}}\RktPn{(}\RktSym{collect{-}local{\hbox{\texttt{!}}}}\mbox{\hphantom{\Scribtexttt{x}}}\RktRdr{\#{\textquotesingle}}\RktSym{v1}\RktPn{)}\RktPn{)}

\mbox{\hphantom{\Scribtexttt{xxxxx}}}\RktRdr{\#{\textasciigrave}}\RktPn{(}\RktSym{r{\hbox{\texttt{:}}}set{\hbox{\texttt{!}}}}\mbox{\hphantom{\Scribtexttt{x}}}\RktSym{v1}\mbox{\hphantom{\Scribtexttt{x}}}\RktPn{(}\RktSym{vector{-}ref}\mbox{\hphantom{\Scribtexttt{x}}}\RktSym{memory}\mbox{\hphantom{\Scribtexttt{x}}}\RktPn{(}\RktSym{op}\mbox{\hphantom{\Scribtexttt{x}}}\RktSym{base}\mbox{\hphantom{\Scribtexttt{x}}}\RktSym{offset}\RktPn{)}\RktPn{)}\RktPn{)}\RktPn{]}

\mbox{\hphantom{\Scribtexttt{xxxx}}}\RktCmt{;}\RktCmt{~}\RktCmt{Else use Racket{\textquotesingle}s set{\hbox{\texttt{!}}}}

\mbox{\hphantom{\Scribtexttt{xxxx}}}\RktPn{[}\RktPn{(}\RktSym{set{\hbox{\texttt{!}}}}\mbox{\hphantom{\Scribtexttt{x}}}\RktSym{v1}\mbox{\hphantom{\Scribtexttt{x}}}\RktSym{v2}\RktPn{)}

\mbox{\hphantom{\Scribtexttt{xxxx}}}\RktCmt{;}\RktCmt{~}\RktCmt{Collect used abstract location}

\mbox{\hphantom{\Scribtexttt{xxxxx}}}\RktPn{(}\RktSym{when}\mbox{\hphantom{\Scribtexttt{x}}}\RktPn{(}\RktSym{aloc{\hbox{\texttt{?}}}}\mbox{\hphantom{\Scribtexttt{x}}}\RktPn{(}\RktSym{syntax{-}{\Stttextmore}datum}\mbox{\hphantom{\Scribtexttt{x}}}\RktRdr{\#{\textquotesingle}}\RktSym{v1}\RktPn{)}\RktPn{)}

\mbox{\hphantom{\Scribtexttt{xxxxxxx}}}\RktPn{(}\RktSym{collect{-}local{\hbox{\texttt{!}}}}\mbox{\hphantom{\Scribtexttt{x}}}\RktRdr{\#{\textquotesingle}}\RktSym{v1}\RktPn{)}\RktPn{)}

\mbox{\hphantom{\Scribtexttt{xxxxx}}}\RktRdr{\#{\textasciigrave}}\RktPn{(}\RktSym{r{\hbox{\texttt{:}}}set{\hbox{\texttt{!}}}}\mbox{\hphantom{\Scribtexttt{x}}}\RktSym{v1}\mbox{\hphantom{\Scribtexttt{x}}}\RktSym{v2}\RktPn{)}\RktPn{]}\mbox{\hphantom{\Scribtexttt{x}}}\RktSym{{\hbox{\texttt{.}}}{\hbox{\texttt{.}}}{\hbox{\texttt{.}}}{\hbox{\texttt{.}}}}\RktPn{)}\RktPn{)}\end{SingleColumn}\end{RktBlk}\end{SCodeFlow}

Abstract locations are written to either when reading from memory, or when
the \RktSym{set{\hbox{\texttt{!}}}} does not involve a memory address at all.
In both cases, if the left{-}hand side is an abstract location, we collect it.

For an abstract location that is already assigned a physical location, we
exclude it from binding in \RktSym{do{-}bind{-}locals} and instead make it a
variable{-}like transformer to its assigned physical location.
For interpreting programs, it isn{'}t necessary to redirect abstract locations to
their assigned physical location.
However, doing so aids debugging by triggering failing tests earlier.
If a bug occurs in one of the register allocation nanopasses, the interpretation
of the program changes as soon as the metadata changes, and not only once the
program text has been rewritten to replace all abstract locations with their
assigned physical locations.

To implement these assigned abstract locations, we use the following
compile{-}time procedures.
\RktSym{make{-}aloc{-}transformer} is an abstraction of
\RktSym{make{-}variable{-}like{-}transformer} for abstract locations that have been
assigned the physical location \RktSym{rloc}, and \RktSym{bind{-}assignments}
generates bindings based on the metadata \RktSym{info} for the program
\RktSym{tail}.

\begin{SCodeFlow}\begin{RktBlk}\begin{SingleColumn}\RktPn{(}\RktSym{begin{-}for{-}syntax}

\mbox{\hphantom{\Scribtexttt{xx}}}\RktPn{(}\RktSym{define}\mbox{\hphantom{\Scribtexttt{x}}}\RktPn{(}\RktSym{make{-}aloc{-}transformer}\mbox{\hphantom{\Scribtexttt{x}}}\RktSym{rloc}\RktPn{)}

\mbox{\hphantom{\Scribtexttt{xxxx}}}\RktPn{(}\RktSym{make{-}variable{-}like{-}transformer}

\mbox{\hphantom{\Scribtexttt{xxxxx}}}\RktSym{rloc}

\mbox{\hphantom{\Scribtexttt{xxxxx}}}\RktPn{(}\RktSym{lambda}\mbox{\hphantom{\Scribtexttt{x}}}\RktPn{(}\RktSym{stx}\RktPn{)}

\mbox{\hphantom{\Scribtexttt{xxxxxxx}}}\RktPn{(}\RktSym{syntax{-}parse}\mbox{\hphantom{\Scribtexttt{x}}}\RktSym{stx}

\mbox{\hphantom{\Scribtexttt{xxxxxxxxx}}}\RktPn{[}\RktPn{(}\RktSym{set{\hbox{\texttt{!}}}}\mbox{\hphantom{\Scribtexttt{x}}}\RktSym{bla}\mbox{\hphantom{\Scribtexttt{x}}}\RktSym{v}\RktPn{)}

\mbox{\hphantom{\Scribtexttt{xxxxxxxxxx}}}\RktRdr{\#{\textasciigrave}}\RktPn{(}\RktSym{r{\hbox{\texttt{:}}}set{\hbox{\texttt{!}}}}\mbox{\hphantom{\Scribtexttt{x}}}\RktRdr{\#,}\RktSym{rloc}\mbox{\hphantom{\Scribtexttt{x}}}\RktSym{v}\RktPn{)}\RktPn{]}\RktPn{)}\RktPn{)}\RktPn{)}\RktPn{)}

\mbox{\hphantom{\Scribtexttt{x}}}

\mbox{\hphantom{\Scribtexttt{xx}}}\RktPn{(}\RktSym{define}\mbox{\hphantom{\Scribtexttt{x}}}\RktPn{(}\RktSym{bind{-}assignments}\mbox{\hphantom{\Scribtexttt{x}}}\RktSym{info}\mbox{\hphantom{\Scribtexttt{x}}}\RktSym{tail}\RktPn{)}

\mbox{\hphantom{\Scribtexttt{xxxx}}}\RktRdr{\#{\textasciigrave}}\RktPn{(}\RktSym{let{-}syntax}\mbox{\hphantom{\Scribtexttt{x}}}\RktRdr{\#,}\RktPn{(}\RktSym{for/list}\mbox{\hphantom{\Scribtexttt{x}}}\RktPn{(}\RktPn{[}\RktSym{assignments}\mbox{\hphantom{\Scribtexttt{x}}}\RktPn{(}\RktSym{dict{-}ref}\mbox{\hphantom{\Scribtexttt{x}}}\RktSym{info}\mbox{\hphantom{\Scribtexttt{x}}}\RktVal{{\textquotesingle}}\RktVal{assignment}\mbox{\hphantom{\Scribtexttt{x}}}\RktVal{{\textquotesingle}}\RktVal{(}\RktVal{)}\RktPn{)}\RktPn{]}\RktPn{)}

\mbox{\hphantom{\Scribtexttt{xxxxxxxxxxxxxxxxxxxxxx}}}\RktPn{(}\RktSym{with{-}syntax}\mbox{\hphantom{\Scribtexttt{x}}}\RktPn{(}\RktPn{[}\RktSym{aloc}\mbox{\hphantom{\Scribtexttt{x}}}\RktPn{(}\RktSym{car}\mbox{\hphantom{\Scribtexttt{x}}}\RktSym{assignments}\RktPn{)}\RktPn{]}

\mbox{\hphantom{\Scribtexttt{xxxxxxxxxxxxxxxxxxxxxxxxxxxxxxxxxxxx}}}\RktPn{[}\RktSym{rloc}\mbox{\hphantom{\Scribtexttt{x}}}\RktPn{(}\RktSym{cadr}\mbox{\hphantom{\Scribtexttt{x}}}\RktSym{assignments}\RktPn{)}\RktPn{]}\RktPn{)}

\mbox{\hphantom{\Scribtexttt{xxxxxxxxxxxxxxxxxxxxxxxx}}}\RktRdr{\#{\textasciigrave}}\RktPn{[}\RktSym{aloc}\mbox{\hphantom{\Scribtexttt{x}}}\RktPn{(}\RktSym{make{-}aloc{-}transformer}\mbox{\hphantom{\Scribtexttt{x}}}\RktRdr{\#{\textquotesingle}}\RktSym{rloc}\RktPn{)}\RktPn{]}\RktPn{)}\RktPn{)}

\mbox{\hphantom{\Scribtexttt{xxxxxxxx}}}\RktRdr{\#,}\RktSym{tail}\RktPn{)}\RktPn{)}\RktPn{)}\end{SingleColumn}\end{RktBlk}\end{SCodeFlow}

\RktSym{bind{-}assignments} is unusal in that it generates syntax with new macro
definitions, \emph{i.e.}, with new compile{-}time code that must be further
expanded.
We use \RktSym{let{-}syntax} to bind each abstract location as a local macro.
This is important since the same abstract location can appear in multiple scopes.
Being defined as a macro, reads and writes to the abstract location are
statically rewritten to the physical location that has been assigned.
The call to \RktSym{make{-}aloc{-}transformer} appears as syntax, not as a value
spliced into the syntax.
We generate a call to the compile{-}time procedure instead of inlining the
\RktSym{make{-}variable{-}like{-}transformer} call in the syntax template to avoid
code duplication.
If t were inlined in the syntax template, the code would be duplicated during
expansion time, allocating a new closure for each binding, and causing
performance problems during macro expansion.

This helper is used in the implementations of scope{-}introducing forms with
attached metadata, such as in the following.

\begin{SCodeFlow}\begin{RktBlk}\begin{SingleColumn}\RktPn{(}\RktSym{define{-}syntax}\mbox{\hphantom{\Scribtexttt{x}}}\RktPn{(}\RktSym{module}\mbox{\hphantom{\Scribtexttt{x}}}\RktSym{stx}\RktPn{)}

\mbox{\hphantom{\Scribtexttt{xx}}}\RktPn{(}\RktSym{syntax{-}parse}\mbox{\hphantom{\Scribtexttt{x}}}\RktSym{stx}

\mbox{\hphantom{\Scribtexttt{xxxx}}}\RktPn{[}\RktPn{(}\RktSym{{\char`\_}}\mbox{\hphantom{\Scribtexttt{x}}}\RktSym{info}\mbox{\hphantom{\Scribtexttt{x}}}\RktSym{defs}\mbox{\hphantom{\Scribtexttt{x}}}\RktSym{{\hbox{\texttt{.}}}{\hbox{\texttt{.}}}{\hbox{\texttt{.}}}}\mbox{\hphantom{\Scribtexttt{x}}}\RktSym{body}\RktPn{)}

\mbox{\hphantom{\Scribtexttt{xxxxx}}}\RktPn{(}\RktSym{define}\mbox{\hphantom{\Scribtexttt{x}}}\RktSym{info{-}dict}\mbox{\hphantom{\Scribtexttt{x}}}\RktPn{(}\RktSym{infostx{-}{\Stttextmore}dict}\mbox{\hphantom{\Scribtexttt{x}}}\RktRdr{\#{\textquotesingle}}\RktSym{info}\RktPn{)}\RktPn{)}

\mbox{\hphantom{\Scribtexttt{xxxxx}}}\RktRdr{\#{\textasciigrave}}\RktPn{(}\RktSym{boundary}

\mbox{\hphantom{\Scribtexttt{xxxxxxxx}}}\RktPn{(}\RktSym{begin}

\mbox{\hphantom{\Scribtexttt{xxxxxxxxxx}}}\RktRdr{\#,}\RktPn{(}\RktSym{bind{-}assignments}

\mbox{\hphantom{\Scribtexttt{xxxxxxxxxxxxx}}}\RktSym{info{-}dict}

\mbox{\hphantom{\Scribtexttt{xxxxxxxxxxxxx}}}\RktRdr{\#{\textasciigrave}}\RktPn{(}\RktSym{local}\mbox{\hphantom{\Scribtexttt{x}}}\RktPn{[}\RktSym{defs}\mbox{\hphantom{\Scribtexttt{x}}}\RktSym{{\hbox{\texttt{.}}}{\hbox{\texttt{.}}}{\hbox{\texttt{.}}}}\RktPn{]}

\mbox{\hphantom{\Scribtexttt{xxxxxxxxxxxxxxxxx}}}\RktPn{(}\RktSym{do{-}bind{-}locals}\mbox{\hphantom{\Scribtexttt{x}}}\RktSym{tail}

\mbox{\hphantom{\Scribtexttt{xxxxxxxxxxxxxxxxxxxxxxxxxxxxxxxxx}}}\RktRdr{\#,@}\RktPn{(}\RktSym{get{-}info{-}bound{-}vars}\mbox{\hphantom{\Scribtexttt{x}}}\RktSym{info{-}dict}\RktPn{)}\RktPn{)}\RktPn{)}\RktPn{)}\RktPn{)}\RktPn{)}\RktPn{]}\RktPn{)}\RktPn{)}\end{SingleColumn}\end{RktBlk}\end{SCodeFlow}

We see the implicit insertion of the \RktSym{boundary} mentioned earlier.
First \RktSym{bind{-}assignments} binds any abstract locations that have been
assigned, then \RktSym{do{-}bind{-}locals} locally expands the body and binds all
other abstract locations.
This order doesn{'}t matter, since \RktSym{do{-}bind{-}locals} excludes binding based
on the \RktSym{info} metadata.

\Ssubsection{Frame Variables}{Frame Variables}\label{t:x28part_x22Framex5fVariablesx22x29}

The intermediate languages contain an unbounded number of global frame variables
corresponding to locations on the current procedure{'}s stack frame.
Frame variables have the form "fv\emph{n}", where \emph{n} is some natural
number.

The first challenge is binding all these variables.
We cannot bind an unbounded number of variables apriori.
For the moment, we cheat: we bind an arbitrary large number, \RktVal{1620}, of
them.\NoteBox{\NoteContent{We would have choosen 42, but needed more than about 50, as students did use
about 50 frame variables in some test programs.
This number is similar to 42, but large enough.}}
We discuss alternatives in \ChapRefLocalUC{t:x28part_x22secx3adiscussionx22x29}{6}{Discussion}.

To bind them, we break hygiene, using \RktSym{syntax{-}local{-}introduce} to bind
frame variables in the right scope, as we don{'}t want to list out all
\RktVal{1620} of them in the interpreter.
We bind each frame variable to a macro that performs the offset calculation into
the stack.

\begin{SCodeFlow}\begin{RktBlk}\begin{SingleColumn}\RktPn{(}\RktSym{begin{-}for{-}syntax}\mbox{\hphantom{\Scribtexttt{x}}}\RktPn{(}\RktSym{define}\mbox{\hphantom{\Scribtexttt{x}}}\RktSym{CURRENT{-}FVARS}\mbox{\hphantom{\Scribtexttt{x}}}\RktVal{1620}\RktPn{)}\RktPn{)}

\mbox{\hphantom{\Scribtexttt{x}}}

\RktPn{(}\RktSym{define{-}syntax}\mbox{\hphantom{\Scribtexttt{x}}}\RktPn{(}\RktSym{define{-}fvars{\hbox{\texttt{!}}}}\mbox{\hphantom{\Scribtexttt{x}}}\RktSym{stx}\RktPn{)}

\mbox{\hphantom{\Scribtexttt{xx}}}\RktPn{(}\RktSym{syntax{-}parse}\mbox{\hphantom{\Scribtexttt{x}}}\RktSym{stx}

\mbox{\hphantom{\Scribtexttt{xxxx}}}\RktPn{[}\RktPn{(}\RktSym{{\char`\_}}\RktPn{)}

\mbox{\hphantom{\Scribtexttt{xxxxx}}}\RktRdr{\#{\textasciigrave}}\RktPn{(}\RktSym{begin}

\mbox{\hphantom{\Scribtexttt{xxxxxxxxx}}}\RktRdr{\#,@}\RktPn{(}\RktSym{for/list}\mbox{\hphantom{\Scribtexttt{x}}}\RktPn{(}\RktPn{[}\RktSym{i}\mbox{\hphantom{\Scribtexttt{x}}}\RktPn{(}\RktSym{in{-}range}\mbox{\hphantom{\Scribtexttt{x}}}\RktVal{0}\mbox{\hphantom{\Scribtexttt{x}}}\RktSym{CURRENT{-}FVARS}\RktPn{)}\RktPn{]}\RktPn{)}

\mbox{\hphantom{\Scribtexttt{xxxxxxxxxxxxxx}}}\RktPn{(}\RktSym{with{-}syntax}\mbox{\hphantom{\Scribtexttt{x}}}\RktPn{(}\RktPn{[}\RktSym{fvar}\mbox{\hphantom{\Scribtexttt{x}}}\RktPn{(}\RktSym{syntax{-}local{-}introduce}

\mbox{\hphantom{\Scribtexttt{xxxxxxxxxxxxxxxxxxxxxxxxxxxxxxxxxxx}}}\RktPn{(}\RktSym{format{-}id}\mbox{\hphantom{\Scribtexttt{x}}}\RktVal{\#f}\mbox{\hphantom{\Scribtexttt{x}}}\RktVal{"fv{\textasciitilde}a"}\mbox{\hphantom{\Scribtexttt{x}}}\RktSym{i}\RktPn{)}\RktPn{)}\RktPn{]}

\mbox{\hphantom{\Scribtexttt{xxxxxxxxxxxxxxxxxxxxxxxxxxxx}}}\RktPn{[}\RktSym{offset}\mbox{\hphantom{\Scribtexttt{x}}}\RktPn{(}\RktSym{*}\mbox{\hphantom{\Scribtexttt{x}}}\RktSym{i}\mbox{\hphantom{\Scribtexttt{x}}}\RktVal{8}\RktPn{)}\RktPn{]}\RktPn{)}

\mbox{\hphantom{\Scribtexttt{xxxxxxxxxxxxxxxx}}}\RktRdr{\#{\textasciigrave}}\RktPn{(}\RktSym{define{-}syntax}\mbox{\hphantom{\Scribtexttt{x}}}\RktSym{fvar}\mbox{\hphantom{\Scribtexttt{x}}}\RktPn{(}\RktSym{make{-}fvar{-}transformer}\mbox{\hphantom{\Scribtexttt{x}}}\RktSym{offset}\RktPn{)}\RktPn{)}\RktPn{)}\RktPn{)}\RktPn{)}\RktPn{]}\RktPn{)}\RktPn{)}

\mbox{\hphantom{\Scribtexttt{x}}}

\RktPn{(}\RktSym{define{-}fvars{\hbox{\texttt{!}}}}\RktPn{)}\end{SingleColumn}\end{RktBlk}\end{SCodeFlow}

The \RktSym{define{-}fvars{\hbox{\texttt{!}}}} form generates frame variables 0 through
\RktVal{1619}, and binds each to a
\RktSym{make{-}fvar{-}transformer} (introduced shortly) at the offset corresponding
to its index into the frame.
We use \RktSym{format{-}id} with \RktVal{\#f}, indicating no scope, to
create the identifier then use \RktSym{syntax{-}local{-}introduce} to prevent the
macro from introducing a fresh scope.
(We thought that providing \RktSym{stx} as the lexical context on the identifier
would have avoided the use of \RktSym{syntax{-}local{-}introduce}, but that doesn{'}t
seem to be the case.)
As before, we define and call \RktSym{make{-}fvar{-}transformer} rather than inline
it in the syntax template.

\RktSym{make{-}fvar{-}transformer} is a wrapper around
\RktSym{make{-}variable{-}like{-}transformer} that performs the offset calculation
into the stack, and otherwise behaves like \RktSym{rbp}.

\begin{SCodeFlow}\begin{RktBlk}\begin{SingleColumn}\RktPn{(}\RktSym{begin{-}for{-}syntax}

\mbox{\hphantom{\Scribtexttt{xx}}}\RktPn{(}\RktSym{define}\mbox{\hphantom{\Scribtexttt{x}}}\RktPn{(}\RktSym{make{-}fvar{-}transformer}\mbox{\hphantom{\Scribtexttt{x}}}\RktSym{frame{-}offset}\RktPn{)}

\mbox{\hphantom{\Scribtexttt{xxxx}}}\RktPn{(}\RktSym{with{-}syntax}\mbox{\hphantom{\Scribtexttt{x}}}\RktPn{(}\RktPn{[}\RktSym{stack{-}index}\mbox{\hphantom{\Scribtexttt{x}}}\RktRdr{\#{\textasciigrave}}\RktPn{(}\RktSym{\mbox{{-}}}\mbox{\hphantom{\Scribtexttt{x}}}\RktPn{(}\RktSym{unbox}\mbox{\hphantom{\Scribtexttt{x}}}\RktVar{{\char`\_}rbp}\RktPn{)}\mbox{\hphantom{\Scribtexttt{x}}}\RktPn{(}\RktSym{+}\mbox{\hphantom{\Scribtexttt{x}}}\RktPn{(}\RktSym{unbox}\mbox{\hphantom{\Scribtexttt{x}}}\RktSym{current{-}fvar{-}offset}\RktPn{)}

\mbox{\hphantom{\Scribtexttt{xxxxxxxxxxxxxxxxxxxxxxxxxxxxxxxxxxxxxxxxxxxxxxxxxxxx}}}\RktRdr{\#,}\RktSym{frame{-}offset}\RktPn{)}\RktPn{)}\RktPn{]}\RktPn{)}

\mbox{\hphantom{\Scribtexttt{xxxxxx}}}\RktPn{(}\RktSym{make{-}variable{-}like{-}transformer}

\mbox{\hphantom{\Scribtexttt{xxxxxxx}}}\RktRdr{\#{\textasciigrave}}\RktPn{(}\RktSym{vector{-}ref}\mbox{\hphantom{\Scribtexttt{x}}}\RktSym{stack}\mbox{\hphantom{\Scribtexttt{x}}}\RktSym{stack{-}index}\RktPn{)}

\mbox{\hphantom{\Scribtexttt{xxxxxxx}}}\RktPn{(}\RktSym{lambda}\mbox{\hphantom{\Scribtexttt{x}}}\RktPn{(}\RktSym{stx}\RktPn{)}

\mbox{\hphantom{\Scribtexttt{xxxxxxxxx}}}\RktPn{(}\RktSym{syntax{-}parse}\mbox{\hphantom{\Scribtexttt{x}}}\RktSym{stx}

\mbox{\hphantom{\Scribtexttt{xxxxxxxxxxx}}}\RktPn{[}\RktPn{(}\RktSym{set{\hbox{\texttt{!}}}}\mbox{\hphantom{\Scribtexttt{x}}}\RktSym{{\char`\_}}\mbox{\hphantom{\Scribtexttt{x}}}\RktSym{v}\RktPn{)}\mbox{\hphantom{\Scribtexttt{x}}}\RktRdr{\#{\textasciigrave}}\RktPn{(}\RktSym{vector{-}set{\hbox{\texttt{!}}}}\mbox{\hphantom{\Scribtexttt{x}}}\RktSym{stack}\mbox{\hphantom{\Scribtexttt{x}}}\RktSym{stack{-}index}\mbox{\hphantom{\Scribtexttt{x}}}\RktSym{v}\RktPn{)}\RktPn{]}\RktPn{)}\RktPn{)}\RktPn{)}\RktPn{)}\RktPn{)}\RktPn{)}\end{SingleColumn}\end{RktBlk}\end{SCodeFlow}

The offset for a frame variable is in two parts: the offset given by the
statically known index on the frame variable, plus a dynamic offset determined
by the run{-}time modifications to the frame base pointer \RktSym{rbp}.
The expression \RktPn{(}\RktSym{unbox}\Scribtexttt{ }\RktSym{current{-}fvar{-}offset}\RktPn{)} occurs under the syntax
template, generating a run{-}time calculation, while \RktSym{frame{-}offset} is an
compile{-}time value spliced into the template.
In the intermediate languages, \RktSym{rbp} is modified by a constant offset
when a new frame is allocated for a non{-}tail call, moving the frame base pointer
to point to the callee{'}s frame.
But a frame variable might be referenced, after this, as an argument for the
call, and must refer to the caller{'}s frame.
The \RktSym{stack{-}index} calculates the offset into the stack in this situation,
using the value \RktSym{current{-}fvar{-}offset} which is updated whenever
\RktSym{rbp} is incremented.

As an example, consider the following low{-}level implementation of a non{-}tail
call.

\noindent \LangBox{\LangBoxLabel{\RktModLink{\RktMod{\#lang}}\RktMeta{}\mbox{\hphantom{\Scribtexttt{x}}}\RktMeta{}\RktModLink{\RktSym{cpsc411/hashlangs/base}}\RktMeta{}}}%
{\begin{SCodeFlow}\begin{RktBlk}\begin{SingleColumn}\Scribtexttt{{\Stttextmore} }\RktPn{(}\RktSym{define}\mbox{\hphantom{\Scribtexttt{x}}}\RktSym{identity}\mbox{\hphantom{\Scribtexttt{x}}}\RktPn{(}\RktSym{lambda}\mbox{\hphantom{\Scribtexttt{x}}}\RktPn{(}\RktPn{)}\mbox{\hphantom{\Scribtexttt{x}}}\RktPn{(}\RktSym{begin}\mbox{\hphantom{\Scribtexttt{x}}}\RktPn{(}\RktSym{set{\hbox{\texttt{!}}}}\mbox{\hphantom{\Scribtexttt{x}}}\RktSym{rax}\mbox{\hphantom{\Scribtexttt{x}}}\RktSym{fv0}\RktPn{)}\mbox{\hphantom{\Scribtexttt{x}}}\RktPn{(}\RktSym{jump}\mbox{\hphantom{\Scribtexttt{x}}}\RktSym{r15}\RktPn{)}\RktPn{)}\RktPn{)}\RktPn{)}

\begin{RktBlk}\begin{SingleColumn}\Scribtexttt{{\Stttextmore} }\RktPn{(}\RktSym{begin}

\mbox{\hphantom{\Scribtexttt{xx}}}\mbox{\hphantom{\Scribtexttt{xx}}}\RktPn{(}\RktSym{set{\hbox{\texttt{!}}}}\mbox{\hphantom{\Scribtexttt{x}}}\RktSym{fv0}\mbox{\hphantom{\Scribtexttt{x}}}\RktVal{41}\RktPn{)}

\mbox{\hphantom{\Scribtexttt{xx}}}\mbox{\hphantom{\Scribtexttt{xx}}}\RktPn{(}\RktSym{set{\hbox{\texttt{!}}}}\mbox{\hphantom{\Scribtexttt{x}}}\RktSym{fv1}\mbox{\hphantom{\Scribtexttt{x}}}\RktVal{1}\RktPn{)}

\mbox{\hphantom{\Scribtexttt{xx}}}\mbox{\hphantom{\Scribtexttt{xx}}}\RktPn{(}\RktSym{set{\hbox{\texttt{!}}}}\mbox{\hphantom{\Scribtexttt{x}}}\RktSym{rbp}\mbox{\hphantom{\Scribtexttt{x}}}\RktPn{(}\RktSym{\mbox{{-}}}\mbox{\hphantom{\Scribtexttt{x}}}\RktSym{rbp}\mbox{\hphantom{\Scribtexttt{x}}}\RktVal{16}\RktPn{)}\RktPn{)}

\mbox{\hphantom{\Scribtexttt{xx}}}\mbox{\hphantom{\Scribtexttt{xx}}}\RktPn{(}\RktSym{return{-}point}\mbox{\hphantom{\Scribtexttt{x}}}\RktSym{l}

\mbox{\hphantom{\Scribtexttt{xx}}}\mbox{\hphantom{\Scribtexttt{xxxx}}}\RktPn{(}\RktSym{begin}

\mbox{\hphantom{\Scribtexttt{xx}}}\mbox{\hphantom{\Scribtexttt{xxxxxx}}}\RktPn{(}\RktSym{set{\hbox{\texttt{!}}}}\mbox{\hphantom{\Scribtexttt{x}}}\RktSym{r15}\mbox{\hphantom{\Scribtexttt{x}}}\RktSym{l}\RktPn{)}

\mbox{\hphantom{\Scribtexttt{xx}}}\mbox{\hphantom{\Scribtexttt{xxxxxx}}}\RktPn{(}\RktSym{set{\hbox{\texttt{!}}}}\mbox{\hphantom{\Scribtexttt{x}}}\RktSym{fv2}\mbox{\hphantom{\Scribtexttt{x}}}\RktSym{fv0}\RktPn{)}

\mbox{\hphantom{\Scribtexttt{xx}}}\mbox{\hphantom{\Scribtexttt{xxxxxx}}}\RktPn{(}\RktSym{jump}\mbox{\hphantom{\Scribtexttt{x}}}\RktSym{identity}\RktPn{)}\RktPn{)}\RktPn{)}

\mbox{\hphantom{\Scribtexttt{xx}}}\mbox{\hphantom{\Scribtexttt{xx}}}\RktPn{(}\RktSym{set{\hbox{\texttt{!}}}}\mbox{\hphantom{\Scribtexttt{x}}}\RktSym{rbp}\mbox{\hphantom{\Scribtexttt{x}}}\RktPn{(}\RktSym{+}\mbox{\hphantom{\Scribtexttt{x}}}\RktSym{rbp}\mbox{\hphantom{\Scribtexttt{x}}}\RktVal{16}\RktPn{)}\RktPn{)}

\mbox{\hphantom{\Scribtexttt{xx}}}\mbox{\hphantom{\Scribtexttt{xx}}}\RktPn{(}\RktSym{set{\hbox{\texttt{!}}}}\mbox{\hphantom{\Scribtexttt{x}}}\RktSym{r10}\mbox{\hphantom{\Scribtexttt{x}}}\RktSym{fv1}\RktPn{)}

\mbox{\hphantom{\Scribtexttt{xx}}}\mbox{\hphantom{\Scribtexttt{xx}}}\RktPn{(}\RktSym{set{\hbox{\texttt{!}}}}\mbox{\hphantom{\Scribtexttt{x}}}\RktSym{rax}\mbox{\hphantom{\Scribtexttt{x}}}\RktPn{(}\RktSym{+}\mbox{\hphantom{\Scribtexttt{x}}}\RktSym{rax}\mbox{\hphantom{\Scribtexttt{x}}}\RktSym{r10}\RktPn{)}\RktPn{)}

\mbox{\hphantom{\Scribtexttt{xx}}}\mbox{\hphantom{\Scribtexttt{xx}}}\RktPn{(}\RktSym{jump}\mbox{\hphantom{\Scribtexttt{x}}}\RktSym{done}\RktPn{)}\RktPn{)}\end{SingleColumn}\end{RktBlk}

\RktRes{42}\end{SingleColumn}\end{RktBlk}\end{SCodeFlow}}

We perform a non{-}tail call to \RktSym{identity} with the argument \RktVal{41},
which is passed on the stack instead of in a register.
We also have one variable live across the call, \RktSym{fv1}.
\RktSym{return{-}point} expands to, essentially, \RktSym{let/cc}, and represents
the start of a non{-}tail call that returns to the continuation \RktSym{l}.
All frame variables are relative to the caller{'}s frame.
However, the frame is allocated just before the non{-}tail call, so the references
to \RktSym{fv0} would be to some uninitialized memory on the callee{'}s frame if
we did not correct the interpretation of frame variables with respect to frame
allocation.
This requires expanding \emph{fvar}s to include a dynamic calculation into
the stack.

We modify the earlier definition of \RktSym{set{\hbox{\texttt{!}}}} to keep track of
\RktSym{current{-}fvar{-}offset} as follows.

\begin{SCodeFlow}\begin{RktBlk}\begin{SingleColumn}\RktPn{(}\RktSym{define{-}syntax}\mbox{\hphantom{\Scribtexttt{x}}}\RktPn{(}\RktSym{set{\hbox{\texttt{!}}}}\mbox{\hphantom{\Scribtexttt{x}}}\RktSym{stx}\RktPn{)}

\mbox{\hphantom{\Scribtexttt{xx}}}\RktPn{(}\RktSym{syntax{-}parse}\mbox{\hphantom{\Scribtexttt{x}}}\RktSym{stx}

\mbox{\hphantom{\Scribtexttt{xxxx}}}\RktCmt{;}\RktCmt{~}\RktCmt{Stack pointer increment}

\mbox{\hphantom{\Scribtexttt{xxxx}}}\RktPn{[}\RktPn{(}\RktSym{{\char`\_}}\mbox{\hphantom{\Scribtexttt{x}}}\RktPn{(}\RktSym{{\textasciitilde}literal}\mbox{\hphantom{\Scribtexttt{x}}}\RktSym{rbp}\RktPn{)}\mbox{\hphantom{\Scribtexttt{x}}}\RktPn{(}\RktSym{binop}\mbox{\hphantom{\Scribtexttt{x}}}\RktPn{(}\RktSym{{\textasciitilde}literal}\mbox{\hphantom{\Scribtexttt{x}}}\RktSym{rbp}\RktPn{)}\mbox{\hphantom{\Scribtexttt{x}}}\RktSym{v}\RktPn{)}\RktPn{)}

\mbox{\hphantom{\Scribtexttt{xxxxx}}}\RktRdr{\#{\textasciigrave}}\RktPn{(}\RktSym{begin}

\mbox{\hphantom{\Scribtexttt{xxxxxxxxx}}}\RktPn{(}\RktSym{set{-}box{\hbox{\texttt{!}}}}\mbox{\hphantom{\Scribtexttt{x}}}\RktSym{current{-}fvar{-}offset}\mbox{\hphantom{\Scribtexttt{x}}}\RktPn{(}\RktSym{binop}\mbox{\hphantom{\Scribtexttt{x}}}\RktPn{(}\RktSym{unbox}\mbox{\hphantom{\Scribtexttt{x}}}\RktSym{current{-}fvar{-}offset}\RktPn{)}\mbox{\hphantom{\Scribtexttt{x}}}\RktSym{v}\RktPn{)}\RktPn{)}

\mbox{\hphantom{\Scribtexttt{xxxxxxxxx}}}\RktPn{(}\RktSym{set{-}box{\hbox{\texttt{!}}}}\mbox{\hphantom{\Scribtexttt{x}}}\RktVar{{\char`\_}rbp}\mbox{\hphantom{\Scribtexttt{x}}}\RktPn{(}\RktSym{binop}\mbox{\hphantom{\Scribtexttt{x}}}\RktSym{rbp}\mbox{\hphantom{\Scribtexttt{x}}}\RktSym{v}\RktPn{)}\RktPn{)}\RktPn{)}\RktPn{]}

\mbox{\hphantom{\Scribtexttt{xxxx}}}\RktSym{{\hbox{\texttt{.}}}{\hbox{\texttt{.}}}{\hbox{\texttt{.}}}{\hbox{\texttt{.}}}}\RktPn{)}\RktPn{)}\end{SingleColumn}\end{RktBlk}\end{SCodeFlow}

Any time the stack pointer is adjusted, we generate a run{-}time update to
\RktSym{current{-}fvar{-}offset} to cooperate with the implementation of frame
variables, in addition to the run{-}time modification of the stack pointer.

\sectionNewpage

\Ssection{Custom Procedures}{Custom Procedures}\label{t:x28part_x22Customx5fProceduresx22x29}

So far, we primarily used Racket{'}s macro system for compile{-}time interposition
on syntactic forms to implement our embedding.
However, one feature relies on impersonators\Autobibref{~(\hyperref[t:x28autobib_x22Tx2e_Stephen_Stricklandx2c_Sam_Tobinx2dHochstadtx2c_Robert_Bruce_Findlerx2c_and_Matthew_FlattChaperones_and_impersonatorsx3a_runx2dtime_support_for_reasonable_interpositionIn_Procx2e_Conference_on_Objectx2dOriented_Programmingx2c_Systemsx2c_Languagesx2c_and_Applications_x28OOPSLAx292012doix3a10x2e1145x2f2384616x2e2384685x22x29]{\AutobibLink{Strickland et al\Sendabbrev{.}}} \hyperref[t:x28autobib_x22Tx2e_Stephen_Stricklandx2c_Sam_Tobinx2dHochstadtx2c_Robert_Bruce_Findlerx2c_and_Matthew_FlattChaperones_and_impersonatorsx3a_runx2dtime_support_for_reasonable_interpositionIn_Procx2e_Conference_on_Objectx2dOriented_Programmingx2c_Systemsx2c_Languagesx2c_and_Applications_x28OOPSLAx292012doix3a10x2e1145x2f2384616x2e2384685x22x29]{\AutobibLink{2012}})}, a feature
for run{-}time interposition on some datatype.

Our compiler implements procedures as a tagged datatype that stores its arity,
to implement dynamic checks, and as a closure, so the label is stored with the
values of all variables free in the procedure{'}s body.
This requries a different procedure interface than Racket{'}s to support
additional operations in intermediate languages.
But, we would like our procedures to expand to Racket procedures, so application
continues to work as expected, and so we can interpret each language primitive
as Racket{'}s equivalent procedure.

To impersonate a procedure, Racket provides \RktSym{impersonate{-}procedure},
which takes an underlying procedure implementation, an optional wrapper around
the result, and a list of optional properties that attach additional information
to the procedure.
Our implementation of procedures is given below.

\begin{SCodeFlow}\begin{RktBlk}\begin{SingleColumn}\RktPn{(}\RktSym{define}\mbox{\hphantom{\Scribtexttt{x}}}\RktPn{(}\RktSym{make{-}procedure}\mbox{\hphantom{\Scribtexttt{x}}}\RktSym{label}\mbox{\hphantom{\Scribtexttt{x}}}\RktSym{arity}\mbox{\hphantom{\Scribtexttt{x}}}\RktSym{env{-}size}\RktPn{)}

\mbox{\hphantom{\Scribtexttt{xx}}}\RktPn{(}\RktSym{impersonate{-}procedure}\mbox{\hphantom{\Scribtexttt{x}}}\RktSym{label}\mbox{\hphantom{\Scribtexttt{x}}}\RktVal{\#f}

\mbox{\hphantom{\Scribtexttt{xxxxxxxxxxxxxxxxxxxxxxxxx}}}\RktSym{proc{-}label{\hbox{\texttt{:}}}prop}\mbox{\hphantom{\Scribtexttt{x}}}\RktSym{label}

\mbox{\hphantom{\Scribtexttt{xxxxxxxxxxxxxxxxxxxxxxxxx}}}\RktSym{proc{-}env{\hbox{\texttt{:}}}prop}\mbox{\hphantom{\Scribtexttt{x}}}\RktPn{(}\RktSym{make{-}vector}\mbox{\hphantom{\Scribtexttt{x}}}\RktSym{env{-}size}\RktPn{)}

\mbox{\hphantom{\Scribtexttt{xxxxxxxxxxxxxxxxxxxxxxxxx}}}\RktSym{proc{-}arity{\hbox{\texttt{:}}}prop}\mbox{\hphantom{\Scribtexttt{x}}}\RktSym{arity}\RktPn{)}\RktPn{)}\end{SingleColumn}\end{RktBlk}\end{SCodeFlow}

In the compiler, procedure datatypes are built from this primitive form
\RktSym{make{-}procedure}. The \RktSym{label} is an assembly label to the code for
the procedure (modelled as a Racket procedure).
The \RktSym{arity} describes the arity of the procedure from the user{'}s
perspective, \emph{i.e.}, the number of parameters the user specified in the
definition.
The \RktSym{env{-}size} declares the size of the procedure{'}s environment.
We embed our procedures as a Racket procedure impersonator, implemented by the
underlying \RktSym{label}, with three properties attaching information needed by
intermediate language interfaces to procedures.

We define the impersonater properties below.

\begin{SCodeFlow}\begin{RktBlk}\begin{SingleColumn}\RktPn{(}\RktSym{define{-}values}\mbox{\hphantom{\Scribtexttt{x}}}\RktPn{(}\RktSym{proc{-}label{\hbox{\texttt{:}}}prop}\mbox{\hphantom{\Scribtexttt{x}}}\RktSym{proc{-}label{\hbox{\texttt{:}}}prop{\hbox{\texttt{?}}}}\mbox{\hphantom{\Scribtexttt{x}}}\RktSym{unsafe{-}procedure{-}label}\RktPn{)}

\mbox{\hphantom{\Scribtexttt{xx}}}\RktPn{(}\RktSym{make{-}impersonator{-}property}\mbox{\hphantom{\Scribtexttt{x}}}\RktVal{{\textquotesingle}}\RktVal{procedure{-}label}\RktPn{)}\RktPn{)}

\mbox{\hphantom{\Scribtexttt{x}}}

\RktPn{(}\RktSym{define{-}values}\mbox{\hphantom{\Scribtexttt{x}}}\RktPn{(}\RktSym{proc{-}env{\hbox{\texttt{:}}}prop}\mbox{\hphantom{\Scribtexttt{x}}}\RktSym{proc{-}env{\hbox{\texttt{:}}}prop{\hbox{\texttt{?}}}}\mbox{\hphantom{\Scribtexttt{x}}}\RktSym{unsafe{-}procedure{-}env}\RktPn{)}

\mbox{\hphantom{\Scribtexttt{xx}}}\RktPn{(}\RktSym{make{-}impersonator{-}property}\mbox{\hphantom{\Scribtexttt{x}}}\RktVal{{\textquotesingle}}\RktVal{procedure{-}env}\RktPn{)}\RktPn{)}

\mbox{\hphantom{\Scribtexttt{x}}}

\RktPn{(}\RktSym{define{-}values}\mbox{\hphantom{\Scribtexttt{x}}}\RktPn{(}\RktSym{proc{-}arity{\hbox{\texttt{:}}}prop}\mbox{\hphantom{\Scribtexttt{x}}}\RktSym{proc{-}arity{\hbox{\texttt{:}}}prop{\hbox{\texttt{?}}}}\mbox{\hphantom{\Scribtexttt{x}}}\RktSym{unsafe{-}procedure{-}arity}\RktPn{)}

\mbox{\hphantom{\Scribtexttt{xx}}}\RktPn{(}\RktSym{make{-}impersonator{-}property}\mbox{\hphantom{\Scribtexttt{x}}}\RktVal{{\textquotesingle}}\RktVal{procedure{-}arity}\RktPn{)}\RktPn{)}

\mbox{\hphantom{\Scribtexttt{x}}}

\RktPn{(}\RktSym{define}\mbox{\hphantom{\Scribtexttt{x}}}\RktPn{(}\RktSym{unsafe{-}procedure{-}ref}\mbox{\hphantom{\Scribtexttt{x}}}\RktSym{p}\mbox{\hphantom{\Scribtexttt{x}}}\RktSym{i}\RktPn{)}

\mbox{\hphantom{\Scribtexttt{xx}}}\RktPn{(}\RktSym{vector{-}ref}\mbox{\hphantom{\Scribtexttt{x}}}\RktPn{(}\RktSym{unsafe{-}procedure{-}env}\mbox{\hphantom{\Scribtexttt{x}}}\RktSym{p}\RktPn{)}\mbox{\hphantom{\Scribtexttt{x}}}\RktSym{i}\RktPn{)}\RktPn{)}

\mbox{\hphantom{\Scribtexttt{x}}}

\RktPn{(}\RktSym{define}\mbox{\hphantom{\Scribtexttt{x}}}\RktPn{(}\RktSym{unsafe{-}procedure{-}set{\hbox{\texttt{!}}}}\mbox{\hphantom{\Scribtexttt{x}}}\RktSym{p}\mbox{\hphantom{\Scribtexttt{x}}}\RktSym{i}\mbox{\hphantom{\Scribtexttt{x}}}\RktSym{v}\RktPn{)}

\mbox{\hphantom{\Scribtexttt{xx}}}\RktPn{(}\RktSym{vector{-}set{\hbox{\texttt{!}}}}\mbox{\hphantom{\Scribtexttt{x}}}\RktPn{(}\RktSym{unsafe{-}procedure{-}env}\mbox{\hphantom{\Scribtexttt{x}}}\RktSym{p}\RktPn{)}\mbox{\hphantom{\Scribtexttt{x}}}\RktSym{i}\mbox{\hphantom{\Scribtexttt{x}}}\RktSym{v}\RktPn{)}\RktPn{)}\end{SingleColumn}\end{RktBlk}\end{SCodeFlow}

Each impersonator property takes a name as a symbol, and returns three values:
the property, a predicate identifying the property, and an accessor which
retrieves the value of the property from an impersonator to which the property
is attached.
We also define the interface for updating the procedure{'}s environment.
In the compiled output, the environment is part of the procedure{'}s
contiguous representation in memory with the label and arity, but in our
embedding, we attach it separately as a vector.

Racket already supports introspecting on a procedure{'}s arity using
\RktSym{procedure{-}arity}.
However, in our embedding, the arity from Racket{'}s perspective is that of the
underlying implementation (\emph{i.e.} the arity of the label), which differs
from the arity of the original procedure.
In the intermediate languages, a procedure takes itself as an argument to have
access to its environment.
We therefore need the arity from the user{'}s perspective, to provide our own
implementation of \RktSym{procedure{-}arity} that is correct with respect to the
user{'}s perspective.

For example, below we define an intermediate language program that corresponds
to the compiled output of the procedure \RktPn{(}\RktSym{let}\Scribtexttt{ }\RktPn{(}\RktPn{[}\RktSym{y}\Scribtexttt{ }\RktVal{21}\RktPn{]}\RktPn{)}\Scribtexttt{ }\RktPn{(}\RktSym{lambda}\Scribtexttt{ }\RktPn{(}\RktSym{x}\RktPn{)}\Scribtexttt{ }\RktPn{(}\RktSym{+}\Scribtexttt{ }\RktSym{x}\Scribtexttt{ }\RktSym{y}\RktPn{)}\RktPn{)}\RktPn{)}, then call it, and inspect its arity.

\LangBox{\LangBoxLabel{\RktModLink{\RktMod{\#lang}}\RktMeta{}\mbox{\hphantom{\Scribtexttt{x}}}\RktMeta{}\RktModLink{\RktSym{cpsc411/hashlangs/base}}\RktMeta{}}}%
{\begin{SCodeFlow}\begin{RktBlk}\begin{SingleColumn}\begin{RktBlk}\begin{SingleColumn}\Scribtexttt{{\Stttextmore} }\RktPn{(}\RktSym{require}\mbox{\hphantom{\Scribtexttt{x}}}\RktPn{(}\RktSym{only{-}in}\mbox{\hphantom{\Scribtexttt{x}}}\RktSym{racket/base}\mbox{\hphantom{\Scribtexttt{x}}}\RktPn{[}\RktSym{lambda}\mbox{\hphantom{\Scribtexttt{x}}}\RktSym{r{\hbox{\texttt{:}}}lambda}\RktPn{]}

\mbox{\hphantom{\Scribtexttt{xx}}}\mbox{\hphantom{\Scribtexttt{xxxxxxxxxxxxxxxxxxxxxxxxxxxxxx}}}\RktPn{[}\RktSym{procedure{-}arity}\mbox{\hphantom{\Scribtexttt{x}}}\RktSym{r{\hbox{\texttt{:}}}procedure{-}arity}\RktPn{]}\RktPn{)}\RktPn{)}\end{SingleColumn}\end{RktBlk}

\begin{RktBlk}\begin{SingleColumn}\Scribtexttt{{\Stttextmore} }\RktPn{(}\RktSym{define}\mbox{\hphantom{\Scribtexttt{x}}}\RktSym{foo}

\mbox{\hphantom{\Scribtexttt{xx}}}\mbox{\hphantom{\Scribtexttt{xx}}}\RktPn{(}\RktSym{make{-}procedure}\mbox{\hphantom{\Scribtexttt{x}}}\RktPn{(}\RktSym{r{\hbox{\texttt{:}}}lambda}\mbox{\hphantom{\Scribtexttt{x}}}\RktPn{(}\RktSym{c}\mbox{\hphantom{\Scribtexttt{x}}}\RktSym{x}\RktPn{)}\mbox{\hphantom{\Scribtexttt{x}}}\RktPn{(}\RktSym{+}\mbox{\hphantom{\Scribtexttt{x}}}\RktSym{x}\mbox{\hphantom{\Scribtexttt{x}}}\RktPn{(}\RktSym{unsafe{-}procedure{-}ref}\mbox{\hphantom{\Scribtexttt{x}}}\RktSym{c}\mbox{\hphantom{\Scribtexttt{x}}}\RktVal{0}\RktPn{)}\RktPn{)}\RktPn{)}\mbox{\hphantom{\Scribtexttt{x}}}\RktVal{1}\mbox{\hphantom{\Scribtexttt{x}}}\RktVal{1}\RktPn{)}\RktPn{)}\end{SingleColumn}\end{RktBlk}

\Scribtexttt{{\Stttextmore} }\RktPn{(}\RktSym{unsafe{-}procedure{-}set{\hbox{\texttt{!}}}}\mbox{\hphantom{\Scribtexttt{x}}}\RktSym{foo}\mbox{\hphantom{\Scribtexttt{x}}}\RktVal{0}\mbox{\hphantom{\Scribtexttt{x}}}\RktVal{21}\RktPn{)}

\Scribtexttt{{\Stttextmore} }\RktPn{(}\RktSym{call}\mbox{\hphantom{\Scribtexttt{x}}}\RktSym{foo}\mbox{\hphantom{\Scribtexttt{x}}}\RktSym{foo}\mbox{\hphantom{\Scribtexttt{x}}}\RktVal{21}\RktPn{)}

\RktRes{42}

\Scribtexttt{{\Stttextmore} }\RktPn{(}\RktSym{unsafe{-}procedure{-}arity}\mbox{\hphantom{\Scribtexttt{x}}}\RktSym{foo}\RktPn{)}

\RktRes{1}

\Scribtexttt{{\Stttextmore} }\RktPn{(}\RktSym{r{\hbox{\texttt{:}}}procedure{-}arity}\mbox{\hphantom{\Scribtexttt{x}}}\RktSym{foo}\RktPn{)}

\RktRes{2}\end{SingleColumn}\end{RktBlk}\end{SCodeFlow}}

The procedure is created using \RktSym{make{-}procedure}.
The embedded representation of a label is a Racket procedure.
The original procedure has one parameter, and has one variable in its
environment.
A low{-}level call to the procedure explicitly passes the procedure itself as the
first argument, and then the arguments corresponding to the parameters.
Variables that were free in the body are now explicitly read from the
procedure{'}s environment.
We can see from \RktSym{r{\hbox{\texttt{:}}}procedure{-}arity} that Racket believes this procedure has
arity \RktVal{2}, but using the accessor defined by our impersonator property,
we get the correct (for the source language definition of the procedure)
\RktVal{1}.

Impersonators are crucial to both reusing, but also extending, existing run{-}time
datatypes.

\sectionNewpage

\Ssection{Interface details: interpositions and eval}{Interface details: interpositions and eval}\label{t:x28part_x22Interfacex5fdetailsx5fx5finterpositionsx5fandx5fevalx22x29}

In most of the intermediate languages, the interpretation of syntax stays the
same.
However, eventually we add tagged datatypes and must reinterpret earlier
primitives.
For example, in the low{-}level languages, \RktSym{+} is the 64{-}bit
twos{-}complement addition.

\noindent \LangBox{\LangBoxLabel{\RktModLink{\RktMod{\#lang}}\RktMeta{}\mbox{\hphantom{\Scribtexttt{x}}}\RktMeta{}\RktModLink{\RktSym{cpsc411/hashlangs/base}}\RktMeta{}}}%
{\begin{SCodeFlow}\begin{RktBlk}\begin{SingleColumn}\Scribtexttt{{\Stttextmore} }\RktPn{(}\RktSym{require}\mbox{\hphantom{\Scribtexttt{x}}}\RktSym{cpsc411/machine{-}ints}\RktPn{)}

\Scribtexttt{{\Stttextmore} }\RktPn{(}\RktSym{+}\RktMeta{}\mbox{\hphantom{\Scribtexttt{x}}}\RktMeta{}\RktPn{(}\RktSym{max{-}int}\RktMeta{}\mbox{\hphantom{\Scribtexttt{x}}}\RktMeta{}\RktVal{64}\RktPn{)}\RktMeta{}\mbox{\hphantom{\Scribtexttt{x}}}\RktMeta{}\RktVal{\#b0}\RktPn{)}\RktMeta{}

\RktRes{9223372036854775807}

\Scribtexttt{{\Stttextmore} }\RktPn{(}\RktSym{+}\RktMeta{}\mbox{\hphantom{\Scribtexttt{x}}}\RktMeta{}\RktPn{(}\RktSym{max{-}int}\RktMeta{}\mbox{\hphantom{\Scribtexttt{x}}}\RktMeta{}\RktVal{64}\RktPn{)}\RktMeta{}\mbox{\hphantom{\Scribtexttt{x}}}\RktMeta{}\RktVal{\#b1}\RktPn{)}\RktMeta{}

\RktRes{\mbox{{-}9}223372036854775808}\end{SingleColumn}\end{RktBlk}\end{SCodeFlow}}

But, after we add tagged data, the surface{-}language \RktSym{+} is defined on
61{-}bit tagged integers.
Intermediate languages further in the pipeline implement this in terms of the
64{-}bit \RktSym{+}.

\LangBox{\LangBoxLabel{\RktModLink{\RktMod{\#lang}}\RktMeta{}\mbox{\hphantom{\Scribtexttt{x}}}\RktMeta{}\RktModLink{\RktSym{cpsc411/hashlangs/v7}}\RktMeta{}}}%
{\begin{SCodeFlow}\begin{RktBlk}\begin{SingleColumn}\Scribtexttt{{\Stttextmore} }\RktPn{(}\RktSym{require}\mbox{\hphantom{\Scribtexttt{x}}}\RktSym{cpsc411/machine{-}ints}\RktPn{)}

\Scribtexttt{{\Stttextmore} }\RktPn{(}\RktSym{+}\RktMeta{}\mbox{\hphantom{\Scribtexttt{x}}}\RktMeta{}\RktPn{(}\RktSym{max{-}int}\RktMeta{}\mbox{\hphantom{\Scribtexttt{x}}}\RktMeta{}\RktVal{61}\RktPn{)}\RktMeta{}\mbox{\hphantom{\Scribtexttt{x}}}\RktMeta{}\RktVal{\#b0}\RktPn{)}\RktMeta{}

\RktRes{1152921504606846975}

\Scribtexttt{{\Stttextmore} }\RktPn{(}\RktSym{+}\RktMeta{}\mbox{\hphantom{\Scribtexttt{x}}}\RktMeta{}\RktPn{(}\RktSym{max{-}int}\RktMeta{}\mbox{\hphantom{\Scribtexttt{x}}}\RktMeta{}\RktVal{61}\RktPn{)}\RktMeta{}\mbox{\hphantom{\Scribtexttt{x}}}\RktMeta{}\RktVal{\#b1}\RktPn{)}\RktMeta{}

\RktRes{\mbox{{-}1}152921504606846976}\end{SingleColumn}\end{RktBlk}\end{SCodeFlow}}

Racket{'}s module system makes importing the existing embedding, redefining
\RktSym{+}, and exposing only the new implementation easy.
In fact, it is no different from extending Racket with an embedding of x64
features; the same features that make embedding x64 into Racket easy make
extending our x64 embedding easy.

We essentially define a module like the following.

\begin{SCodeFlow}\begin{RktBlk}\begin{SingleColumn}\RktPn{(}\RktSym{module}\mbox{\hphantom{\Scribtexttt{x}}}\RktSym{tagged{-}+}\mbox{\hphantom{\Scribtexttt{x}}}\RktSym{racket}

\mbox{\hphantom{\Scribtexttt{xx}}}\RktPn{(}\RktSym{require}\mbox{\hphantom{\Scribtexttt{x}}}\RktPn{(}\RktSym{except{-}in}\mbox{\hphantom{\Scribtexttt{x}}}\RktVal{"base{\hbox{\texttt{.}}}rkt"}\mbox{\hphantom{\Scribtexttt{x}}}\RktSym{+}\RktPn{)}

\mbox{\hphantom{\Scribtexttt{xxxxxxxxxxx}}}\RktSym{cpsc411/machine{-}ints}

\mbox{\hphantom{\Scribtexttt{xxxxxxxxxxx}}}\RktPn{(}\RktSym{rename{-}in}\mbox{\hphantom{\Scribtexttt{x}}}\RktSym{racket}\mbox{\hphantom{\Scribtexttt{x}}}\RktPn{[}\RktSym{define}\mbox{\hphantom{\Scribtexttt{x}}}\RktSym{r{\hbox{\texttt{:}}}define}\RktPn{]}\RktPn{)}\RktPn{)}

\mbox{\hphantom{\Scribtexttt{xx}}}\RktPn{(}\RktSym{provide}\mbox{\hphantom{\Scribtexttt{x}}}\RktPn{(}\RktSym{all{-}from{-}out}\mbox{\hphantom{\Scribtexttt{x}}}\RktVal{"base{\hbox{\texttt{.}}}rkt"}\RktPn{)}

\mbox{\hphantom{\Scribtexttt{xxxxxxxxxxx}}}\RktSym{+}\RktPn{)}

\mbox{\hphantom{\Scribtexttt{x}}}

\mbox{\hphantom{\Scribtexttt{xx}}}\RktPn{(}\RktSym{r{\hbox{\texttt{:}}}define}\mbox{\hphantom{\Scribtexttt{x}}}\RktSym{+}\mbox{\hphantom{\Scribtexttt{x}}}\RktPn{(}\RktSym{curry}\mbox{\hphantom{\Scribtexttt{x}}}\RktSym{twos{-}complement{-}add}\mbox{\hphantom{\Scribtexttt{x}}}\RktVal{61}\RktPn{)}\RktPn{)}\RktPn{)}\end{SingleColumn}\end{RktBlk}\end{SCodeFlow}

We import everything from the base embedding, except \RktSym{+}.
Then, we export all the base definitions we imported, and export our new 61{-}bit
two{'}s complement version of \RktSym{+}.
The new module can then be used as a language.
This is the language{-}oriented approach to building a new interpreter.

While using these embedded languages in the REPL and as a \RktModLink{\RktMod{\#lang}} is
useful, recall that our original goal was to write \emph{interpreters}.
The embedded language provides an interpretation of the programs, but we have
not yet defined an interpreter, \emph{i.e.}, a function that consumes the syntax
of the program and returns the result.

To do this, we reuse \RktSym{eval}, which takes a representation of program
syntax (either as a syntax object or as a quoted list) and evaluates it.
In Racket, however, \RktSym{eval} also takes an optional second argument,
specifying the namespace in which to evaluate the program.
We can use this to specify evaluating the program in our language, rather than
in Racket.

\LangBox{\LangBoxLabel{\RktModLink{\RktMod{\#lang}}\RktMeta{}\mbox{\hphantom{\Scribtexttt{x}}}\RktMeta{}\RktModLink{\RktSym{racket}}\RktMeta{}}}%
{\begin{SCodeFlow}\begin{RktBlk}\begin{SingleColumn}\Scribtexttt{{\Stttextmore} }\RktPn{(}\RktSym{require}\mbox{\hphantom{\Scribtexttt{x}}}\RktSym{cpsc411/machine{-}ints}\RktPn{)}

\begin{RktBlk}\begin{SingleColumn}\Scribtexttt{{\Stttextmore} }\RktPn{(}\RktSym{define}\mbox{\hphantom{\Scribtexttt{x}}}\RktPn{(}\RktSym{base{-}eval}\mbox{\hphantom{\Scribtexttt{x}}}\RktSym{x}\RktPn{)}

\mbox{\hphantom{\Scribtexttt{xx}}}\mbox{\hphantom{\Scribtexttt{xx}}}\RktPn{(}\RktSym{local{-}require}\mbox{\hphantom{\Scribtexttt{x}}}\RktSym{cpsc411/hashlangs/base}\RktPn{)}

\mbox{\hphantom{\Scribtexttt{xx}}}\mbox{\hphantom{\Scribtexttt{xx}}}\RktPn{(}\RktSym{eval}\mbox{\hphantom{\Scribtexttt{x}}}\RktSym{x}\mbox{\hphantom{\Scribtexttt{x}}}\RktPn{(}\RktSym{module{-}{\Stttextmore}namespace}\mbox{\hphantom{\Scribtexttt{x}}}\RktVal{{\textquotesingle}}\RktVal{cpsc411/hashlangs/base}\RktPn{)}\RktPn{)}\RktPn{)}\end{SingleColumn}\end{RktBlk}

\begin{RktBlk}\begin{SingleColumn}\Scribtexttt{{\Stttextmore} }\RktPn{(}\RktSym{define}\mbox{\hphantom{\Scribtexttt{x}}}\RktPn{(}\RktSym{v7{-}eval}\mbox{\hphantom{\Scribtexttt{x}}}\RktSym{x}\RktPn{)}

\mbox{\hphantom{\Scribtexttt{xx}}}\mbox{\hphantom{\Scribtexttt{xx}}}\RktPn{(}\RktSym{local{-}require}\mbox{\hphantom{\Scribtexttt{x}}}\RktSym{cpsc411/hashlangs/v7}\RktPn{)}

\mbox{\hphantom{\Scribtexttt{xx}}}\mbox{\hphantom{\Scribtexttt{xx}}}\RktPn{(}\RktSym{eval}\mbox{\hphantom{\Scribtexttt{x}}}\RktSym{x}\mbox{\hphantom{\Scribtexttt{x}}}\RktPn{(}\RktSym{module{-}{\Stttextmore}namespace}\mbox{\hphantom{\Scribtexttt{x}}}\RktVal{{\textquotesingle}}\RktVal{cpsc411/hashlangs/v7}\RktPn{)}\RktPn{)}\RktPn{)}\end{SingleColumn}\end{RktBlk}

\Scribtexttt{{\Stttextmore} }\RktPn{(}\RktSym{base{-}eval}\mbox{\hphantom{\Scribtexttt{x}}}\RktVal{{\textasciigrave}}\RktVal{(}\RktVal{+}\mbox{\hphantom{\Scribtexttt{x}}}\RktRdr{,}\RktPn{(}\RktSym{max{-}int}\mbox{\hphantom{\Scribtexttt{x}}}\RktVal{64}\RktPn{)}\mbox{\hphantom{\Scribtexttt{x}}}\RktVal{0}\RktVal{)}\RktPn{)}

\RktRes{9223372036854775807}

\Scribtexttt{{\Stttextmore} }\RktPn{(}\RktSym{base{-}eval}\mbox{\hphantom{\Scribtexttt{x}}}\RktVal{{\textasciigrave}}\RktVal{(}\RktVal{+}\mbox{\hphantom{\Scribtexttt{x}}}\RktRdr{,}\RktPn{(}\RktSym{max{-}int}\mbox{\hphantom{\Scribtexttt{x}}}\RktVal{64}\RktPn{)}\mbox{\hphantom{\Scribtexttt{x}}}\RktVal{1}\RktVal{)}\RktPn{)}

\RktRes{{-}9223372036854775808}

\Scribtexttt{{\Stttextmore} }\RktPn{(}\RktSym{v7{-}eval}\mbox{\hphantom{\Scribtexttt{x}}}\RktVal{{\textasciigrave}}\RktVal{(}\RktVal{+}\mbox{\hphantom{\Scribtexttt{x}}}\RktRdr{,}\RktPn{(}\RktSym{max{-}int}\mbox{\hphantom{\Scribtexttt{x}}}\RktVal{61}\RktPn{)}\mbox{\hphantom{\Scribtexttt{x}}}\RktVal{0}\RktVal{)}\RktPn{)}

\RktRes{1152921504606846975}

\Scribtexttt{{\Stttextmore} }\RktPn{(}\RktSym{v7{-}eval}\mbox{\hphantom{\Scribtexttt{x}}}\RktVal{{\textasciigrave}}\RktVal{(}\RktVal{+}\mbox{\hphantom{\Scribtexttt{x}}}\RktRdr{,}\RktPn{(}\RktSym{max{-}int}\mbox{\hphantom{\Scribtexttt{x}}}\RktVal{61}\RktPn{)}\mbox{\hphantom{\Scribtexttt{x}}}\RktVal{1}\RktVal{)}\RktPn{)}

\RktRes{{-}1152921504606846976}\end{SingleColumn}\end{RktBlk}\end{SCodeFlow}}

Now, when we need an interpreter for one of the intermediate languages, we
wrap \RktSym{eval} with the namespace derived from the module that embeds the
right set of features.
To expose all the right identifiers, we still did have to do that about 126
times, though.

This method of implementing our interpreters has some limitations.
Since almost all the features of all the languages are implemented in a single
module, invalid programs have an interpretation.
For example, we managed to evaluate the expression \RktVal{{\textasciigrave}}\RktVal{(}\RktVal{+}\Scribtexttt{ }\RktRdr{,}\RktPn{(}\RktSym{max{-}int}\Scribtexttt{ }\RktVal{64}\RktPn{)}\Scribtexttt{ }\RktVal{0}\RktVal{)}, although in none of the grammars above is that considered a program.

We separate the implementation of the interpreters from validating the input
programs.
We use a DSL to generate PLT Redex\Autobibref{~(\hyperref[t:x28autobib_x22Matthias_Felleisenx2c_Robert_Bruce_Findlerx2c_and_Matthew_FlattSemantics_Engineering_with_PLT_Redex2009httpsx3ax2fx2fmitpressx2emitx2eedux2fbooksx2fsemanticsx2dengineeringx2dpltx2dredexx22x29]{\AutobibLink{Felleisen et al\Sendabbrev{.}}} \hyperref[t:x28autobib_x22Matthias_Felleisenx2c_Robert_Bruce_Findlerx2c_and_Matthew_FlattSemantics_Engineering_with_PLT_Redex2009httpsx3ax2fx2fmitpressx2emitx2eedux2fbooksx2fsemanticsx2dengineeringx2dpltx2dredexx22x29]{\AutobibLink{2009}}; \hyperref[t:x28autobib_x22Casey_Kleinx2c_John_Clementsx2c_Christos_Dimoulasx2c_Carl_Eastlundx2c_Matthias_Felleisenx2c_Matthew_Flattx2c_Jay_Ax2e_McCarthyx2c_Jon_Rafkindx2c_Sam_Tobinx2dHochstadtx2c_and_Robert_Bruce_FindlerRun_your_researchx3a_on_the_effectiveness_of_lightweight_mechanizationIn_Procx2e_Symposium_on_Principles_of_Programming_Languages_x28POPLx292012doix3a10x2e1145x2f2103656x2e2103691x22x29]{\AutobibLink{Klein et al\Sendabbrev{.}}} \hyperref[t:x28autobib_x22Casey_Kleinx2c_John_Clementsx2c_Christos_Dimoulasx2c_Carl_Eastlundx2c_Matthias_Felleisenx2c_Matthew_Flattx2c_Jay_Ax2e_McCarthyx2c_Jon_Rafkindx2c_Sam_Tobinx2dHochstadtx2c_and_Robert_Bruce_FindlerRun_your_researchx3a_on_the_effectiveness_of_lightweight_mechanizationIn_Procx2e_Symposium_on_Principles_of_Programming_Languages_x28POPLx292012doix3a10x2e1145x2f2103656x2e2103691x22x29]{\AutobibLink{2012}})} grammar
definitions from our typeset language grammars.
We then use Redex{'}s \RktSym{redex{-}match} to generate predicates for checking
that terms are syntactically well formed for each language; the \emph{p}
non{-}terminal defines a valid program in the language.

As an example, consider the languages involved in adding tagged data.
Similar to the previous grammar, in \ChapRefLocalUC{t:x28part_x22secx3ailx22x29}{3}{Intermediate language abstractions and metadata}, the surface language
requires programs to be a module with optional top{-}level procedure definitions
and a top{-}level expression.
Since expressions are over tagged data, integer literals are restricted to 61
bits.

\noindent \LangBox{\LangBoxLabel{\RktModLink{\RktMod{\#lang}}\RktMeta{}\mbox{\hphantom{\Scribtexttt{x}}}\RktMeta{}\RktModLink{\RktSym{racket}}\RktMeta{}}}%
{\begin{SCodeFlow}\begin{RktBlk}\begin{SingleColumn}\Scribtexttt{{\Stttextmore} }\RktPn{(}\RktSym{require}\mbox{\hphantom{\Scribtexttt{x}}}\RktSym{cpsc411/langs/v7}\mbox{\hphantom{\Scribtexttt{x}}}\RktSym{cpsc411/machine{-}ints}\RktPn{)}

\Scribtexttt{{\Stttextmore} }\RktPn{(}\RktSym{exprs{-}lang{-}v7{\hbox{\texttt{?}}}}\mbox{\hphantom{\Scribtexttt{x}}}\RktVal{{\textasciigrave}}\RktVal{(}\RktVal{+}\mbox{\hphantom{\Scribtexttt{x}}}\RktRdr{,}\RktPn{(}\RktSym{max{-}int}\mbox{\hphantom{\Scribtexttt{x}}}\RktVal{64}\RktPn{)}\mbox{\hphantom{\Scribtexttt{x}}}\RktVal{0}\RktVal{)}\RktPn{)}

\RktRes{\#f}

\Scribtexttt{{\Stttextmore} }\RktPn{(}\RktSym{interp{-}exprs{-}lang{-}v7}\mbox{\hphantom{\Scribtexttt{x}}}\RktVal{{\textasciigrave}}\RktVal{(}\RktVal{+}\mbox{\hphantom{\Scribtexttt{x}}}\RktRdr{,}\RktPn{(}\RktSym{max{-}int}\mbox{\hphantom{\Scribtexttt{x}}}\RktVal{64}\RktPn{)}\mbox{\hphantom{\Scribtexttt{x}}}\RktVal{0}\RktVal{)}\RktPn{)}

\RktRes{\mbox{{-}1}}

\Scribtexttt{{\Stttextmore} }\RktPn{(}\RktSym{exprs{-}lang{-}v7{\hbox{\texttt{?}}}}\mbox{\hphantom{\Scribtexttt{x}}}\RktVal{{\textasciigrave}}\RktVal{(}\RktVal{module}\mbox{\hphantom{\Scribtexttt{x}}}\RktVal{(}\RktVal{call}\mbox{\hphantom{\Scribtexttt{x}}}\RktVal{+}\mbox{\hphantom{\Scribtexttt{x}}}\RktRdr{,}\RktPn{(}\RktSym{max{-}int}\mbox{\hphantom{\Scribtexttt{x}}}\RktVal{61}\RktPn{)}\mbox{\hphantom{\Scribtexttt{x}}}\RktVal{0}\RktVal{)}\RktVal{)}\RktPn{)}

\RktRes{\#t}

\Scribtexttt{{\Stttextmore} }\RktPn{(}\RktSym{interp{-}exprs{-}lang{-}v7}\mbox{\hphantom{\Scribtexttt{x}}}\RktVal{{\textasciigrave}}\RktVal{(}\RktVal{module}\mbox{\hphantom{\Scribtexttt{x}}}\RktVal{(}\RktVal{call}\mbox{\hphantom{\Scribtexttt{x}}}\RktVal{+}\mbox{\hphantom{\Scribtexttt{x}}}\RktRdr{,}\RktPn{(}\RktSym{max{-}int}\mbox{\hphantom{\Scribtexttt{x}}}\RktVal{61}\RktPn{)}\mbox{\hphantom{\Scribtexttt{x}}}\RktVal{0}\RktVal{)}\RktVal{)}\RktPn{)}

\RktRes{1152921504606846975}\end{SingleColumn}\end{RktBlk}\end{SCodeFlow}}

The expression \RktVal{{\textasciigrave}}\RktVal{(}\RktVal{+}\Scribtexttt{ }\RktRdr{,}\RktPn{(}\RktSym{max{-}int}\Scribtexttt{ }\RktVal{64}\RktPn{)}\Scribtexttt{ }\RktVal{0}\RktVal{)} is invalid, since this language
requires an explicit \RktSym{call} form to call a procedure, and because the
language only supports 61{-}bit integer literals.
But the interpreter happily does something anyway.
It is useful to expose this unsafe interface to discuss things such as undefined
behaviour, but we may want to expose another interface to warn the user that a
test failed because the program was syntactically invalid, and not because the
behaviour changed.

We can combine the interpreter and these predicates at our desired interface to
ensure that our interpreters only receive well{-}formed terms, using
contracts\Autobibref{~(\hyperref[t:x28autobib_x22Robert_Bruce_Findler_and_Matthias_FelleisenContracts_for_higherx2dorder_functionsIn_Procx2e_International_Conference_on_Functional_Programming_x28ICFPx292002doix3a10x2e1145x2f581478x2e581484x22x29]{\AutobibLink{Findler and Felleisen}} \hyperref[t:x28autobib_x22Robert_Bruce_Findler_and_Matthias_FelleisenContracts_for_higherx2dorder_functionsIn_Procx2e_International_Conference_on_Functional_Programming_x28ICFPx292002doix3a10x2e1145x2f581478x2e581484x22x29]{\AutobibLink{2002}})}.
Below, we attach the predicate as a contract to the interpreter as it is
exported from the module \RktSym{checked{-}interp{-}v7}.
The previous, uncontracted implementation is still available from the original
module.
The contract requires the input to the interpreter to satsify the predicate
\RktSym{exprs{-}lang{-}v7{\hbox{\texttt{?}}}}, and output from the interpreter to be a 61{-}bit integer.

\noindent \LangBox{\LangBoxLabel{\RktModLink{\RktMod{\#lang}}\RktMeta{}\mbox{\hphantom{\Scribtexttt{x}}}\RktMeta{}\RktModLink{\RktSym{racket}}\RktMeta{}}}%
{\begin{SCodeFlow}\begin{RktBlk}\begin{SingleColumn}\begin{RktBlk}\begin{SingleColumn}\Scribtexttt{{\Stttextmore} }\RktPn{(}\RktSym{module}\mbox{\hphantom{\Scribtexttt{x}}}\RktSym{checked{-}interp{-}v7}\mbox{\hphantom{\Scribtexttt{x}}}\RktSym{racket}

\mbox{\hphantom{\Scribtexttt{xx}}}\mbox{\hphantom{\Scribtexttt{xx}}}\RktPn{(}\RktSym{require}\mbox{\hphantom{\Scribtexttt{x}}}\RktSym{cpsc411/langs/v7}\mbox{\hphantom{\Scribtexttt{x}}}\RktSym{cpsc411/compiler{-}lib}\RktPn{)}

\mbox{\hphantom{\Scribtexttt{xx}}}

\mbox{\hphantom{\Scribtexttt{xx}}}\mbox{\hphantom{\Scribtexttt{xx}}}\RktPn{(}\RktSym{provide}\mbox{\hphantom{\Scribtexttt{x}}}\RktPn{(}\RktSym{contract{-}out}

\mbox{\hphantom{\Scribtexttt{xx}}}\mbox{\hphantom{\Scribtexttt{xxxxxxxxxxxx}}}\RktPn{[}\RktSym{interp{-}exprs{-}lang{-}v7}\mbox{\hphantom{\Scribtexttt{x}}}\RktPn{(}\RktSym{\mbox{{-}{\Stttextmore}}}\mbox{\hphantom{\Scribtexttt{x}}}\RktSym{exprs{-}lang{-}v7{\hbox{\texttt{?}}}}\mbox{\hphantom{\Scribtexttt{x}}}\RktSym{int61{\hbox{\texttt{?}}}}\RktPn{)}\RktPn{]}\RktPn{)}\RktPn{)}\RktPn{)}\end{SingleColumn}\end{RktBlk}

\Scribtexttt{{\Stttextmore} }\RktPn{(}\RktSym{require}\mbox{\hphantom{\Scribtexttt{x}}}\RktVal{{\textquotesingle}}\RktVal{checked{-}interp{-}v7}\RktPn{)}

\Scribtexttt{{\Stttextmore} }\RktPn{(}\RktSym{interp{-}exprs{-}lang{-}v7}\mbox{\hphantom{\Scribtexttt{x}}}\RktVal{{\textasciigrave}}\RktVal{(}\RktVal{+}\mbox{\hphantom{\Scribtexttt{x}}}\RktRdr{,}\RktPn{(}\RktSym{max{-}int}\mbox{\hphantom{\Scribtexttt{x}}}\RktVal{64}\RktPn{)}\mbox{\hphantom{\Scribtexttt{x}}}\RktVal{0}\RktVal{)}\RktPn{)}

\RktErr{interp{-}exprs{-}lang{-}v7: contract violation}

\RktErr{}\mbox{\hphantom{\Scribtexttt{xx}}}\RktErr{expected: exprs{-}lang{-}v7?}

\RktErr{}\mbox{\hphantom{\Scribtexttt{xx}}}\RktErr{given: {\textquotesingle}(+ 9223372036854775807 0)}

\RktErr{}\mbox{\hphantom{\Scribtexttt{xx}}}\RktErr{in: the 1st argument of}

\RktErr{}\mbox{\hphantom{\Scribtexttt{xx}}}\RktErr{}\mbox{\hphantom{\Scribtexttt{xx}}}\RktErr{}\mbox{\hphantom{\Scribtexttt{xx}}}\RktErr{({-}$>$ exprs{-}lang{-}v7? int61?)}

\RktErr{}\mbox{\hphantom{\Scribtexttt{xx}}}\RktErr{contract from: checked{-}interp{-}v7}

\RktErr{}\mbox{\hphantom{\Scribtexttt{xx}}}\RktErr{blaming: program}

\RktErr{}\mbox{\hphantom{\Scribtexttt{x}}}\RktErr{}\mbox{\hphantom{\Scribtexttt{xx}}}\RktErr{(assuming the contract is correct)}

\RktErr{}\mbox{\hphantom{\Scribtexttt{xx}}}\RktErr{at: eval:6:0}

\Scribtexttt{{\Stttextmore} }\RktPn{(}\RktSym{interp{-}exprs{-}lang{-}v7}\mbox{\hphantom{\Scribtexttt{x}}}\RktVal{{\textasciigrave}}\RktVal{(}\RktVal{module}\mbox{\hphantom{\Scribtexttt{x}}}\RktVal{(}\RktVal{call}\mbox{\hphantom{\Scribtexttt{x}}}\RktVal{+}\mbox{\hphantom{\Scribtexttt{x}}}\RktRdr{,}\RktPn{(}\RktSym{max{-}int}\mbox{\hphantom{\Scribtexttt{x}}}\RktVal{64}\RktPn{)}\mbox{\hphantom{\Scribtexttt{x}}}\RktVal{0}\RktVal{)}\RktVal{)}\RktPn{)}

\RktErr{interp{-}exprs{-}lang{-}v7: contract violation}

\RktErr{}\mbox{\hphantom{\Scribtexttt{xx}}}\RktErr{expected: exprs{-}lang{-}v7?}

\RktErr{}\mbox{\hphantom{\Scribtexttt{xx}}}\RktErr{given: {\textquotesingle}(module (call + 9223372036854775807 0))}

\RktErr{}\mbox{\hphantom{\Scribtexttt{xx}}}\RktErr{in: the 1st argument of}

\RktErr{}\mbox{\hphantom{\Scribtexttt{xx}}}\RktErr{}\mbox{\hphantom{\Scribtexttt{xx}}}\RktErr{}\mbox{\hphantom{\Scribtexttt{xx}}}\RktErr{({-}$>$ exprs{-}lang{-}v7? int61?)}

\RktErr{}\mbox{\hphantom{\Scribtexttt{xx}}}\RktErr{contract from: checked{-}interp{-}v7}

\RktErr{}\mbox{\hphantom{\Scribtexttt{xx}}}\RktErr{blaming: program}

\RktErr{}\mbox{\hphantom{\Scribtexttt{x}}}\RktErr{}\mbox{\hphantom{\Scribtexttt{xx}}}\RktErr{(assuming the contract is correct)}

\RktErr{}\mbox{\hphantom{\Scribtexttt{xx}}}\RktErr{at: eval:6:0}

\Scribtexttt{{\Stttextmore} }\RktPn{(}\RktSym{interp{-}exprs{-}lang{-}v7}\mbox{\hphantom{\Scribtexttt{x}}}\RktVal{{\textasciigrave}}\RktVal{(}\RktVal{module}\mbox{\hphantom{\Scribtexttt{x}}}\RktVal{(}\RktVal{call}\mbox{\hphantom{\Scribtexttt{x}}}\RktVal{+}\mbox{\hphantom{\Scribtexttt{x}}}\RktRdr{,}\RktPn{(}\RktSym{max{-}int}\mbox{\hphantom{\Scribtexttt{x}}}\RktVal{61}\RktPn{)}\mbox{\hphantom{\Scribtexttt{x}}}\RktVal{0}\RktVal{)}\RktVal{)}\RktPn{)}

\RktRes{1152921504606846975}\end{SingleColumn}\end{RktBlk}\end{SCodeFlow}}

The error message isn{'}t precise since the predicate we use in the contract is
na\"{i}vely generated, but it does tell a user whose test fails that the problem is
an invalid source program provided to the interpreter, rather than a compiler
pass producing a valid a program that runs to an unexpected result.

\sectionNewpage

\Ssection{Discussion}{Discussion}\label{t:x28part_x22secx3adiscussionx22x29}

We presented an extended case study of language{-}oriented design to macro{-}embed a
family of compiler intermediate languages into Racket.
The same abstractions used to implement new general{-}purpose languages and
various high{-}level DSLs prove equally useful for low{-}level languages and
exposing a family of languages quickly.

The full implementation differs slightly from what we described, although the
typeset interactive examples all run as above.
The main difference a reader might notice is that global state, such as
registers and memory, are implicitly reset at module boundaries, except during
REPL interactions.
Resetting global state makes interoperability with Racket more difficult, but
simplifies our testing framework.

Aside from the engineering benefits of the embedding, one persistant thought
during this experience is that the implementation by local, compositional
macro{-}embedding essentially forces us to construct a compositional \emph{model}
of the language, in a way that writing a recursive interpreter does not.
This view of constructing{-}a{-}model{-}as{-}implementation has been beneficial
in helping us understand the semantics of the languages and identifying
weaknesses.
Invariably, when we cannot find an easy local embedding, one of two things is
going on: either the language is poorly designed, or we don{'}t understand the
language as well as we thought we did.
One instance that stood out and is discussed earlier is our implementation of
abstract locations, which requires inverting expansion order and performing an
analysis.
This analysis felt unnecessary, and upon reflection, we realized the compiler
already had the binding information we were trying to reconstruct, but threw it
away.
If we redesigned the intermediate languages to declare the scope of abstract
locations, this would improve our embedding, but also have pedagogical benefits
for the course{---}students are often confused about the scope of abstract
locations, and of course they are when no form explicitly declares their scope.
A similar thought occurs about frame variables; instead of doing either an
analysis or binding an arbitrary number of them \emph{a priori}, wouldn{'}t it be
better if we could embed a new variable{-}like form?
If the syntax were \RktPn{(}\RktSym{fv}\Scribtexttt{ }\RktVal{1}\RktPn{)} instead of \RktSym{fv1}, and variable{-}like
forms could be compound rather than just identifiers, then embedding frame
variables would be as easy as registers.

During this project, we struggled in a couple of places that suggest the need
for new abstractions.

The primary one discussed in this article is the use of a more general
definition of variable by \RktSym{set{\hbox{\texttt{!}}}}, so that addresses like \RktPn{(}\RktSym{rbp}\Scribtexttt{ }\RktSym{\mbox{{-}}}\Scribtexttt{ }\RktVal{0}\RktPn{)} can be interpreted as assignable.
We seem to need a generalization of \RktSym{make{-}variable{-}like{-}transformer}
that supports a variable{-}like \emph{expression}, that can appear either as an
identifier or as an operator.
Such "variables" could be referenced as expressions, or appear as the left{-}hand
side of a \RktSym{set{\hbox{\texttt{!}}}} to indicate assignment.
Then addresses could be macro{-}embedded modularly, like registers, although
because registers in operator position would be part of an address, such an
implementation would interact with the embedding of registers.
The abstraction should be a simple extension of \RktSym{set{\hbox{\texttt{!}}}}.
Although this feature is desirable in our implementation, it is unclear whether
it would have more general use.
As is, \RktSym{set{\hbox{\texttt{!}}}} implements multiple cross{-}cutting functionality, and
is the least modular aspect of our implementation.

One headache not discussed in this article is the transition from a module
language to a \RktModLink{\RktMod{\#lang}}.
Any module can easily be interpreted as an s{-}expression{-}based language, with its
exported identifiers forming the language.
For example, the module defining our macro embedding is
\RktModLink{\RktSym{cpsc411/langs/base}}; to use it as a language we need only start a
language line with \RktModLink{\RktMod{\#lang}}\RktMeta{}\mbox{\hphantom{\Scribtexttt{x}}}\RktMeta{}\RktModLink{\RktSym{s{-}exp}}\RktMeta{}\mbox{\hphantom{\Scribtexttt{x}}}\RktMeta{cpsc411/langs/base}.
However, excluding the \RktSym{s{-}exp} meta{-}language to get a language that has
the same status as \RktModLink{\RktMod{\#lang}}\RktMeta{}\mbox{\hphantom{\Scribtexttt{x}}}\RktMeta{}\RktModLink{\RktSym{racket}}\RktMeta{} takes work.
We must create a particular file and module structure for each language.
As a result, we only create two honest{-}to{-}goodness \RktModLink{\RktMod{\#lang}}s, and normally
use \RktModLink{\RktMod{\#lang}}\RktMeta{}\mbox{\hphantom{\Scribtexttt{x}}}\RktMeta{}\RktModLink{\RktSym{s{-}exp}}\RktMeta{}\mbox{\hphantom{\Scribtexttt{x}}}\RktMeta{cpsc411/langs/base} instead.
This kind of boilerplate outside the linguistic abstractions is in opposition to
The Racket Manifesto\Autobibref{~(\hyperref[t:x28autobib_x22Matthias_Felleisenx2c_Robert_Bruce_Findlerx2c_Matthew_Flattx2c_Shriram_Krishnamurthix2c_Eli_Barzilayx2c_Jay_Ax2e_McCarthyx2c_and_Sam_Tobinx2dHochstadtThe_Racket_ManifestoIn_Procx2e_SNAPL2015doix3a10x2e4230x2fLIPIcsx2eSNAPLx2e2015x2e113x22x29]{\AutobibLink{Felleisen et al\Sendabbrev{.}}} \hyperref[t:x28autobib_x22Matthias_Felleisenx2c_Robert_Bruce_Findlerx2c_Matthew_Flattx2c_Shriram_Krishnamurthix2c_Eli_Barzilayx2c_Jay_Ax2e_McCarthyx2c_and_Sam_Tobinx2dHochstadtThe_Racket_ManifestoIn_Procx2e_SNAPL2015doix3a10x2e4230x2fLIPIcsx2eSNAPLx2e2015x2e113x22x29]{\AutobibLink{2015}})}, although we don{'}t know how to improve
it.

\sectionNewpage

\Ssection{Acknowledgements}{Acknowledgements}\label{t:x28part_x22Acknowledgementsx22x29}

For their time and thoughtful feedback on this paper, I gratefully acknowledge
the anonymous reviewers, Jonathan Chan, Yanze Li, Nico Ritschel, Eric Conlon,
and Ivan Beschastnikh.
I owe many thanks to Andy Keep, Roshan James, and R. Kent Dybvig for teaching me
compiler design and implementation, and providing the course notes on which the
design of these intermediate languages are based.
Thanks also to Ron Garcia, Lily Bryant, Justin Hu, Yuchong Pan, Benjamin Lerner,
and David Ewert for providing support and feedback on this project.
Finally, I owe a big thank you to all the students of UBC{'}s CPSC 411 for testing
and reporting bugs in this software.

\sectionNewpage

\Ssectionstarx{Bibliography}{Bibliography}\label{t:x28part_x22docx2dbibliographyx22x29}

\begin{AutoBibliography}\begin{SingleColumn}\Autobibtarget{\label{t:x28autobib_x22Ryan_CulpepperFortifying_macrosJournal_of_Functional_Programming_x28JFPx29_22x284x2d5x29x2c_ppx2e_439x2dx2d4762012doix3a10x2e1017x2fS0956796812000275x22x29}\Autobibentry{Ryan Culpepper. Fortifying macros. \textit{Journal of Functional Programming (JFP)} 22(4{-}5), pp. 439{--}476, 2012. \pseudodoi{doi:\href{https://doi.org/10.1017/S0956796812000275}{10{\hbox{\texttt{.}}}1017/S0956796812000275}}}}

\Autobibtarget{\label{t:x28autobib_x22Ryan_Culpepper_and_Matthias_FelleisenFortifying_macrosIn_Procx2e_International_Conference_on_Functional_Programming_x28ICFPx292010doix3a10x2e1145x2f1932681x2e1863577x22x29}\Autobibentry{Ryan Culpepper and Matthias Felleisen. Fortifying macros. In \textit{Proc. International Conference on Functional Programming (ICFP)}, 2010. \pseudodoi{doi:\href{https://doi.org/10.1145/1932681.1863577}{10{\hbox{\texttt{.}}}1145/1932681{\hbox{\texttt{.}}}1863577}}}}

\Autobibtarget{\label{t:x28autobib_x22Matthias_Felleisenx2c_Robert_Bruce_Findlerx2c_and_Matthew_FlattSemantics_Engineering_with_PLT_Redex2009httpsx3ax2fx2fmitpressx2emitx2eedux2fbooksx2fsemanticsx2dengineeringx2dpltx2dredexx22x29}\Autobibentry{Matthias Felleisen, Robert Bruce Findler, and Matthew Flatt. \textit{Semantics Engineering with PLT Redex}. 2009. \href{https://mitpress.mit.edu/books/semantics-engineering-plt-redex}{\Snolinkurl{https://mitpress.mit.edu/books/semantics-engineering-plt-redex}}}}

\Autobibtarget{\label{t:x28autobib_x22Matthias_Felleisenx2c_Robert_Bruce_Findlerx2c_Matthew_Flattx2c_Shriram_Krishnamurthix2c_Eli_Barzilayx2c_Jay_Mccarthyx2c_and_Sam_Tobinx2dHochstadtA_Programmable_Programing_LanguageCommunications_of_the_ACM_61x283x29x2c_ppx2e_62x2dx2d712018doix3a10x2e1145x2f3127323x22x29}\Autobibentry{Matthias Felleisen, Robert Bruce Findler, Matthew Flatt, Shriram Krishnamurthi, Eli Barzilay, Jay Mccarthy, and Sam Tobin{-}Hochstadt. A Programmable Programing Language. \textit{Communications of the ACM} 61(3), pp. 62{--}71, 2018. \pseudodoi{doi:\href{https://doi.org/10.1145/3127323}{10{\hbox{\texttt{.}}}1145/3127323}}}}

\Autobibtarget{\label{t:x28autobib_x22Matthias_Felleisenx2c_Robert_Bruce_Findlerx2c_Matthew_Flattx2c_Shriram_Krishnamurthix2c_Eli_Barzilayx2c_Jay_Ax2e_McCarthyx2c_and_Sam_Tobinx2dHochstadtThe_Racket_ManifestoIn_Procx2e_SNAPL2015doix3a10x2e4230x2fLIPIcsx2eSNAPLx2e2015x2e113x22x29}\Autobibentry{Matthias Felleisen, Robert Bruce Findler, Matthew Flatt, Shriram Krishnamurthi, Eli Barzilay, Jay A. McCarthy, and Sam Tobin{-}Hochstadt. The Racket Manifesto. In \textit{Proc. SNAPL}, 2015. \pseudodoi{doi:\href{https://doi.org/10.4230/LIPIcs.SNAPL.2015.113}{10{\hbox{\texttt{.}}}4230/LIPIcs{\hbox{\texttt{.}}}SNAPL{\hbox{\texttt{.}}}2015{\hbox{\texttt{.}}}113}}}}

\Autobibtarget{\label{t:x28autobib_x22Robert_Bruce_Findler_and_Matthias_FelleisenContracts_for_higherx2dorder_functionsIn_Procx2e_International_Conference_on_Functional_Programming_x28ICFPx292002doix3a10x2e1145x2f581478x2e581484x22x29}\Autobibentry{Robert Bruce Findler and Matthias Felleisen. Contracts for higher-order functions. In \textit{Proc. International Conference on Functional Programming (ICFP)}, 2002. \pseudodoi{doi:\href{https://doi.org/10.1145/581478.581484}{10{\hbox{\texttt{.}}}1145/581478{\hbox{\texttt{.}}}581484}}}}

\Autobibtarget{\label{t:x28autobib_x22Matthew_Flattx2c_Eli_Barzilayx2c_and_Robert_Bruce_FindlerScribblex3a_Closing_the_Book_on_Ad_Hoc_Documentation_ToolsIn_Procx2e_International_Conference_on_Functional_Programming_x28ICFPx292009doix3a10x2e1145x2f1596550x2e1596569x22x29}\Autobibentry{Matthew Flatt, Eli Barzilay, and Robert Bruce Findler. Scribble: Closing the Book on Ad Hoc Documentation Tools. In \textit{Proc. International Conference on Functional Programming (ICFP)}, 2009. \pseudodoi{doi:\href{https://doi.org/10.1145/1596550.1596569}{10{\hbox{\texttt{.}}}1145/1596550{\hbox{\texttt{.}}}1596569}}}}

\Autobibtarget{\label{t:x28autobib_x22Abdulaziz_GhuloumAn_Incremental_Approach_to_Compiler_ConstructionIn_Procx2e_Scheme_Workshop2006httpx3ax2fx2fscheme2006x2ecsx2euchicagox2eedux2f11x2dghuloumx2epdfx22x29}\Autobibentry{Abdulaziz Ghuloum. An Incremental Approach to Compiler Construction. In \textit{Proc. Scheme Workshop}, 2006. \href{http://scheme2006.cs.uchicago.edu/11-ghuloum.pdf}{\Snolinkurl{http://scheme2006.cs.uchicago.edu/11-ghuloum.pdf}}}}

\Autobibtarget{\label{t:x28autobib_x22Andrew_Wx2e_Keep_and_Rx2e_Kent_DybvigA_nanopass_framework_for_commercial_compiler_developmentIn_Procx2e_International_Conference_on_Functional_Programming_x28ICFPx292013doix3a10x2e1145x2f2500365x2e2500618x22x29}\Autobibentry{Andrew W. Keep and R. Kent Dybvig. A nanopass framework for commercial compiler development. In \textit{Proc. International Conference on Functional Programming (ICFP)}, 2013. \pseudodoi{doi:\href{https://doi.org/10.1145/2500365.2500618}{10{\hbox{\texttt{.}}}1145/2500365{\hbox{\texttt{.}}}2500618}}}}

\Autobibtarget{\label{t:x28autobib_x22Casey_Kleinx2c_John_Clementsx2c_Christos_Dimoulasx2c_Carl_Eastlundx2c_Matthias_Felleisenx2c_Matthew_Flattx2c_Jay_Ax2e_McCarthyx2c_Jon_Rafkindx2c_Sam_Tobinx2dHochstadtx2c_and_Robert_Bruce_FindlerRun_your_researchx3a_on_the_effectiveness_of_lightweight_mechanizationIn_Procx2e_Symposium_on_Principles_of_Programming_Languages_x28POPLx292012doix3a10x2e1145x2f2103656x2e2103691x22x29}\Autobibentry{Casey Klein, John Clements, Christos Dimoulas, Carl Eastlund, Matthias Felleisen, Matthew Flatt, Jay A. McCarthy, Jon Rafkind, Sam Tobin{-}Hochstadt, and Robert Bruce Findler. Run your research: on the effectiveness of lightweight mechanization. In \textit{Proc. Symposium on Principles of Programming Languages (POPL)}, 2012. \pseudodoi{doi:\href{https://doi.org/10.1145/2103656.2103691}{10{\hbox{\texttt{.}}}1145/2103656{\hbox{\texttt{.}}}2103691}}}}

\Autobibtarget{\label{t:x28autobib_x22Daniel_Pattersonx2c_Jamie_Percontix2c_Christos_Dimoulasx2c_and_Amal_AhmedFunTALx3a_Reasonably_Mixing_a_Functional_Language_with_AssemblyIn_Procx2e_International_Conference_on_Programming_Language_Design_and_Implementation_x28PLDIx292017doix3a10x2e1145x2f3062341x2e3062347x22x29}\Autobibentry{Daniel Patterson, Jamie Perconti, Christos Dimoulas, and Amal Ahmed. FunTAL: Reasonably Mixing a Functional Language with Assembly. In \textit{Proc. International Conference on Programming Language Design and Implementation (PLDI)}, 2017. \pseudodoi{doi:\href{https://doi.org/10.1145/3062341.3062347}{10{\hbox{\texttt{.}}}1145/3062341{\hbox{\texttt{.}}}3062347}}}}

\Autobibtarget{\label{t:x28autobib_x22Tx2e_Stephen_Stricklandx2c_Sam_Tobinx2dHochstadtx2c_Robert_Bruce_Findlerx2c_and_Matthew_FlattChaperones_and_impersonatorsx3a_runx2dtime_support_for_reasonable_interpositionIn_Procx2e_Conference_on_Objectx2dOriented_Programmingx2c_Systemsx2c_Languagesx2c_and_Applications_x28OOPSLAx292012doix3a10x2e1145x2f2384616x2e2384685x22x29}\Autobibentry{T. Stephen Strickland, Sam Tobin{-}Hochstadt, Robert Bruce Findler, and Matthew Flatt. Chaperones and impersonators: run-time support for reasonable interposition. In \textit{Proc. Conference on Object{-}Oriented Programming, Systems, Languages, and Applications (OOPSLA)}, 2012. \pseudodoi{doi:\href{https://doi.org/10.1145/2384616.2384685}{10{\hbox{\texttt{.}}}1145/2384616{\hbox{\texttt{.}}}2384685}}}}

\Autobibtarget{\label{t:x28autobib_x22Sam_Tobinx2dHochstadt_and_Matthias_FelleisenThe_design_and_implementation_of_typed_schemeIn_Procx2e_Symposium_on_Principles_of_Programming_Languages_x28POPLx292008doix3a10x2e1145x2f1328438x2e1328486x22x29}\Autobibentry{Sam Tobin{-}Hochstadt and Matthias Felleisen. The design and implementation of typed scheme. In \textit{Proc. Symposium on Principles of Programming Languages (POPL)}, 2008. \pseudodoi{doi:\href{https://doi.org/10.1145/1328438.1328486}{10{\hbox{\texttt{.}}}1145/1328438{\hbox{\texttt{.}}}1328486}}}}

\Autobibtarget{\label{t:x28autobib_x22Sam_Tobinx2dHochstadt_and_Matthias_FelleisenThe_Design_and_Implementation_of_Typed_Schemex3a_From_Scripts_to_ProgramsCoRR_absx2f1106x2e25752011doix3a10x2e48550x2farXivx2e1106x2e2575x22x29}\Autobibentry{Sam Tobin{-}Hochstadt and Matthias Felleisen. The Design and Implementation of Typed Scheme: From Scripts to Programs. \textit{CoRR} abs/1106.2575, 2011. \pseudodoi{doi:\href{https://doi.org/10.48550/arXiv.1106.2575}{10{\hbox{\texttt{.}}}48550/arXiv{\hbox{\texttt{.}}}1106{\hbox{\texttt{.}}}2575}}}}

\Autobibtarget{\label{t:x28autobib_x22Sam_Tobinx2dHochstadtx2c_Vincent_Stx2dAmourx2c_Ryan_Culpepperx2c_Matthew_Flattx2c_and_Matthias_FelleisenLanguages_As_LibrariesIn_Procx2e_International_Conference_on_Programming_Language_Design_and_Implementation_x28PLDIx292011doix3a10x2e1145x2f1993498x2e1993514x22x29}\Autobibentry{Sam Tobin{-}Hochstadt, Vincent St{-}Amour, Ryan Culpepper, Matthew Flatt, and Matthias Felleisen. Languages As Libraries. In \textit{Proc. International Conference on Programming Language Design and Implementation (PLDI)}, 2011. \pseudodoi{doi:\href{https://doi.org/10.1145/1993498.1993514}{10{\hbox{\texttt{.}}}1145/1993498{\hbox{\texttt{.}}}1993514}}}}\end{SingleColumn}\end{AutoBibliography}

\postDoc
\end{document}